\documentclass[11pt]{article}
\usepackage{caption}
\usepackage{scalerel}
\usepackage{amsthm,latexsym,amsxtra,graphicx,appendix,epstopdf,feynmf,hyperref,setspace,fix-cm}
\usepackage[normalem]{ulem}
\usepackage{makeidx}
\usepackage{fullpage}
\usepackage{amsmath}
\usepackage{amssymb}
\usepackage{setspace}
\usepackage{bbm}
\usepackage{dsfont}
\usepackage{epsfig}
\usepackage[font=footnotesize,labelfont=bf,justification=centerlast,width=.94\textwidth]{caption}
\usepackage{cite}
 \usepackage{multirow}
\usepackage{array,booktabs}
\usepackage{subfigure}
\usepackage{mathtools}
\usepackage{color}
\usepackage[makeroom]{cancel}
\usepackage{tikz}
\usepackage{pgfplots}
\usetikzlibrary{intersections, pgfplots.fillbetween}
\usepackage{tensor}
\usepackage{simplewick}
\usepackage{frcursive}

\usepackage{pgfornament,multicol}

\usepackage{caption}
\usepackage{hyperref}

\usepackage[skip=0.2cm, indent=0pt]{parskip}

\hypersetup{
    bookmarks=true,%
    colorlinks,%
    citecolor=blue,%
    filecolor=blue,%
    linkcolor=blue,%
    urlcolor=blue
}
\usepackage{float}
\usepackage{slashed}
\usepackage{tikz}
\usetikzlibrary{decorations.pathmorphing}

\def\r{\rightarrow}

\newcommand{\bi}{\begin{itemize}}
\newcommand{\ei}{\end{itemize}}
\newcommand{\bea}{\begin{eqnarray}}
\newcommand{\eea}{\end{eqnarray}}
\newcommand{\be}{\begin{equation}}
\newcommand{\ee}{\end{equation}}

\def\r{\mathfrak{r}}

\def\bomega{{\boldsymbol{\omega}}}

\def\XXint#1#2#3{{\setbox0=\hbox{$#1{#2#3}{\int}$}
     \vcenter{\hbox{$#2#3$}}\kern-.5\wd0}}

\def\={\, = \,}

\numberwithin{equation}{section}
\allowdisplaybreaks  

\begin{document}

\vspace*{2.5cm}
\begin{center}
{ \LARGE {Gravitational Observatories in AdS$_4$} \\ \vspace*{1cm}}

\end{center}

\begin{center}
Dionysios Anninos,$^{1,2}$ Raúl Arias,$^{3}$ Dami\'an A. Galante,$^{1}$ and Chawakorn Maneerat$^{1}$
\end{center}
\begin{center}
{
\footnotesize
\vspace{0.4cm}
$^1$Department of Mathematics, King's College London, Strand, London WC2R 2LS, UK
\\
{$^2$ Instituut voor Theoretische Fysica, KU Leuven, Celestijnenlaan 200D, B-3001 Leuven, Belgium} 
\\
{$^3$ Instituto de F\'isica La Plata - CONICET and 
Departamento de F\'isica, Universidad Nacional de La Plata C.C. 67, 1900, La Plata, Argentina}}
\end{center}
\begin{center}
{\textsf{\footnotesize{
dionysios.anninos@kcl.ac.uk, rarias@fisica.unlp.edu.ar, damian.galante@kcl.ac.uk, chawakorn.maneerat@kcl.ac.uk}} } 
\end{center}

\vspace*{0.5cm}

\vspace*{1.5cm}
\begin{abstract}
\noindent
We consider four-dimensional general relativity with a negative cosmological constant in the presence of a finite size boundary, $\Gamma$, for both Euclidean and Lorentzian signature. As our boundary condition, we consider the `conformal' boundary condition that fixes the conformal class of the induced metric at $\Gamma$ and the trace of the extrinsic curvature, $K(x^m)$. In Lorentzian signature, we must supplement these with appropriate initial data comprising the standard Cauchy data along a spatial slice and, in addition, initial data for a boundary mode that appears due to the presence of the finite size boundary. We perform a linearised analysis of the gravitational field equations for both an $S^2\times \mathbb{R}$ as well as a Minkowskian, $\mathbb{R}^{2,1}$, boundary. In the $S^2\times \mathbb{R}$ case, in addition to the usual AdS$_4$ normal modes, we uncover a novel  linearised perturbation, $\bomega(x^m)$, which can exhibit complex frequencies at sufficiently large angular momentum. Upon moving $\Gamma$ toward the infinite asymptotic AdS$_4$ boundary, the complex frequencies appear at increasingly large angular momentum and vanish altogether in the strict limit. In the $\mathbb{R}^{2,1}$ case, although we uncover an analogous novel perturbation, we show it does not exhibit complex frequencies. In Euclidean signature, we show that $K(x^m)$ plays the role of a source for $\bomega(x^m)$. {When close to the AdS$_4$ asymptotic boundary, we speculate on the holographic interpretation of $\bomega(x^m)$.}

\end{abstract}

\newpage

\setcounter{tocdepth}{2}
\tableofcontents

\newpage

\section{Introduction}
\label{section: Introduction}

The class of four-dimensional asymptotically anti-de Sitter solutions to the Einstein field equations with negative cosmological constant $\Lambda=-\tfrac{3}{\ell^2}$  admit the systematic expansion 
\begin{equation}\label{fg}
    \frac{ds^2}{\ell^2} = d\rho^2 + e^{2\rho}\left(g^{(0)}_{mn} + e^{-2\rho}g^{(2)}_{mn} + e^{-3\rho} g^{(3)}_{mn}  + \mathcal{O}(e^{-4\rho}) \right)dx^m dx^n \, ,
\end{equation}
first uncovered by Fefferman and Graham \cite{fefferman1985conformal}. The asymptotic AdS$_4$ boundary, $\Gamma_\infty$, is located at ${\rho} = \infty$. The asymptotic boundary, $\Gamma_\infty$, is endowed with asymptotic data  composed of an equivalence class under conformal transformations
\begin{equation}\label{Omega}
\{g^{(0)}_{mn},g^{(3)}_{mn} \} \sim \{\Omega^2(x^m) g^{(0)}_{mn},\Omega^{-1}(x^m) g^{(3)}_{mn} \}~,
\end{equation}
where $\Omega(x^m)$ is a smooth positive function of the boundary coordinates $x^m$. From the perspective of the AdS$_4$/CFT$_3$ correspondence, this is a reflection of the fact that a three-dimensional conformal field theory naturally lives on a conformal metric, which we occasionally denote as $[g^{(0)}_{mn}]$ to indicate it contains $g^{(0)}_{mn}$ as an instance. Moreover, the Einstein equations impose that $g^{(3)}_{mn}$ is transverse and traceless with respect to $g^{(0)}_{mn}$, and that $g^{(2)}_{mn}$ is fixed entirely in terms of  $g^{(0)}_{mn}$. Another general feature of the expansion (\ref{fg}), which is somewhat less often emphasised, is that the trace of the extrinsic curvature, $K(x^m)$, along $\Gamma_\infty$ takes the fixed value $K \ell = 3$ regardless of the choice $\{g^{(0)}_{mn},g^{(3)}_{mn} \}$. 

In Lorentzian signature, physically distinct configurations carrying different stress-energy are registered by changes in $g^{(3)}_{mn}$. The two independent functions in $g^{(3)}_{mn}$ encode the two locally propagating gravitational degrees of freedom. Going from one conformal frame in (\ref{Omega}) to another can be achieved by a coordinate transformation that preserves the form of (\ref{fg}). As such, $\Omega(x^m)$ does not carry physical information about the configuration --- it is pure gauge. To develop a particular initial condition, we must fix Cauchy data along a complete spatial slice $\Sigma$, which can be taken to be the set $\mathcal{C}_\Sigma = \{\tilde{g}_{ij},\tilde{K}_{ij}\}$ given by the induced metric and extrinsic curvature along the spatial slice $\Sigma$ equipped with coordinates $x^i$. The Cauchy data $\mathcal{C}_\Sigma$ must further satisfy the gravitational constraint equations along $\Sigma$, and is identified under tangential diffeomorphisms mapping $\Sigma$ to itself. The conformal structure of the boundary metric $g^{(0)}_{mn}$ is fixed data specific to a  particular theory. In Euclidean signature, one views  $g^{(0)}_{mn}$ as a source for the boundary stress-tensor \cite{Witten:1998qj}, and one seeks instead for a smooth solution of the Euclidean equations with the prescribed asymptotic data.

In this article, we are concerned with the question of how to move the asymptotic boundary $\Gamma_\infty$ into the interior of four-dimensional anti-de Sitter space (AdS$_4$). Our analysis is inherently gravitational, with the goal of acquiring sufficient theoretical ground that might guide us toward a finite-size generalisation of the AdS$_4$/CFT$_3$ dictionary. Previous works on AdS$_4$/CFT$_3$ at finite boundary include ideas about the holographic renormalisation group \cite{Akhmedov:1998vf,deBoer:1999tgo,Lee:2013dln,Heemskerk:2010hk,Skenderis:2002wp}, as well as more recent discussions on a  higher-dimensional version of the $T\bar{T}$ deformation \cite{Hartman:2018tkw,Taylor:2018xcy,Araujo-Regado:2022gvw,Silverstein:2024xnr}. 

At the face of it, and as originally envisioned by York for the Minkowskian case \cite{York:1986it}, one might be inclined to bring in the asymptotic boundary $\Gamma_\infty$ to a large but finite value of $\rho$ and fixing the entire induced metric $g_{mn}$ along a finite-size timelike boundary $\Gamma$, along with some Cauchy data along a spatial slice $\Sigma$ that intersects $\Gamma$ along a two-dimensional spatial boundary $\partial \Sigma$ \cite{Marolf:2012dr,Andrade:2015gja}. In doing so, and following \cite{An:2021fcq}, we should ensure that the Einstein constraint equation projected  along $\Gamma$ is satisfied, namely,
\begin{equation}\label{cG}
\mathcal{R} - K^2 + K_{mn} K^{mn} - 2\Lambda |_{\Gamma} = 0~,
\end{equation}
where $\mathcal{R}$ is the Ricci scalar with respect to the induced metric $g_{mn}$. 
Parameterically close to the AdS$_4$ boundary, there is no immediate obstruction to the Dirichlet problem.  An argument for this is presented in the linearised approximation about a planar AdS$_4$ background in appendix \ref{app: Dirichlet problem}.  On the other hand, it was established in \cite{Avramidi:1997sh,Anderson_2008,An:2021fcq} that when the boundary $\Gamma$ resides in an approximately Minkwoskian or Rindler corner, the constraint equation (\ref{cG}) will not be satisfied for generic $g_{mn}$. (It should be noted, that although not generic, there do exist many solutions to the Dirichlet problem as well. These include the important case of an $S^2\times \mathbb{R}$ boundary.)

Instead of fixing Dirichlet data along $\Gamma$, bearing in mind that the entire Fefferman-Graham configuration space obeys  $K\ell=3$ at $\Gamma_\infty$, one may consider \cite{York:1986lje} fixing the conformal class of the induced metric, $[g_{mn}]$, at $\Gamma$ along with the trace of the extrinsic curvature, $K(x^m)$. Such a boundary condition, which we will refer to as the conformal boundary condition, has been the subject of recent work \cite{Witten:2018lgb,Anninos:2023epi,Anninos:2024wpy,Liu:2024ymn,Banihashemi:2024yye}, whilst also having appeared previously in the fluid-gravity literature \cite{Bredberg:2011xw,Anninos:2011zn}. From the mathematical side, conformal boundary conditions were proven in  \cite{Anderson_2008} to constitute a well-posed elliptic problem in  Euclidean signature. In Lorentzian signature conformal boundary conditions, accompanied by appropriate Cauchy data along some initial time slice $\Sigma$, have been conjectured  \cite{An:2021fcq} to constitute a well-posed initial boundary value problem (see also \cite{An:2024taa,Capoferri:2024sgo} for recent developments). The corresponding gravitational phase space and Brown-York boundary stress-tensor, $T_{mn}$, have been considered in \cite{Odak:2021axr}. Interestingly, $T_{mn}$ is traceless and, in the case of constant $K$, conserved. 

In what follows we will analyse  general relativity with a negative cosmological constant $\Lambda=-\tfrac{3}{\ell^2}$ on a manifold with a finite size timelike boundary subject to conformal boundary conditions. Schematically, the location of the boundary is controlled by $K\ell$. The  shape of the boundary $\Gamma$ is instead controlled by the conformal class $[g_{mn}]$, which is fixed, along with a novel dynamical boundary degree of freedom $\bomega(x^m)$ that comes to life upon bringing in the asymptotic AdS$_4$ boundary.\footnote{The appearance of a dynamical boundary mode is somewhat reminiscent of what happens in topological quantum field theories when quantised on a manifold with a boundary \cite{Elitzur:1989nr}. In the context at hand, a linearised analysis around the Minkowski corner already indicates the existence of a boundary/corner mode that cannot be undone with a permissible diffeomorphism. This was studied in section 4.1.2 of \cite{Anninos:2023epi} for the de Donder gauge. Relatedly, in appendix B of \cite{Anninos:2022ujl} it was shown in the de Donder gauge and about the Minkowski corner, that upon fixing the entire initial value data set along a Cauchy surface (including the initial data of the boundary/corner mode) to vanish fixes the linearised evolution uniquely for conformal boundary conditions. These properties can be confirmed for other gauge choices also. Relatedly, for a spherical spatial boundary, it was shown in \cite{Anninos:2023epi,Anninos:2024wpy,Liu:2024ymn} that there exists a spherically symmetric physical diffeomorphism (see (5.23) of \cite{Anninos:2023epi} for the infinitesimal case) whose initial data must further be fixed to ensure a unique evolution of the initial boundary value problem. The spherically symmetric diffeomorphisms can be treated at the non-linear level also (see footnote 2 of \cite{Anninos:2024wpy} and the more detailed section 4.1 of \cite{Liu:2024ymn}). A complete proof of well-posedness (or lack thereof) of the conformal boundary conditions must necessarily incorporate the boundary/corner mode at the non-linear level. It remains a very interesting open problem. We would like to thank M. Anderson and L. Lehner for illuminating discussions on this point.} Interestingly, although bringing in the AdS$_4$ boundary might be envisioned as the result integrating out degrees of freedom of the dual CFT$_3$, a new degree of freedom appears in the gravitational phase space. Moreover, $\bomega(x^m)$ interacts non-trivially with the bulk gravitational degrees of freedom. From the perspective of AdS$_4$/CFT$_3$, based on our analysis we anticipate that bringing in the asymptotic boundary invokes at least two novel features. Firstly,  a new dynamical field $\bomega(x^m)$, that is not part of the original CFT$_3$, appears in the theory. Secondly, the conformal invariance of the CFT$_3$ on $S^2\times \mathbb{R}$ appears to be broken due to the presence of growing modes in the linearised spectrum about global AdS$_4$. From the gravitational side, the interaction strength between $\bomega(x^m)$ and the CFT$_3$ fields --- or, similarly, the degree of conformal symmetric breaking --- is governed by the deviation, $\delta K = K-\tfrac{3}{\ell}$, in the trace of the extrinsic curvature away from its asymptotic boundary value. An alternative somewhat more specialised boundary condition which has been shown to have good properties for $\Lambda \le 0$, is the `umbilical' boundary condition \cite{schlenker2001einstein,Fournodavlos:2021eye} whereby the induced metric at the boundary is taken to be proportional to the extrinsic curvature.

From a thermodynamic point of view, one can consider computing the horizon  entropy of a large Schwarzschild AdS$_4$ black hole subject to conformal boundary conditions, as analysed in \cite{Coleman:2020jte,Anninos:2023epi,Banihashemi:2024yye}. In a high-temperature limit, and taking $K\ell > 3$ to be constant,  the leading order contribution to the Bekenstein-Hawking relation reads \cite{Anninos:2023epi,Banihashemi:2024yye}
\begin{equation}\label{BH}
S_{\text{BH}} = \frac{16\pi^3 \ell^2}{81  G_N} \left( K \ell - \sqrt{K^2\ell^2-9} \right)^2 {\beta^{-2}}~,
\end{equation} 
where ${\beta}$ is the (conformally invariant) inverse Hawking temperature as measured by the boundary clock. As a function of $\beta$, the above behaves as the thermal entropy of a three-dimensional conformal field theory at high-temperature. The pre-factor of $\beta^{-2}$ is a monotonically decreasing function of $K\ell$, that tends to the Minkowskian value $\tfrac{4\pi^3}{G_N K^2}$ \cite{Anninos:2023epi} at large $K\ell$, and to the standard AdS$_4$/CFT$_3$ pre-factor $ \tfrac{16 \ell^2 \pi^3}{9 G_N}$ in the limit $K\ell \to 3$. The pre-factor is unaltered for the planar AdS$_4$ black brane with conformal boundary conditions also. We might view the pre-factor, then, as a generalised measure of the number of degrees of freedom, at least in some semiclassical sense, for the putative theory dual to the gravitational theory with finite size conformal boundary. Unlike a Dirichlet boundary, there is no obstruction in making the size of the horizon arbitrarily large for the case at hand.\footnote{Interestingly, a similar expression to (\ref{BH}) holds for a de Sitter horizon surrounding a timelike boundary subject to conformal boundary conditions \cite{Anninos:2024wpy}, yielding a positive prefactor irrespective of the sign of $K \ell_{\text{dS}}$. In contrast to the Dirichlet case, this indicates that the de Sitter horizon subject to conformal boundary conditions can have a positive specific heat.}

As an alternative measure of the number of degrees of freedom, we might consider the semiclassical gravitational path integral for Euclidean AdS$_4$ with an $S^3$ boundary (see for example \cite{Jafferis:2011zi}). For conformal boundary data given by the conformal class containing the round metric on $S^3$, and a $K$ which is independent of $x^m$, a straightforward computation for the saddle point approximation of the gravitational path integral yields
\begin{equation}\label{Zs3}
\log \mathcal{Z}[S^3,K] =  \frac{\pi \ell^2}{2 G_N}   \left(\frac{{K\ell}-\sqrt{K^2 \ell^2-9}}{\sqrt{K^2 \ell^2-9}}\right) \approx \frac{\pi \ell^2}{2 G_N}   \left( \sqrt{\frac{3}{2(K\ell-3)}}-1 + \ldots \right)~.
\end{equation}
We note that $\log \mathcal{Z}[S^3,K]$ is a monotonically decreasing function of $K\ell$. Once we deviate from $K\ell=3$, the divergences that are ordinarily removed by local counterterms in AdS$_4$/CFT$_3$ are subsumed into the unified expression (\ref{Zs3}). In ordinary AdS$_4$/CFT$_3$, the divergent term in the final expansion of (\ref{Zs3}) corresponds to an ultraviolet divergence that can be removed with a standard local counterterm \cite{Skenderis:2002wp}. Perhaps away from $K\ell=3$, there is a different perspective on this, that no longer allows us to discard such terms and instead they are combined into a unified picture. This might tie with the question of whether the holographic theory describing an anti-de Sitter universe with a finite-size is a non-local theory --- if so, attempting to disentangle (would be) ultraviolet divergences from other effects might be somewhat unfounded.

The outline of the paper is as follows. In section \ref{sec: framework}, we introduce the general framework for our analysis. We discuss the Brown-York stress energy tensor, $T_{mn}$, pertinent to the conformal boundary conditions \cite{Odak:2021axr}, and make contact with the boundary stress tensor \cite{Balasubramanian:1999re} at the asymptotic AdS$_4$ boundary. In section \ref{sec: lin dynamics global AdS}, we analyse the linearised gravitational equations subject to conformal boundary conditions about the global AdS$_4$ spacetime. In addition to a collection of ordinary gravitational normal modes, we uncover a boundary mode, ${\bomega(x^m)}$. Along with its time derivative, ${\bomega(x^m)}$ is part of the phase space of the gravitational theory. For sufficiently large angular momentum, we find a collection of modes with complex frequency that generalise those uncovered in the Minkowskian analysis of \cite{Anninos:2023epi}. When the boundary, $\Gamma$, is pushed closer to the asymptotic boundary, $\Gamma_\infty$, the complex modes appear at increasingly large angular momentum. In section \ref{sec: planar AdS}, we analyse the linearised gravitational equations subject to conformal boundary conditions about the planar AdS$_4$ spacetime. In this case, we uncover a massless boundary mode, ${\bomega(x^m)}$,  which can now be expressed (locally) as a diffeomorphism. Unlike the case of global AdS$_4$, we find that there are no exponentially growing modes for the planar case. We end the section by analysing the $\bomega(x^m)$ mode about an AdS$_4$ black brane background, and show it can dress the black hole in a way that modifies its energy. In section \ref{sec: source}, we consider the Euclidean problem. In particular, we show that the boundary datum $K({x}^m)$ acts as a source for $\bomega(x^m)$ --- extending the usual Euclidean AdS$_4$/CFT$_3$ dictionary. The output of the Euclidean gravitational path integral will now be a functional, $\mathcal{Z}[[g_{mn}(x^m)], K(x^m)]$, of the conformal structure, $[g_{mn}]$, and the mean curvature, $K(x^m)$, along $\Gamma$. Details of various computations and technical derivations, can be found in the several appendices.

\section{General framework}
\label{sec: framework}

In this section we give a brief overview of the general framework for the conformal boundary condition problem. Our discussion follows that of \cite{Anninos:2023epi, Anninos:2024wpy}, where a more detailed account can be found. 

\subsection{Initial boundary value problem}

Our Lorentzian spacetime action is given by
\begin{equation}\label{bulkS}
I = \frac{1}{16\pi G_N} \int_{\mathcal{M}} d^4 x \sqrt{-\det g_{\mu\nu}} \left( R  - 2\Lambda \right) + \frac{1}{24\pi G_N} \int_\Gamma d^3 x \sqrt{-\det g_{mn}} K~.
\end{equation}
The timelike boundary is denoted by $\Gamma$, and is endowed with the induced metric $g_{mn}$. The quantity $K = g^{mn} K_{mn}$, is the trace of the extrinsic curvature $K_{mn} = \tfrac{1}{2} \mathcal{L}_{n^\mu} g_{mn}$ with respect to the outward-pointing unit normal $n^\mu$. Our cosmological constant $\Lambda=-\tfrac{3}{\ell^2}$ is taken to be negative in what follows.  The boundary conditions we will impose fix the conformal class of the induced metric at the timelike boundary, $[g_{mn}]|_\Gamma$, as well as $K$ \cite{An:2021fcq}, and we denote these as conformal boundary conditions in what follows. The boundary term in (\ref{bulkS}), which differs from that of the standard Gibbons-Hawking-York term by a coefficient,  has been adjusted so as to permit for a well-defined variational problem \cite{An:2021fcq, Odak:2021axr} with respect to the conformal boundary conditions. Coordinates along the timelike boundary $\Gamma$ will often be denoted by $x^m$. 

The conformal boundary conditions along $\Gamma$ must be supplied with Cauchy data along a spatial initial time slice $\Sigma$. Part of the Cauchy data is comprised of the induced metric, $\tilde{g}_{ij}$ along $\Sigma$, as well as the extrinsic curvature, $\tilde{K}_{ij}$, with respect to the unit normal $\tilde{n}^\mu$ of $\Sigma$. The Cauchy data $\mathcal{C}_\Sigma = \{\tilde{g}_{ij},\tilde{K}_{ij}\}_{\Sigma}$ must be subjected to the gravitational constraint equations along $\Sigma$, and forms an equivalence class under tangential diffeomorphisms that preserve the conformal boundary conditions and map $\Sigma$ to itself. We must further ensure that the fields at the boundary $\partial\Sigma$ of $\Sigma$ are compatible with the boundary data along $\Gamma$ that intersects with $\partial\Sigma$.

In addition, we must specify the initial conditions for any independent boundary data residing $\partial \Sigma$. As will appear in our linearised analysis below, the boundary data is encoded by the volume element $\sqrt{\det \tilde{g}_{ij}}|_{\partial{\Sigma}}$ and the corner angle $\beta \equiv \sinh^{-1}{n^\mu \tilde{n}_\mu|_{\partial\Sigma}}$.  To properly do so, as will be detailed below, we must supplement the Cauchy data along $\Sigma$ with the initial data of a boundary mode, $\bomega(x^m)$, residing at $\partial\Sigma$ \cite{Odak:2021axr,elsewhere}.\footnote{The reason that there is independent data at $\partial\Sigma$ --- itself not part of $\mathcal{C}_\Sigma$ --- is that unlike the case of a complete Cauchy surface, we are now quotienting configurations along $\Sigma$ by a restriction of the tangential diffeomorphisms. This allows for the possibility of non-trivial contributions to the configuration space that may locally appear as a pure diffeomorphism. Explicit examples of such configurations will be presented in the following section.} Generically, when $\Sigma$ intersects $\Gamma$ at a non-perpendicular angle, an additional corner term (known as the Hayward term \cite{Hayward:1990zm}) must be supplemented to the action \cite{Lehner:2016vdi}. The coefficient of the Hayward term was shown to vanish for the case of conformal boundary conditions \cite{Odak:2021axr}.  A careful examination of these boundary degrees of freedom will be presented elsewhere \cite{elsewhere}, and previous work can be found in \cite{Odak:2021axr,An:2021fcq}.

\textbf{Linearised initial boundary value problem.} In much of what follows, we will often be considering the problem at the linearised level. To this end, consider a metric that is slightly perturbed away from a background $\bar{g}_{\mu\nu}$, such that
\begin{equation} \label{metric_pert_lin}
g_{\mu\nu} = \bar{g}_{\mu\nu} + \varepsilon h_{\mu\nu}~, \quad\quad |\varepsilon| \ll 1~.
\end{equation}
Unless otherwise specified, our background metric is taken to be of the form
\begin{equation}\label{gbar}
ds^2 = -f(r) dt^2 + \frac{dr^2}{f(r)} + r^2 d\Sigma_k~,
\end{equation}
where $\Sigma_k$ is a maximally symmetric two dimensional space of curvature $k= \{-1,0,1\}$. We will only focus on $k=\{0,1\}$ in what follows, leaving the case $k=-1$ for future work. In the case of $k=+1$, $d\Sigma_{+1} = d\Omega_2^2$ is the round metric on the unit two-sphere.

We take the boundary to be located along a surface of constant coordinate $r=\mathfrak{r}$. As boundary conditions, we take that the induced metric at the boundary is of the form
\begin{equation}\label{indmetric}
ds^2|_\Gamma =  e^{2\boldsymbol{\omega}(x^m)}\left(-f(\r) dt^2  + \r^2 d\Sigma_k \right)~,
\end{equation}
where $\boldsymbol{\omega}(x^m)$ is an unspecified function of the boundary coordinates, whilst the trace of the extrinsic curvature at the boundary is fixed to be $K$.

To preserve the conformal structure of the induced metric at the boundary, we must enforce that
\begin{equation}\label{eqn: bdry cond 1}
	\left.h_{mn} - \frac{1}{3}\bar{g}_{mn} h^p{}_p \right|_{\Gamma} \= 0 \, .
\end{equation}
The variation of the trace of the extrinsic curvature to first order in $\varepsilon$ is given by
\begin{equation}\label{eqn: bdry cond 2}
	\left.\frac{\delta K (h_{\mu\nu})}{\sqrt{f(r)}}\right|_{\Gamma} \equiv \left.\frac{K(\bar{g}_{\mu\nu}+\varepsilon h_{\mu\nu}) - K(\bar{g}_{\mu\nu})}{\varepsilon\sqrt{f(r)}} \right|_{\Gamma}= \left.\frac{1}{2} \partial_r h^m{}_m - \mathcal{D}^m h_{rm} - \frac{\sqrt{f(r)}}{2}K h_{rr} \right|_{\Gamma} = 0 \, ,
\end{equation}
where $\mathcal{D}_n$ denotes the covariant derivative with respect to the background metric $\bar{g}_{mn}$ induced at the timelike boundary. 

In order to fully specify the initial boundary value problem, we must also specify boundary conditions for the gauge parameter of the diffeomorphisms. At the linear level one has a vector field $\xi^\mu$ transforming
\begin{equation}
\begin{cases}
    x^\mu \to x^\mu - \varepsilon \, \xi^\mu~, \\
h_{\mu\nu} \to h_{\mu\nu} + \nabla_\mu \xi_\nu + \nabla_\nu \xi_\mu~,
\end{cases}
\end{equation}
where $\nabla_\mu$ denotes the standard covariant derivative. In order for the gauge parameter to be consistent with the conformal boundary conditions, we must impose that  
\begin{equation}\label{eqn: bdry conds diffeo}
\begin{cases}
    \left.2\left(K_{mn} - \frac{K}{3}\bar{g}_{mn}\right)\left(\sqrt{f(r)}\xi_r\right) + \left(\mathcal{D}_m \xi_n + \mathcal{D}_n \xi_m - \frac{2\bar{g}_{mn}}{3}\mathcal{D}^p \xi_p\right)\right|_{\Gamma} &=\, 0 \, , \\
    \left.\left(\sqrt{f(r)}\partial_r K - \mathcal{D}^m \mathcal{D}_m \right) \left(\sqrt{f(r)}\xi_r\right) + \xi_m \mathcal{D}^m K \right|_{\Gamma} &=\, 0 \, .
\end{cases}
\end{equation}
It is worth noting that the second term in the first equation corresponds to the conformal Killing equation with respect to the boundary metric $\bar{g}_{mn}$. Additionally, allowed diffeomorphisms should satisfy $\xi^r|_{\Gamma} = 0$, such that the location of the boundary is not disturbed.

\subsection{Conformal Brown-York boundary stress tensor} \label{sec: conf by}

One can compute the Brown-York boundary stress tensor $T_{m n}$ for our boundary value problem by varying the classical on-shell action with respect to the boundary conformal metric $[g]_{mn}|_\Gamma$, whilst keeping $K$ fixed. Here,  $[g]_{mn}$ denotes a particular representative of the conformal class $[g_{mn}]$. This exercise was undertaken in \cite{Odak:2021axr,An:2021fcq}, where it is shown that the resulting stress tensor takes the following form\footnote{The precise variation of the on-shell action with respect to the conformal metric gives the following conformal Brown-York tensor density \cite{Banihashemi:2024yye}
\begin{equation}
    T_{mn}^{\text{CBY}} \equiv -\frac{2}{\sqrt{-\text{det} [g]_{mn}}} \frac{\delta I}{\delta [g]^{mn}} = - \frac{e^{\bomega}}{8\pi G_N} \left(K_{mn} - \frac{1}{3} K g_{mn} \right) \left(-\det{[g]_{mn}}\right)^{1/6} \,,
\end{equation}
where $[g]_{mn}$ is the conformal metric at the boundary, defined via  $g_{mn} \equiv e^{2\bomega} [g]_{mn}$. Note that this definition for the stress tensor density differs from our definition by an overall factor of $\left(-\det{[g]_{mn}}\right)^{1/6}$.}
\begin{equation}\label{cby}
T_{m n} = \left. -\frac{e^{\boldsymbol{\omega}}}{8\pi G_N}\left( K_{m n} - \frac{1}{3} K g_{mn} \right) \right|_\Gamma ~.
\end{equation}
We note that $T_{mn}$ is manifestly traceless, and it transforms with conformal weight $-1$ under a Weyl transformation of the conformal representative.

The divergence of $T_{mn}$ (with respect to  (\ref{indmetric}) with $\bomega=0$) follows from the momentum constraint at $\Gamma$. One finds 
\begin{equation} \label{eq: divergence Tmn}
    \mathcal{D}^m T_{mn} = -\frac{1}{12\pi G_N}e^{3\bomega}\mathcal{D}_{n}K \, ,
\end{equation}
which renders $T_{mn}$ divergenceless when $K$ is a constant. For completeness, the Hamiltonian constraint at $\Gamma$ reads 
\begin{equation} \label{eq: Hamiltonian constraint}
    \mathcal{D}_m \mathcal{D}^m \bomega + \frac{1}{2}\ \mathcal{D}_m\bomega \mathcal{D}^m\bomega - \frac{1}{4}R - 16 \pi^2 G_N^2 T^{mn} T_{mn}e^{-4\bomega} + \frac{1}{6}\left(K^2-\frac{9}{\ell^2}\right) e^{2\bomega} = 0 \, ,
\end{equation}
where $R$ is the Ricci scalar of the conformal representative of $[g]_{mn}|_{\Gamma}$ with $\bomega=0$.\footnote{{For $(d+1)$-dimensional pure Einstein gravity with a cosmological constant $\Lambda$, the Hamiltonian constraint along a timelike boundary $\Gamma$ is given by
\begin{equation}
    \mathcal{D}_m\mathcal{D}^m \bomega + \frac{d-2}{2}\mathcal{D}_m \bomega \mathcal{D}^m \bomega - \frac{1}{2(d-1)}R - \frac{64 \pi^2G_N^2}{2(d-1)}T^{mn}T_{mn} e^{-2(d-1)\bomega} + \left(K^2 + \frac{2d}{d-1}\Lambda\right)\frac{e^{2\bomega}}{2d}=0 \, ,
\end{equation}
where the conformal Brown-York stress tensor is $T_{mn} =\left.- \frac{e^{(d-2)\bomega}}{8\pi G_N}\left(K_{mn}-\frac{1}{d}K g_{mn}\right)\right|_{\Gamma}$.}}
Both \eqref{eq: divergence Tmn} and \eqref{eq: Hamiltonian constraint} form {a hyperbolic version of Lichnerowicz-York equations} \cite{JMPA_1944_9_23__37_0, PhysRevLett.28.1082, choquet}.

Given the stress tensor \eqref{cby}, along with \eqref{eq: divergence Tmn}, one can construct a conserved quantity associated with tangential diffeomorphisms (\ref{eqn: bdry conds diffeo}) that preserve the conformal boundary conditions as follows. Let us take the case of constant $K$, and let $\xi^m$ be a conformal Killing vector of the induced metric \eqref{indmetric} with $\bomega=0$. If we further set $\xi^r$ to zero, it follows from \eqref{eqn: bdry conds diffeo} that $\xi^m$ preserves the conformal boundary conditions. The current $J_m \equiv \xi^n T_{mn}$ obeys the conservation equation $\mathcal{D}^m J_m = 0$, as implied by \eqref{eq: divergence Tmn} and the tracelessness of $T_{mn}$.\footnote{To obtain a conserved current $J_m$, one could relax the constant $K$ condition to $\xi^m \mathcal{D}_m K = 0$ for the $\xi^m$ of interest. Since the orthogonal component of $\xi^m$ is zero, this condition is equivalent to the trace of the extrinsic curvature-preserving condition \eqref{eqn: bdry conds diffeo}.} Let $u^m$ be a timelike unit-normal vector with respect to the same metric, \eqref{indmetric} with $\bomega=0$, whose explicit form is given by $u^m \partial_m = f(\r)^{-1/2}\partial_t$. It follows that 
\begin{equation}
    Q_\xi \equiv \r^2 \left. \int  d\Sigma_k  \, u^m J_m \right|_{\Gamma}\, ,
\end{equation} 
where $d\Sigma_k$ is the volume form of $\Sigma_k$, is a conserved charge with respect to $\xi^m$. Taking, for example, a time translation, $\xi^m  \partial_m = \partial_t$, the associated conserved quantity, $E_{\text{conf}}\equiv Q_{\partial_t}$, which we denote by the conformal energy, is given by
\begin{equation}
    E_\text{conf} = \left.\frac{\r^2}{\sqrt{f(\r)}} \int d\Sigma_k \, T_{tt} \right|_{\Gamma}\, .
\end{equation}
We note that $ E_\text{conf}$ is invariant under Weyl transformations of the conformal representative.

{\textbf{Fefferman-Graham limit.}} It is interesting to note here that in AdS/CFT, the Brown-York stress tensor is ordinarily computed \cite{Balasubramanian:1999re} with the standard Gibbons-Hawking-York boundary term that appears in the Dirichlet problem. Here, we have instead used an action (\ref{bulkS}) with a boundary term whose prefactor differs from the standard one by a factor of $\tfrac{2}{3}$. Let us consider a general asymptotically AdS$_4$ configuration in the following coordinates:
\begin{equation} \label{eq: fg limit}
\frac{ds^2}{\ell^2} = d\rho^2 + \frac{e^{2\rho}}{\ell^2} \left( g^{(0)}_{mn}  + \ell^2 e^{-2\rho} g^{(2)}_{mn} + \ell^3e^{-3\rho} g^{(3)}_{mn} + \ldots \right) dx^m dx^n~,
\end{equation}
where $g^{(2)}_{mn}$ is minus the Schouten tensor built from $g^{(0)}_{mn}$
\begin{equation}\label{schouten}
g^{(2)}_{mn} \equiv \frac{1}{4}R[g^{(0)}_{mn}] g^{(0)}_{mn}-R_{mn}[g^{(0)}_{mn}]~,
\end{equation}
and $g^{(3)}_{mn}$ is a transverse-traceless tensor, also with respect to $g^{(0)}_{mn}$. As a simple example, consider the case $g^{(0)}_{mn} = \eta_{mn}$. We impose conformal boundary conditions at a large $\rho = \rho_c$, such that
\begin{equation}
g_{mn}|_\Gamma =  e^{2\bomega(x^m) } \left(\eta_{mn} + \mathcal{O}\left(e^{-\rho_c}\right) \right)~, \quad\quad K \ell = 3 + \mathcal{O}\left(e^{-2\rho_c}\right)~,  
\end{equation}
where $\bomega=\rho_c$ and $[g]_{mn} = \eta_{mn}$. Given \eqref{eq: fg limit}, it is easy to compute the extrinsic curvature at the boundary (see appendix \ref{app: near infinity}). We then find that the conformal Brown-York stress tensor (\ref{cby}) evaluates to
\begin{equation}
T_{mn} = \frac{3\ell^2}{16\pi G_N } \, g^{(3)}_{mn} +... \,,
\end{equation}
as we approach the asymptotic boundary. The above expression is indeed proportional to the standard boundary stress-tensor in AdS$_4$/CFT$_3$. 

For general $g^{(0)}_{mn}$, and assuming that $R[g_{mn}^{(0)}]$ is non-vanishing, it is convenient to first bring in the boundary infinitesimally by moving $K\ell$ slightly  away from its AdS$_4$ boundary value. Details of this can be found in appendix \ref{app: near infinity}. Upon doing so, our conformal Brown-York stress tensor, to leading order in the deviation, reads
\begin{equation} \label{eq: brown-york near boundary}
    T_{mn} = -\frac{\ell^2}{8\pi G_N} \sqrt{\frac{R [g_{mn}^{(0)}]}{4(K\ell-3)}} \left( R_{mn}[g_{mn}^{(0)}] - \frac{1}{3}R [g_{mn}^{(0)}] g_{mn}^{(0)} \right) + \frac{3\ell^2}{16 \pi G_N} g^{(3)}_{mn} +...\, .
\end{equation}
The first term is a divergent quantity in the strict AdS$_4$ boundary limit, whereby $K\ell \to 3$. The divergent term is fixed in terms of the asymptotic AdS$_4$ boundary data, and is often removed via some renormalisation scheme \cite{Skenderis:2002wp}. Interestingly, the divergent term depends on the conformal structure of $g^{(0)}_{mn}$ and thus, for $K\ell \neq 3$, it contributes to the boundary stress-energy tensor (\ref{cby}) in a physically meaningful way. In particular, we will see that it can receive non-trivial contributions from the field $\bomega$, which is now part of the dynamical phase space. 

What happens in the strict asymptotically AdS$_4$ limit, is that the boundary mode effectively decouples from the remaining dynamics, and it becomes reasonable to remove it from the phase space altogether. The finite term, proportional to $g^{(3)}_{mn}$, is the standard boundary stress-tensor in AdS$_4$/CFT$_3$ \cite{Balasubramanian:1999re}. As such, the standard AdS$_4$/CFT$_3$ boundary stress-tensor is recovered from the conformal boundary conditions, even though the coefficient of the Gibbons-Hawking-York term is different from that employed in the original literature.\footnote{We would like to acknowledge H. Liu, D. Harlow, and E. Shaghoulian for valuable discussions regarding this point.} 

The difference in the approach we have discussed, as compared to the standard Dirichlet approach, is how one controls the deviation away from the asymptotic boundary. Here, instead of fixing a finite size induced metric at the boundary $\Gamma$, and varying its size, we tune the value $K\ell$. That the choice of conformal boundary conditions is argued \cite{Anderson_2008,An:2021fcq} to have, generically, better well-posedness properties than the Dirichlet one, might be viewed as adding further value to this approach. 

In any case, in order to further quantify the physical content of the conformal boundary conditions, it is important to understand their dynamical implications. We will proceed to do so,  at the linearised level, in the next section.

\textbf{Example: Schwarzschild AdS$_4$ mass.} To exemplify the above expressions,  we calculate the conformal energy of a Schwarszchild AdS$_4$ black hole. The  corresponding a background metric $\bar{g}_{\mu\nu}$ in (\ref{gbar}) has $k=+1$ and
\begin{equation} \label{eq: ads bh}
f(r) = 1- \frac{2 G_N M}{r} + \frac{r^2}{\ell^2}~.
\end{equation}
We take the conformal boundary conditions to be%
\begin{equation}\label{omegaM}
    \left.ds^2\right|_{r=e^\bomega\ell} = e^{2\bomega}\left(-du^2+\ell^2d\Omega_2^2\right) \, , \qquad \left.K\right|_{r=e^\bomega \ell} = \frac{3 e^{3\bomega}\ell + \left(2 e^\bomega \ell - 3 G_N M\right)}{ e^{2\bomega}\ell^2 \sqrt{f(e^\bomega \ell)}} \, ,
\end{equation}
where, for the AdS$_4$ black hole, $\bomega$ is constant. For the sake of convenience, and to make clear the comparison to the AdS/CFT literature, we have rescaled the boundary clock to the time coordinate labeled by $u\in\mathbb{R}$. For the same purpose, we also set $\r$ to $\ell$.

Given some $M>0$ and $K\ell>3$, there exists a unique real $\bomega$ that satisfies (\ref{omegaM}). Our Cauchy surface $\Sigma$ is given by the spatial slice $t=\frac{e^\bomega \, u}{\sqrt{f(e^\bomega \ell)}}=0$ that orthogonally intersects the boundary at $r=e^\bomega\ell$. In the $K\ell\to 3$ limit, the trace of the extrinsic curvature equation can be inverted, resulting in the expansion
\begin{equation}
    e^\bomega = \frac{1}{\sqrt{2 (K\ell-3)}} \left(1+ \frac{K\ell-3}{4} + \mathcal{O}(K\ell-3)^{3/2}\right)\,.
\end{equation}
The conformal Brown-York stress energy tensor (\ref{cby}) evaluates to
\begin{equation}
T_{uu} = \frac{2}{\ell^2}T_{\theta \theta} = \frac{2 }{\ell^2\sin^2{\theta}}T_{\phi \phi}  = \frac{e^\bomega \left(3 G_N M - e^\bomega \ell\right)}{12 \pi G_N \ell^2 \sqrt{f(e^\bomega \ell)}} ~,
\end{equation}
with all other components vanishing. We thus find 
\begin{equation}\label{divE}
    E_\text{conf}(M) = \ell^2  \,  \int d\Omega_2  T_{uu} = \frac{e^{\bomega}M}{\sqrt{f(e^\bomega \ell )}} - \frac{e^{2\bomega}\ell}{3 G_N \sqrt{f(e^\bomega \ell )}} \, ,
\end{equation}
where $\bomega$ is implicitly defined in terms of $K$. As $K\ell \to 3$,  we find an $\bomega$-dependent expression that diverges in the limit. This is a general feature of conserved charges for asymptotic boundaries. Instead, we can compute the difference between the black hole energy and that of empty global AdS$_4$, which yields
\begin{equation}
    E_\text{conf}(M)-E_\text{conf}(M=0) = M -\frac{5}{3}M (K\ell-3)+ \mathcal{O}(K\ell-3)^{3/2} \, .
\end{equation}
Comparing to the standard expression in AdS$_4$/CFT$_3$, as calculated for instance in \cite{Witten:1998zw,Balasubramanian:1999re}, we find agreement for the leading contribution to the energy.

\section{Linearised dynamics about global AdS$_4$}\label{sec: lin dynamics global AdS}

In this section, we consider linearised perturbations about a fixed global AdS$_4$ spacetime. We take the background metric to be
\begin{equation}\label{globalAdS4}
    ds^2=-\left(1+\frac{r^2}{\ell^2}\right)dt^2+\frac{dr^2}{1+\frac{r^2}{\ell^2}}+r^2d\Omega_2^2,\,\,\,\,\,\,\,\,\,\,\,\,\,\,\,\,   d\Omega_2^2\equiv d\theta^2+\sin^2\theta d\varphi^2 \,,
\end{equation}
where the coordinate ranges are $t\in \mathbb{R}$, $r \in (0,\infty)$, $\theta\in (0,\pi)$, and $\varphi\in (0,2\pi)$. 

We consider a three-dimensional timelike boundary $\Gamma$ located at $r=\r$. By appropriately selecting the sign of $K\ell$, we are able to consider gravitational dynamics for either the interior region, $r \in (0,\r)$, or for the exterior region, $r\in (\r, \infty)$. We will focus on the dynamics inside the tube. 
The induced metric at $r=\r$ is given by
\begin{equation}\label{gbar_global}
     \left. \bar{g}_{mn}dx^mdx^n \right|_\Gamma=-\left(1+\frac{\r^2}{\ell^2}\right)dt^2+\r^2 d\Omega_2^2 \,,
\end{equation}
while the extrinsic curvature is given by
\begin{equation}
  \left.  K_{mn}dx^mdx^n\right|_\Gamma=\frac{\r}{\ell^2}\sqrt{1+\frac{\r^2}{\ell^2}}\left(-dt^2+\ell^2 d\Omega_2^2\right) \,.
\end{equation}
The trace of the extrinsic curvature at $\Gamma$ is 
\begin{equation} \label{eq: K function of r}
K \ell|_{\Gamma}=\frac{2\ell^2+3\r^2}{\r\sqrt{\ell^2+\r^2}}.    
\end{equation}
Note that $K \ell$ is a positive real number satisfying $K \ell \geq 3$, which is saturated as $\r$ approaches the asymptotic boundary of AdS$_4$. In the small $\r$ limit, $K\ell$ diverges as $\tfrac{2}{\r}$, retrieving the Minkowski behavior in \cite{Anninos:2023epi}. 


We consider perturbations of the form \eqref{metric_pert_lin},
subject to conformal boundary conditions at $\Gamma$. As boundary conditions, we fix the conformal class of the metric to be (\ref{gbar_global}) and the trace of the extrinsic curvature to be constant. Furthermore, we require the leading perturbation of the trace of the extrinsic curvature to vanish at the boundary, $\delta K|_\Gamma = 0$. In addition, we must specify conformal boundary conditions for the set of allowed diffeomorphisms $\xi^\mu$. Finally, we require $\xi^r |_\Gamma=0$, so that allowed diffeomorphisms do not move the location of the boundary. 

We follow a procedure analogous to the one for $\Lambda = 0$ \cite{Anninos:2023epi} and $\Lambda>0$ \cite{Anninos:2024wpy}. Namely, we decompose the metric perturbations into irreducible representations of the $S^2$, which are labeled by the total angular momentum $l \in \mathbb{Z}^+$ and the integer valued axial angular momentum, $m \in (-l,l)$. For $l=0$, and $l=1$, only modes that are locally diffeomorphic are present. We study $l=0$ modes in section \ref{l0modes}, while in section \ref{sec: non-linear diffeos} we provide their non-linear extension. Modes with $l=1$ are studied in \ref{l1modes}. %
In section \ref{sec: bulk modes}, 
we analyse linearised modes with $l\geq 2$, following the Kodama-Ishibashi procedure \cite{Kodama:2000fa,Kodama:2003jz}.

\subsection{Locally diffeomorphic modes with $l=0$} \label{l0modes}

There are no local excitations of the metric field with $l=0$ at the linearised level. Nevertheless, in the presence of a boundary, we can ask whether there exist physical modes that can be expressed as diffeomorphisms in a given small neighborhood. Concretely, we are interested in linearised perturbations $h_{\mu\nu}$ taking the following form
\begin{equation}\label{eqn: physical diffeo perturbation}
    h_{\mu \nu} \= \nabla_\mu \xi_\nu + \nabla_\nu \xi_\mu \, ,
\end{equation}
for an arbitrary vector field $\xi^\mu$. The above perturbation automatically satisfies the linearised Einstein field equation.

The general coordinate transformation that preserves the spherical symmetry of the background metric is given by
\begin{equation}
\xi^\mu(t,r) \partial_\mu = \xi^r(t,r) \partial_r + \xi^t(t,r) \partial_t~.
\end{equation}
Upon imposing that the trace of the extrinsic curvature remains unchanged by (\ref{eqn: physical diffeo perturbation}) we find that the vector field components must satisfy
\begin{equation}
\left.\left( r^2 \partial_t^2 \xi^r (t,r)-\left(2+\frac{r^2}{\ell^2}\right) \xi^r(t,r) \right)\right|_{\Gamma} =0 \,.
\end{equation}
From the above we obtain
\begin{equation}
\xi^r(t,r) = e^{-i \omega^{(0)} t} f_+ (\r) + e^{i \omega^{(0)} t} f_- (\r) + f_1 (t,r) ~,
\end{equation}
where $\omega^{(0)} \r= i\sqrt{2+\frac{\r^2}{\ell^2}}$, $f_\pm(\r)$ are arbitrary constants, and $f_1 (t,r)$ is a function of both $t$ and $r$ such that at the boundary $f_1(t,r=\r)$ goes to zero. We note that at the boundary we have exponential growth/decay, as in the case of non-negative cosmological constant \cite{Anninos:2023epi,Anninos:2024wpy}, and moreover in the Minkowskian limit $\tfrac{\r}{\ell} \to 0$ we retrieve the result in \cite{Anninos:2023epi}.

Further imposing that the boundary metric is conformally equivalent to the Lorentzian cylinder, we find
\begin{equation}
\left. \left( \frac{\xi^r(t,r)}{r}  - \left(1+\frac{r^2}{\ell^2}\right) \partial_t \xi^t (t,r) \right)\right|_{\Gamma}= 0 ~,
\end{equation}
yielding
\begin{equation}\label{xit}
\xi^t(t, r) = \frac{f_+(\r) \, e^{-i \omega^{(0)} t}}{\left(1+\r^2/\ell^2\right) \sqrt{2+\r^2/\ell^2}}  -\frac{f_-(\r) \, e^{i \omega^{(0)} t}}{\left(1+\r^2/\ell^2\right) \sqrt{2+\r^2/\ell^2}} + f_2(t, r) ~,
\end{equation}
where $f_2(t, r)$ is another smooth function of $t$ and $r$, such that at the boundary it becomes independent of time, $f_2(t,r=\r) = f_2(\r)$. Now we can look at the general form of the linearised metric after imposing both boundary conditions. The correction to the Weyl factor at the boundary simply becomes,
\begin{equation}\label{omegadiff}
(\r \, \delta \bomega) |_{\Gamma} = f_+(\r) \, e^{-i \omega^{(0)} t} + f_-(\r) \, e^{i \omega^{(0)} t} \,.
\end{equation}
Note again that the correction to the Weyl factor grows exponentially with time. We can also consider the corner component of the metric at the boundary,
\begin{equation}
 \delta \beta|_{\Gamma} = h_{tr}|_{\Gamma}  =  \frac{\sqrt{2+\r^2/\ell^2} }{\r \left(1+\r^2/\ell^2\right)} \left(f_+(\r) e^{-i \omega^{(0)} t} - f_-(\r) e^{i \omega^{(0)} t} \right) -\left(1+\frac{\r^2}{\ell^2}\right) \partial_r f_2(t,\r) \,.   \label{eq: deltab} 
\end{equation} 
From (\ref{xit}) it follows that $f_2(t,r)$ can be set to zero using an ordinary diffeomorphism. The first term in \eqref{eq: deltab}, instead, is proportional to $\partial_t \delta \boldsymbol{\omega} |_{\Gamma}$. Together with $\delta \boldsymbol{\omega} |_{\Gamma}$, the pair $\{ \delta \boldsymbol{\omega}, \partial_t \delta \boldsymbol{\omega} \} |_{\Gamma}$ constitutes independent initial data along the $t=0$
 initial slice.

We observe that spherically symmetric physical diffeomorphisms can vary both the Weyl factor of the boundary metric as well as the corner component. The boundary Weyl factor and corner can be viewed as conjugate variables of the gravitational phase space at the boundary, upon imposing conformal boundary conditions \cite{Odak:2021axr}. It is worth contrasting the above with linearised perturbations about the standard Minkowski corner, as analysed in \cite{Anninos:2023epi}, where physical diffeomorphisms perturb the corner but not the boundary Weyl factor.\footnote{In \cite{Anninos:2023epi} this was shown in the harmonic gauge. It is straightforward to show this independently of the gauge choice. %
}

An analogous analysis can be performed for the $l=0$ mode about the AdS$_4$ black hole metric \eqref{eq: ads bh} in the Fefferman-Graham gauge, see appendix \ref{app: FG l=0}.



\subsection{Non-linear diffeomorphic modes with $l=0$} \label{sec: non-linear diffeos} 
One can also treat the diffeomorphism non-linearly \cite{Liu:2024ymn,Anninos:2024wpy}. Here, we seek a surface embedded in global AdS$_4$, which preserves our boundary conditions. That is, the trace of the extrinsic curvature at the boundary is fixed to $K$, where we allow $K$ to be an arbitrary function of the boundary time, and the induced metric at the boundary is conformal to \eqref{gbar_global},
\begin{equation}\label{eqn: non-linear conf rep}
    ds^2|_\Gamma = e^{2\bomega (x^m)} \left( -\left(1+\frac{\r^2}{\ell^2}\right)dt^2+\r^2 d\Omega_2^2 \right) \,.
\end{equation}
As a consequence, $\r$ is not a function of $K$ and should be treated as a part of the conformal representative of the induced metric. 

The resulting non-linear equation governing the Weyl factor upon imposing our boundary conditions is given by %
\begin{equation}\label{eqn: brane eqn}
\r^2\partial^2_t \bomega = -\left(1+\frac{\r^2}{\ell^2}\right)\left(2  + \frac{3 \r^2 e^{2\bomega}}{\ell^2}\right)- 2 \left(\r \partial_t\bomega\right)^2 + K(t) \, \r  e^\bomega\left(1+\frac{\r^2}{\ell^2}\right) \sqrt{1  + \frac{\r^2}{\ell^2} e^{2\bomega}+ \frac{\left(\r \partial_t\bomega\right)^2}{1+\frac{\r^2}{\ell^2}} }~.
\end{equation}
Upon linearising the Weyl mode about $\bomega(t)=0$ and further setting $K$ to be \eqref{eq: K function of r}, we are left with %
\begin{equation}\label{linear}
\r^2 \partial^2_t \delta \bomega(t) =  \left( \frac{\r ^2}{\ell^2}+2 \right) \delta \bomega(t)~,
\end{equation}
from which we retrieve  the exponential solutions (\ref{omegadiff}).

The conformal energy for the solutions to (\ref{omegadiff}) can be obtained using the conformal Brown-York stress energy tensor \eqref{cby}. 
Assuming that $K$ is a constant, a direct computation gives
\begin{equation}\label{Econfglob}
    E_\text{conf} = \frac{\r\sqrt{1+\frac{\r^2}{\ell^2}}}{3 G_N}\left(K\r  e^{3\bomega}- 3e^{2\bomega}\sqrt{1  + \frac{\r^2}{\ell^2} e^{2\bomega}+ \frac{\left(\r \partial_t\bomega\right)^2}{1+\frac{\r^2}{\ell^2}} }\right) \, .
\end{equation}
To obtain this expression, we use \eqref{eqn: brane eqn}. One can show that $\frac{dE_\text{conf}}{dt}=0$ by imposing \eqref{eqn: brane eqn}. {Upon using this conformal Brown-York stress tensor, we find that \eqref{eqn: brane eqn} is consistent with \eqref{eq: Hamiltonian constraint}.}

\textbf{A non-linear example.} An inherently  non-linear time-dependent solution to (\ref{eqn: brane eqn}) is given by
\begin{equation}\label{nonlinear}
e^{2\bomega(t)} = \frac{\alpha_{\text{n.l.}}}{\cos^2 \beta_{\text{n.l.}} t}~, \quad \text{where} \quad  \alpha_{\text{n.l.}} \equiv \frac{9 \ell^2}{\r^2 \left(K^2 \ell^2-9\right)} ~  , \quad \beta_{\text{n.l.}} \equiv \frac{1}{\r} \sqrt{1+\tfrac{\r^2}{\ell^2}}~.
\end{equation}

For this non-linear solution the conformal Brown-York stress-tensor vanishes identically. As such, it is distinguishable from the constant $\bomega(t)=\bomega_0$ solution to (\ref{eqn: brane eqn}) which reads
\begin{equation}
e^{2\bomega_0} = \frac{\ell^2}{\r^2} \frac{ \left(\sqrt{K^2\ell^2-8}-K\ell\right)K\ell+12}{2 \left(K^2\ell^2-9\right)}~,
\end{equation}
and has negative conformal energy. 

\textbf{Asymptotic boundary analysis.} Another interesting limit is the large Weyl factor limit. Consider $\bomega = - \frac{1}{2}\log(K\ell-3) + \delta \bomega$. Taking $K\ell \to 3$, the equation \eqref{eqn: brane eqn} becomes
\begin{equation}\label{eqn: brane dynamics near AdS bdry}
    \r^2\partial_t^2\delta\bomega = -\left(1+\frac{\r^2}{\ell^2}\right)\left(\frac{1}{2} - \frac{\r^2}{\ell^2}e^{2\delta\bomega}\right) - \frac{1}{2}\left(\r \partial_t\delta\bomega\right)^2 \, ,
\end{equation}
to leading order in the small $K\ell-3$ expansion. An exact solution to this equation is given by a constant Weyl factor $e^{2\delta\bomega_0} = \frac{\ell^2}{2\r^2}$.

We would like to compare the Brown-York stress tensor this limit with the standard one from the Fefferman-Graham expansion  \eqref{eq: brown-york near boundary}. For this, we set the Weyl factor to be $e^{2\bomega}=\tfrac{1}{K\ell-3}$ and the conformal representative to be $g^{(0)}_{mn} = e^{2\delta\bomega} \bar{g}_{mn}$ where $\bar{g}_{mn}$ is (\ref{gbar_global}) and $\delta\bomega$ is purely time-dependent.

Then, the Brown-York stress tensor \eqref{cby} in this limit becomes 
\begin{equation}
    T_{tt} = \frac{2}{\r^2}\left(1+\frac{\r^2}{\ell^2}\right)T_{\theta \theta} = \frac{2}{\r^2\sin^2\theta}\left(1+\frac{\r^2}{\ell^2}\right)T_{\phi \phi}= \frac{\left(1+\frac{\r^2}{\ell^2} \right)\left(\frac{2}{3}e^{2\delta\bomega}-\frac{\ell^2}{\r^2}\right)-(\ell\partial_t\delta\bomega)^2}{8\pi G_N\ell\sqrt{K\ell-3}}\,.
\end{equation}
Note that this term diverges in the strict $K\ell\to3$ limit. Upon implementing (\ref{eqn: brane dynamics near AdS bdry}), this  stress tensor agrees with the first term in \eqref{eq: brown-york near boundary}. Moreover, \eqref{eqn: brane dynamics near AdS bdry} is equivalent to the condition that $R[g^{(0)}_{mn}]=\tfrac{4}{\ell^2}$ for a purely time-dependent $\bomega$. We can also compute the conformal energy, which in this limit simplifies to 
\begin{equation}
E_\text{conf} = \frac{4\pi \r^2 e^{\delta\bomega}}{\sqrt{1+\frac{\r^2}{\ell^2}}} T_{tt} \,.
\end{equation}

The conformal energy takes the form of a single particle with negative kinetic term and is unbounded from below. It is interesting to note that standard positivity notions of energy may no longer hold in the presence of a timelike boundary subject to conformal boundary conditions. {It is also the case that wrong-sign kinetic terms can appear in the kinetic term of the conformal mode of the induced metric, upon performing a classical ADM analysis of gravitational dynamics. Perhaps, then, $\bomega$ is carrying some of this flavour.}

{\textbf{Euclidean uniqueness.}} As a side remark, one might suspect that the equation analogous to (\ref{eqn: brane eqn}) in  Euclidean signature could signal the violation of uniqueness for conformal boundary conditions established in the theorem of \cite{Anderson_2008}. However, we recall that the uniqueness results in Euclidean signature assume the boundary to be a compact manifold. Consider an $S^2 \times S^1$ boundary, where the Euclidean time $\tau = i t$ is a periodic variable. In this way, one cannot generically generate multiple regular Euclidean solutions given the conformal boundary data, restoring the good uniqueness properties of \cite{Anderson_2008}. For concreteness, consider the linearised expression (\ref{linear}). Taking $t\to-i\tau$ and requiring $\tau$ to have a given periodicity $\tau \sim \tau+\beta$, we see that generically there are no regular solutions to the Euclidean version of (\ref{linear}) since this equation would enforce a different periodicity in $\tau$. In Lorentzian signature, instead, care must be taken in imposing appropriate initial data at the corner to ensure a unique and well-posed evolution. Relatedly, had we considered a Euclidean manifold of with boundary of the type $S^2 \times \mathcal{I}$, where $\mathcal{I}$ is a closed interval in $\mathbb{R}$, one should be cautious about additional Euclidean zero modes \cite{Capoferri:2024sgo}.

\subsection{Locally diffeomorphic modes with $l=1$}
\label{l1modes}

For the case of diffeomorphisms built from $l=1$ spherical harmonics, we solve again for a perturbation of the form \eqref{eqn: physical diffeo perturbation}. The general diffeomorphism built from the scalar spherical harmonics with $l=1$, $\mathbb{S}_1$, can be written as%
\footnote{For $l=1$ modes, the functions $\mathbb{S}_1$ are angular-dependent functions defined by $(\tilde{\nabla}^2 +2) \mathbb{S}_1 = 0$, with $\tilde{\nabla}^2$ being the Laplacian of the unit two-sphere. From $\mathbb{S}_1$, the $\mathbb{S}_{1,i}$ functions can be simply defined by $\mathbb{S}_{1,i} = -\frac{1}{\sqrt{2}} \tilde{\nabla}_i \mathbb{S}_1$, where the $i$ represent indices on the two-sphere.}
\begin{equation}\label{eqn: l=1 sol AdS}
    \xi_\mu^{\pm} dx^\mu \, = \, \frac{\alpha_\pm \, e^{\mp i \omega^{(1)} t}}{\sqrt{1+r^2/\ell^2}} \left(\mathbb{S}_1 \, dr \pm i\, \omega^{(1)}\, r\left(1+r^2/\ell^2\right) \,\mathbb{S}_1 \, dt - \sqrt{2}\left(1+r^2/\ell^2\right) \,\mathbb{S}_{1,i} \,r \,d\Omega^i\right) + \tilde{\xi}_\mu^{\pm}dx^\mu \, ,
\end{equation}
where $\omega^{(1)}= \ell^{-1}$, $\alpha_\pm$ are constants, and $\tilde{\xi}_\mu^{\pm}$ are arbitrary diffeomorphisms that vanish at the boundary, $\tilde{\xi}_\mu^{\pm}(t,\r)=0$. As opposed to $\tilde{\xi}_\mu^{\pm}$, the $\alpha_\pm$ modes have a non-vanishing $\xi^r$ at the boundary which means that they are physical and cannot be gauged away.

To conclude, the $l=1$ diffeomorphism built from the scalar spherical harmonics yields the subset of Killing vectors on AdS$_4$ that become conformal Killing vectors (with nonvanishing conformal factor) when projected onto the boundary metric. A similar analysis holds for diffeomorphisms built from $l=1$ vectorial spherical harmonics, where up to trivial diffeomorphisms, one finds the subset of AdS$_4$ Killing vectors that project onto Killing vectors of the boundary metric itself. Thus, there are no physical perturbations of the metric that correspond to $l=1$ diffeomorphisms.

\subsection{Bulk linearised modes with $l\geq2$} \label{sec: bulk modes}

For $l\geq 2$, there are no physical diffeomorphisms (see appendix \ref{sec: phys_diffeos_l}) and the only allowed perturbations are bulk modes. We employ the Kodama-Ishibashi method to analyse them \cite{Kodama:2000fa, Kodama:2003jz}. This procedure is analogous to the one done for positive and vanishing cosmological constant in \cite{Anninos:2023epi, Anninos:2024wpy}, so we will keep the discussion brief here. We refer the reader to appendix C in \cite{Anninos:2023epi} for a detailed account of definitions and conventions of the Kodama-Ishibashi method.

Using the spherical symmetry of the background metric, we can decompose the metric perturbations into a scalar and a vector sector,
\begin{equation}
    h_{\mu \nu} = h_{\mu\nu}^{(S)} + h_{\mu\nu}^{(V)} \,.
\end{equation}

The scalar sector can be described in terms of the scalar spherical harmonics $\mathbb{S}_l$, while the vector perturbation depends on the vectorial spherical harmonics $\mathbb{V}_i$.\footnote{The spherical harmonics of total angular momentum $l$, denoted as $\mathbb{S}_l$, are given by the solutions to $(\tilde{\nabla}^2 + l (l+1)) \mathbb{S}_l = 0$. It is also useful to define $\mathbb{S}_{l,i} \equiv -(l(l+1))^{-1/2} \tilde{\nabla}_i \mathbb{S}_l$. The vectorial spherical harmonics $\mathbb{V}_i$ are defined as solutions to the equations $(\tilde\nabla^2+l(l+1)-1)\mathbb{V}_i=0$ and $\tilde\nabla_i\mathbb{V}^i=0$.} The Kodama-Ishibashi formalism reduces the linearised Einstein field equations into a set of differential equations for scalar master fields $\Phi^{(S/V)}$, from which one can directly infer a solution for the metric perturbation $h_{\mu\nu}^{(S/V)}$. In the absence of boundaries, this procedure can be performed in a gauge-invariant way. 

For our analysis, we select a gauge in which the boundary conditions only act on either $h_{\mu\nu}^{(S)}$ or $h_{\mu\nu}^{(V)}$ \cite{Anninos:2023epi, Anninos:2024wpy}, so that the general metric perturbation can be written as
\begin{eqnarray}
\begin{cases}
	h_{mn} &= - \bar{g}_{mn} \frac{1}{2r}\left[ l\left(l+1\right)\left(1+ \frac{2r^2}{\ell^2}\right)+2 r^2 \partial_t^2 + 2  \left(1+\frac{r^2}{\ell^2}\right)^2  r \partial_r\right]\Phi^{(S)} \mathbb{S}_l  \\
 & \quad + \left(\delta^i_m\delta^t_n + \delta^i_n \delta^t_m\right) \left(1+\frac{r^2}{\ell^2}\right)\partial_r \left(r \Phi^{(V)}\right) \mathbb{V}_i \, , \\
    h_{rr} &= -\frac{1}{r\left(1+\frac{r^2}{\ell^2}\right)^2}\Bigg[\frac{l(l+1)}{2}\left(3+\frac{7r^2}{\ell^2}+\frac{4r^4}{\ell^4}\right)+ \left(3+\frac{2 r^2}{\ell^2}\right)r^2\partial_t^2 \\
    		&\quad + \left(1+\frac{r^2}{\ell^2}\right)\left(\left(1+\frac{r^2}{\ell^2}\right)\left(l(l+1) + 1+2\frac{r^2}{\ell^2} \right)+ r^2\partial_t^2\right) r\partial_r\Bigg]\Phi^{(S)} \mathbb{S}_l \, , \\
    h_{tr} &=  -\frac{1}{2\left(1+\frac{r^2}{\ell^2}\right)}\partial_t \left[l(l+1)\left(1+\frac{r^2}{\ell^2}\right) - 2 + r^2\partial_t^2 + \left(1+\frac{r^2}{\ell^2}\right)\frac{r^2}{\ell^2}r\partial_r\right]\Phi^{(S)} \mathbb{S}_l \, ,  \\
    h_{ri} &=  \frac{\sqrt{l(l+1)}}{2\left(1+\frac{r^2}{\ell^2}\right)}\left[l(l+1)\left(1+\frac{r^2}{\ell^2}\right)+r^2\partial_t^2+\left(2+\left(3+\frac{r^2}{\ell^2}\right)\frac{r^2}{\ell^2}\right)r\partial_r\right]\Phi^{(S)} \mathbb{S}_{l,i} + \frac{r}{1+\frac{r^2}{\ell^2}} \partial_t \Phi^{(V)} \mathbb{V}_i \, .	
    \end{cases}\label{eqn: spherical l>2 ansatz}
\end{eqnarray}
The indices $m$ and $n$ denote indices with respect to coordinates tangential to the boundary, $(t,\theta,\phi)$, while the index $i$ denotes indices on the two-sphere. In order to satisfy the Einstein field equation, the master fields $\Phi^{(S/V)}$ should satisfy,
\begin{equation}\label{AdS2KG}
\left(-\boldsymbol{\nabla}^2  + \frac{l(l+1)}{r^2} \right) \Phi^{(S/V)}(t,r) = 0~.
\end{equation}
In the above, $\boldsymbol{\nabla}^2$ denotes the Laplacian on a two-dimensional anti-de Sitter space,
\begin{equation}
-\boldsymbol{\nabla}^2 \equiv  \left(1+\frac{r^2}{\ell^2}\right)^{-1}{\partial_t^2} - \partial_r \left(1+\frac{r^2}{\ell^2}\right) \partial_r~. 
\end{equation}
Since we are interested in the gravitational dynamics inside the worldtube, we require solutions to be regular as $r\to0$. Then, the general solution for each frequency, both for the scalar and vector master fields can be written as 
\begin{equation}\label{PhiS AdS}
    \Phi^{(S/V)} (t,r) \= -\text{Re} \left[ \, e^{-i\omega t}\, \left(\frac{r}{\ell}\right)^{l+1} \left(1+\frac{r^2}{\ell^2}\right)^{\frac{\omega \ell}{2}} \, _2F_1\left(\frac{1}{2} (l+\omega \ell
   +1),\frac{1}{2} (l+\omega \ell +2);l+\frac{3}{2};-\frac{r^2}{\ell^2}\right) \right]\, .
\end{equation}

As mentioned, the metric perturbation \eqref{eqn: spherical l>2 ansatz} is given in a gauge where the linearised conformal boundary conditions act separately on $h_{\mu\nu}^{(S)}$ and $h_{\mu\nu}^{(V)}$, so we can study each contribution independently. 

\subsubsection{Vector sector}

\begin{figure}[h!]
        \centering
         \subfigure[$K \ell = 5$]{
                \includegraphics[scale = 0.5]{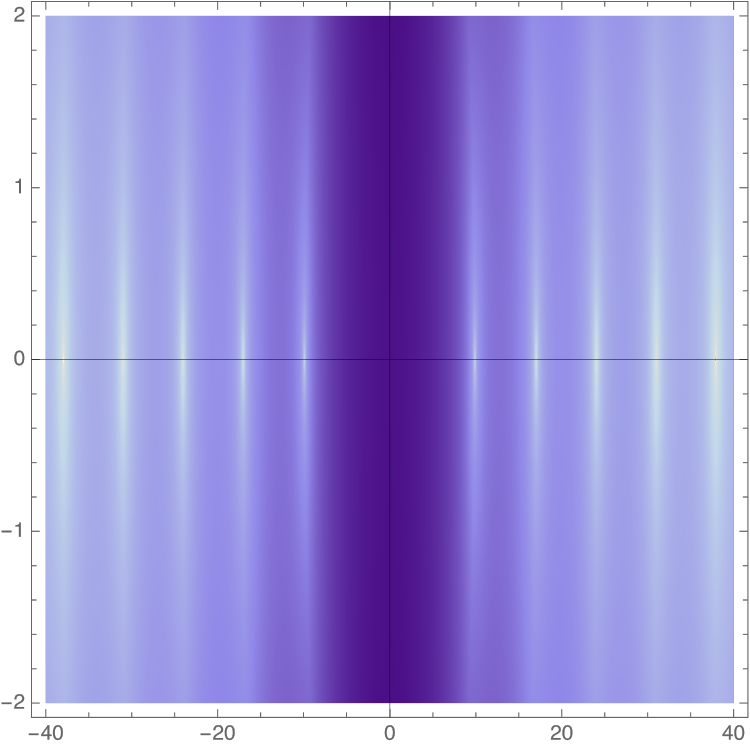}\label{fig: contour 1_V}}  \quad\quad
                 \subfigure[$K \ell = 3.01$]{
                \includegraphics[scale = 0.5]{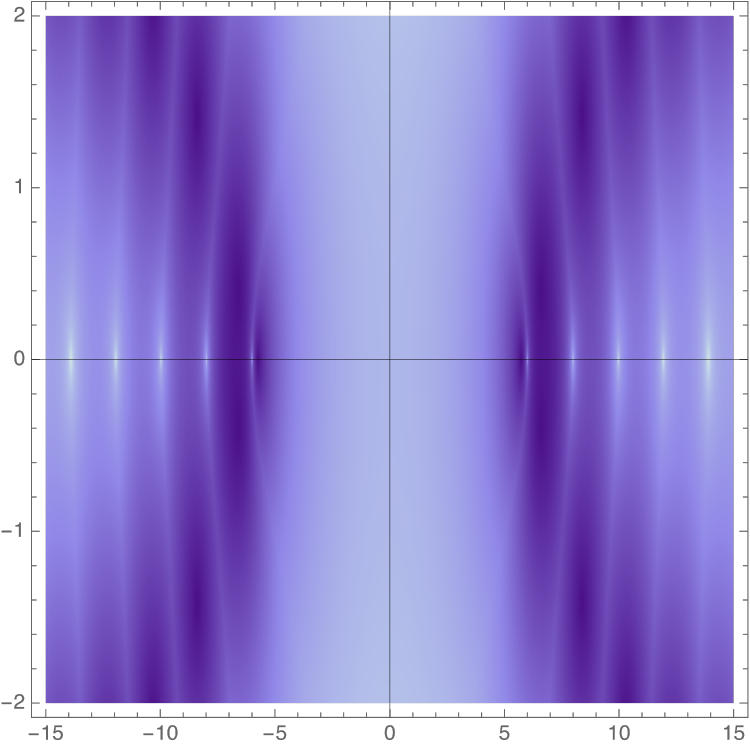} \label{fig: contour 2_V}}                           
                \caption{Density plot of absolute value of $\log \mathcal{F}^{(V)}_l(K \ell,  \omega \ell)^2$ in the complex $\omega \ell$ plane for $l=2$, at $t=0$. Note that all allowed frequencies are real.} \label{fig: contour_vector}
\end{figure}

In the case of $h_{\mu\nu}^{(V)}$, the boundary condition \eqref{eqn: bdry cond 2} is automatically satisfied while the boundary condition \eqref{eqn: bdry cond 1} imposes
\begin{equation}
 \mathcal{F}_l^{(V)} (K \ell, \omega \ell) \equiv   \left.\frac{\Phi^{(V)}}{r} + \partial_r \Phi^{(V)}\right|_{\Gamma}\=0 \,.
\end{equation}
Given the solution for the master field \eqref{PhiS AdS}, we numerically scan for solutions in the complex frequency plane for different values of $l$ and $\r$ and find that all allowed frequencies that satisfy that boundary conditions are real. Two examples for $l=2$ can be found in figure \ref{fig: contour_vector}. This behaviour mimics that of the full asymptotically AdS$_4$ case, and is essentially that of a `particle in a box'. As $K\ell \to 3$ the spectrum reproduces a subset of the normal mode frequencies,  $|\omega_n \ell| = l + 4 + 2n$ with $n \in \mathbb{Z}^+$, in global AdS$_4$. 

\subsubsection{Scalar sector}

For the scalar sector, it is straightforward to verify that \eqref{eqn: bdry cond 1} is automatically satisfied at $r=\r$. Then we only need to impose that the change in the mean curvature is vanishing at the boundary of the tube. This gives,
\begin{eqnarray}\label{eqn: deltaK scalar}
	\mathcal{F}^{(S)}_l(K\ell\,,  \omega\ell) \,\equiv\,\left.\left(\frac{a_1}{r^4}+\frac{a_2}{r^2}\partial_t^2 - 2 \partial_t^4\right)\Phi^{(S)}  + \left(\frac{a_3}{r^2} -2 \partial_t^2\right)\left(1+\frac{r^2}{\ell^2}\right)^2\frac{\partial_r \Phi^{(S)}}{r} \right|_{\Gamma} \= 0 \, ,
 \end{eqnarray}
 where we should use \eqref{eq: K function of r} to write $\r$ in terms of $K \ell$ and
 \begin{eqnarray}
     \begin{cases}
	& a_1  \= l(l+1)\left(1+\frac{r^2}{\ell^2}\right)\left(3+\frac{2r^2}{\ell^2}-2l(l+1)\left(1+\frac{r^2}{\ell^2}\right)\right) \, , \\
	& a_2  \= 4+\frac{2r^2}{\ell^2} - 4 l(l+1)\left(1+\frac{r^2}{\ell^2}\right) \, , \\
	& a_3  \= 4+\frac{2r^2}{\ell^2}-l(l+1)\left(3+\frac{2r^2}{\ell^2}\right) \, .
    \end{cases}
\end{eqnarray}

\begin{figure}[H]
        \centering
         \subfigure[$K \ell = 5$]{
                \includegraphics[scale = 0.5]{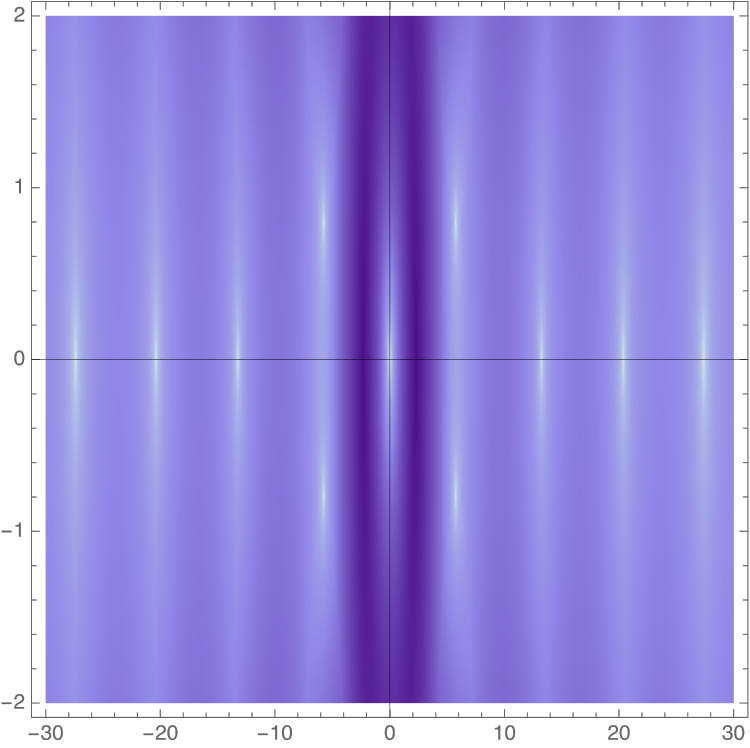}\label{fig: contour 1}}  \quad\quad
                 \subfigure[$K \ell = 3.01$]{
                \includegraphics[scale = 0.5]{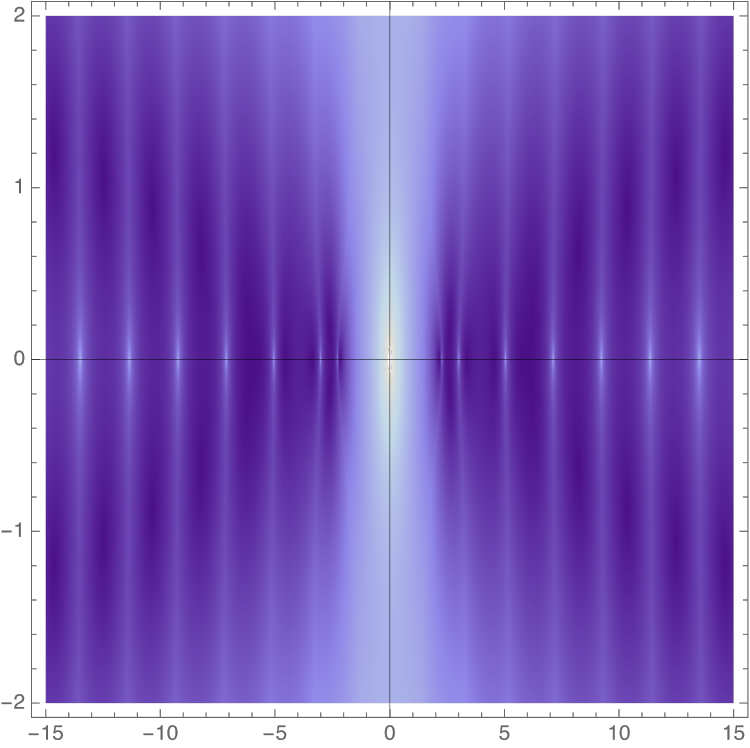} \label{fig: contour 2}}                           
                \caption{Density plot of absolute value of $\log \mathcal{F}^{(S)}_l(K \ell,  \omega \ell)^2$ in the complex $\omega \ell$ plane for $l=2$, at $t=0$. We multiply $\mathcal{F}^{(S)}_l(K \ell,  \omega \ell)$ by  $|\omega \ell|^{-2}$ to highlight the location of the zeros. Note that when $K \ell \approx 3$, all the zeros are in the real line, while for larger $K\ell$, there are, additionally, two pairs of complex conjugate frequencies.} \label{fig: contour}
\end{figure}

Equation \eqref{eqn: deltaK scalar} selects the set of allowed frequencies in the scalar sector that satisfy the conformal boundary conditions. In general these frequencies can be complex. As an example, in figure \ref{fig: contour} we provide a density plot of $\mathcal{F}^{(S)}_l (K \ell, \omega \ell)$ in the complex $\omega \ell$ plane for the $l=2$ modes for two different values of $K \ell$. In both cases, we observe a set of normal modes with real frequencies. In addition, when $K \ell$ is large enough, there are four additional complex frequencies, two of which have positive imaginary part.

The allowed modes give rise to a non-vanishing Weyl factor on the timelike boundary,
\begin{equation}
    \delta \bomega|_\Gamma = \left.-\frac{1}{4r}\left[ l\left(l+1\right)\left(1+ \frac{2r^2}{\ell^2}\right)+2 r^2 \partial_t^2 + 2  \left(1+\frac{r^2}{\ell^2}\right)^2  r \partial_r\right]\Phi^{(S)} \mathbb{S}_l\right|_{\Gamma}~.
\end{equation}
We can also compute the Brown-York stress tensor \eqref{cby} of these modes, whose $tt$-component is given by
\begin{equation}
    T_{tt}=\left.\frac{\sqrt{1+\frac{r^2}{\ell^2}}}{16 \pi G_Nr^2}\left(-\frac{4r}{3}+l(l+1)\left(\left(1+\frac{r^2}{\ell^2}\right)l(l+1)-1+r^2\partial_t^2+\left(1+\frac{r^2}{\ell^2}\right)^2r\partial_r \right)\Phi^{(S)}\mathbb{S}_l\right)\right|_{\Gamma}\,. 
\end{equation}
The conformal energy computed from this stress tensor is vanishing to leading order in the perturbative expansion. This is because, for any $l\geq 2$, the integral of any spherical harmonic $\mathbb{S}_l$ over the two-sphere  is zero. Alternatively, the energy carried by the linearised modes might follow the construction of a gravitational energy-momentum pseudo-stress tensor, as in \cite{Abbott:1981ff}, or a careful consideration of higher order terms. We leave this for future investigation. 

\begin{figure}[h!]
        \centering
         \subfigure[Re($\omega \ell$)]{
                \includegraphics[scale = 0.5]{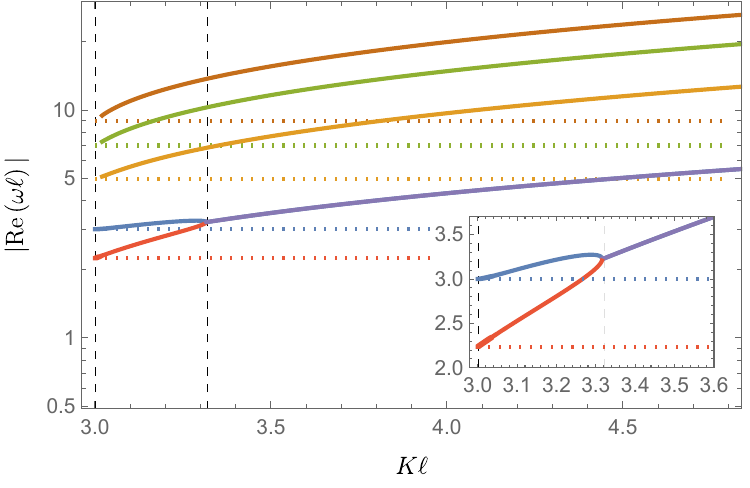}\label{fig: l2a}}  \quad\quad
                 \subfigure[Im($\omega \ell$)]{
                \includegraphics[scale= 0.5]{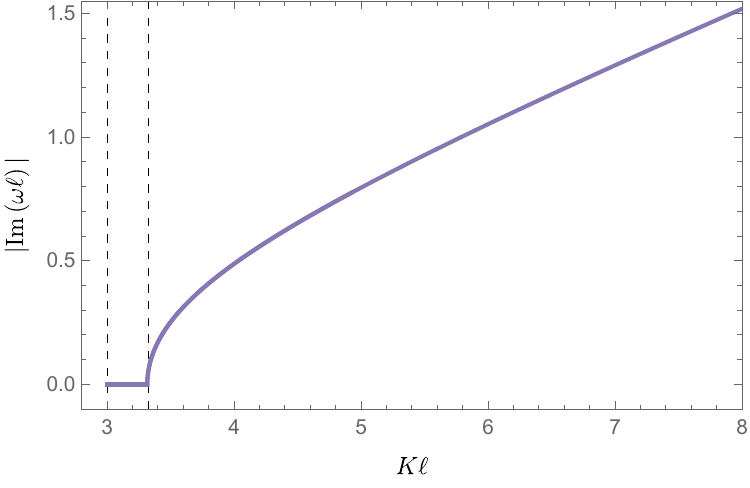} \label{fig: Re l2 plot}}                           
                \caption{Real and imaginary part of allowed frequencies in the scalar sector for $l=2$. The location of the conformal boundary of AdS$_4$ is given by $K \ell = 3$, and we observe that for $l=2$, $K_c \ell \approx 3.32$. Both are shown in vertical dashed black lines. For $K \ell < K_c \ell$, the lowest frequency asymptotes the value of $|\omega \ell| = \sqrt{5}$, followed by frequencies that asymptote to $|\omega \ell| = 3, 5, 7, 9, \cdots$. These values are marked with horizontal dotted lines in the figure. At $K\ell = K_c \ell$, the two lowest frequencies coalesce into one --- see inset in (a), which also develops an imaginary part, shown in (b). 
                After the critical value, the four lowest real frequencies, that correspond to plus and minus the blue and red curves, become four complex frequencies with $\omega \ell = \pm |\text{Re} \, (\omega \ell)| \pm i |\text{Im} \,(\omega \ell)|$, colored in purple. All the other frequencies remain real. At large enough $K \ell$, both the real and imaginary part become linear in $K \ell$, reproducing the flat space limit.
                } \label{fig: Im l2 plot}
\end{figure}

At a fixed value of $l$, the magnitude of the imaginary part of the complex frequencies decreases as $K \ell$ decreases, i.e., as the boundary of the tube moves towards the conformal boundary of AdS. In fact, for each $l$, we find a critical value of $K$, that we call $K_{c}$, such that $\text{Im} (\omega \ell) = 0$, for all $K \ell \in [3, K_{c} \ell ]$. We show this behaviour for $l=2$ in figure \ref{fig: Im l2 plot}.

Note that for large $K \ell$, both the real and imaginary parts of the allowed frequencies become linear in $K \ell$. This is consistent with the fact that in flat space, the frequencies scale with $\r$ instead of $\ell$ (for a spherical boundary in flat space, $K = \tfrac{2}{\r}$). In the limit $K \ell \to \infty$, the values of these frequencies precisely coincide with those found for the analogous problem in flat space for any $l$, including the large $l$ behavior, $\omega \r \approx \pm l \pm 0.34 i \, l^{1/3}$ \cite{Anninos:2023epi, Liu:2024ymn}. 

The imaginary part of $\omega$ is observed, numerically, to depend on $K \ell$ (see appendix \ref{app: large l}). To obtain an approximation at large $l$, we can employ a WKB approximation of (\ref{AdS2KG}). We can do this by placing (\ref{AdS2KG}) in Schr\"odinger form, by defining $x=\cos^{-1}(1+\tfrac{r^2}{\ell^2})^{-\tfrac{1}{2}}$ with $x\in(0,\tfrac{\pi}{2})$. For a given frequency, $\omega$, the Schr\"odinger equation governing the scalar mode is 
\begin{equation}
\partial^2_x \Phi^{(S)}(x) =   Q_l(x) \Phi^{(S)}(x)~, \quad\quad  Q_l(x) \equiv  \frac{l(l+1)}{\sin^{2} x}-\omega^2 \ell^2~.
\end{equation}
The WKB approximation, in the limit of parameterically large potential, is found to be 
\begin{equation}
\Phi^{(S)}(x) \approx \sqrt[4]{Q_l(x)} \exp {\int^x dx  \sqrt{Q_l(x)}}~. 
\end{equation}
The integral of $Q(x)$ can be expressed in terms of elementary functions, and we have chosen here the solution that becomes smooth near the origin. Implementing (\ref{eqn: deltaK scalar}), and expanding near the AdS$_4$ boundary by taking $K\ell \to 3$, one finds only real frequency solutions compatible with our preceding analysis. 

\begin{figure}[H]
        \centering
                \includegraphics[scale = 0.6]{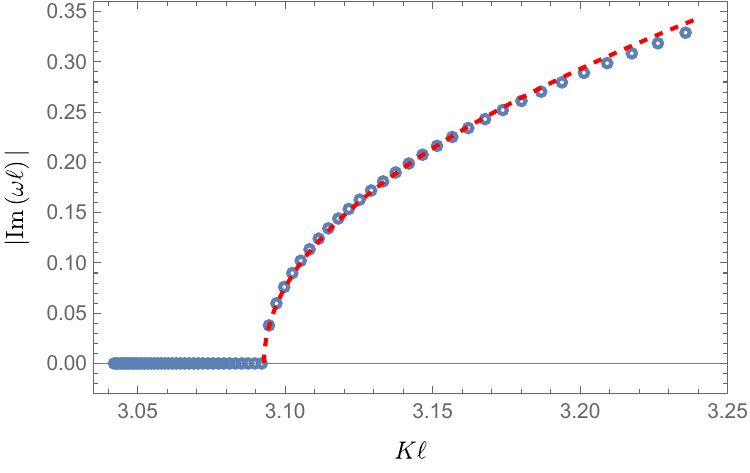}\label{fig: l4}
                \caption{Imaginary part of the allowed complex frequencies with $l=4$ in the scalar sector close to the critical value $K_c \ell \approx 3.09$. The dots are numerical, while the red dashed line shows the best fit close to the critical point, $\text{Im} (\omega \ell) = 0.89 \sqrt{K\ell - K_c \ell}$. } \label{fig: Im vs R near Rc plot}
\end{figure}

{\textbf{The transition from complex-to-real frequencies.}}
As we approach  $K \to K_{c}$, we  numerically find that for any fixed $l$,
\begin{equation}
    \text{Im} (\omega \ell) = \begin{cases}
        \alpha(l) \sqrt{K \ell - K_c \ell} + \mathcal{O}(K \ell - K_c \ell )^{3/2}~, \quad &  \quad K \ell \gtrsim K_c \ell \,, \\
        0~, \quad  &  \quad 3 \leq K \ell < K_c \ell \,.
        \end{cases}
\end{equation}
for some function $\alpha(l)$ that can be found numerically for each $l$. An example near the critical value for $l=4$ is shown in figure \ref{fig: Im vs R near Rc plot}.

\begin{figure}[H]
        \centering
                \includegraphics[scale=0.6]{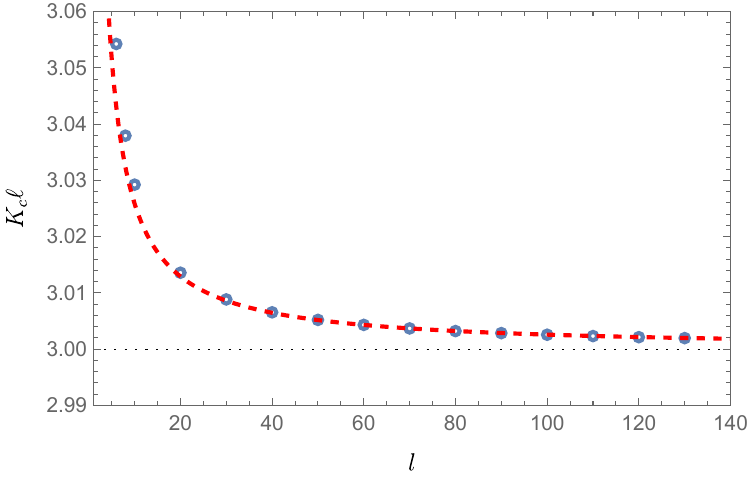}
                \caption{Critical value for the extrinsic curvature $K_c \ell$ for different values of $l$. The dots were obtained numerically for each $l$, while the dashed red curve shows the best fit at large $l$, which gives $K_c 
                \ell = 3 + 0.26 \, l^{-1}$. As a reference, we show the black dotted line of $K \ell =3$, the conformal boundary of AdS$_4$ space.} \label{fig: Rc vs l plot}
\end{figure}

We can ask how the value of $K_c \ell$ depends on the angular momentum number $l$. For this, we numerically find the value of $K_c \ell$ for different values of $l$, ranging from $l=2$ to $l = 130$. The results can be seen in figure \ref{fig: Rc vs l plot}, where we numerically find that for large $l$,
\begin{equation}
    K_c \ell = 3 + a \, l^{-1}+ \mathcal{O} (l^{-2}) \,, \quad \quad a \approx 0.26 \,.
\end{equation}
Conversely, this result can be inverted to establish that at any given $K \ell \gtrsim 3$, modes with angular momentum satisfying $l \gtrsim \tfrac{0.26}{(K\ell-3)}$ will  exhibit exponential growth in time at the linearised level.

{\textbf{Structure of the scalar normal frequencies.}} For $K < K_c$, the four complex frequencies split into four real frequencies, that we can follow as the boundary approaches the AdS$_4$ boundary. In the limit $K\ell \to 3$, the absolute value of the lowest two real frequencies is found to approach
\begin{equation}\label{eqn: bdry dispersion}
    |\omega^{(l)} \ell| = \sqrt{l(l+1) -1} \,.
\end{equation}
As an example,  in figure \ref{fig: Im l2 plot}  we plot the case of $l=2$ with $|\omega^{(2)} \ell|= \sqrt{5}$. These frequencies can be thought as near physical diffeomorphisms as the timelike boundary approaches the AdS$_4$ boundary (see appendix \ref{sec: phys_diffeos_l}).
For these modes, the variation of the boundary Weyl factor becomes,
\begin{equation}
    \delta \bomega = \left.\left(\frac{l(l+1)(l+2)(l-1)}{3l(l+1)-4}\frac{\ell}{\r}+\mathcal{O}\left(\frac{\ell}{\r}\right)^{2}\right)\Phi^{(S)}\mathbb{S}_l\right|_{\Gamma} \, .
\end{equation}

The other two real frequencies saturate to $|\omega \ell| = l+1$ as $K\ell \to 3$, and they are accompanied by a tower of normal (real) frequencies,
\begin{equation}\label{eqn: bulk dispersion}
    |\omega_n^{(l)} \ell| = (l+1) + 2n \,,\, \quad n \in \mathbb{N}_0 \,,
\end{equation}
that are the standard frequencies for metric perturbations about AdS$_4$. By evaluating the Weyl factor at the boundary for large $\r$, we find that for the frequencies (\ref{eqn: bulk dispersion}),
\begin{equation}
    \delta \bomega = \left.\left(\frac{l(l+1)(l+2)(l-1)}{8n (n+l+1)+2l+4}\frac{\ell^3}{\r^3}+\mathcal{O}\left(\frac{\ell}{\r}\right)^{4}\right)\Phi^{(S)}\mathbb{S}_l\right|_{\Gamma} \, .
\end{equation}
We note that the Weyl factor for these modes decays faster in $\r$ as compared to that of the modes in \eqref{eqn: bdry dispersion}. Given that $\sqrt{l(l+1)-1}$ is strictly less than $l+1$ for positive $l$, the lowest frequencies are always the $\pm \omega^{(l)}$ of (\ref{eqn: bdry dispersion}).

In the asymptotic limit, both sets of allowed frequencies can be found analytically by looking at the asymptotic form of $\mathcal{F}^{(S)}_l(K\ell\,,  \omega\ell)$ in \eqref{eqn: deltaK scalar}. As $K \ell \to 3$, $\tfrac{\r}{\ell} \to \infty$ and we obtain,
\begin{equation}
   \mathcal{F}^{(S)}_l(K\ell\,,  \omega\ell) \,=\,  \frac{4 \sqrt{\pi } \left(l(l+1)-1-\omega^2 \ell^2\right) \Gamma \left(l+\frac{3}{2}\right) }{\Gamma \left(\frac{1}{2} (l +1 -\omega \ell)\right) \Gamma \left(\frac{1}{2} (\omega \ell + l+1)\right)} \frac{\r}{\ell} \cos \left(\omega t\right)+\mathcal{O}\left(1\right) \,,
\end{equation}
which indeed has zeros at real frequencies with $\omega^{(l)} \ell = \pm \sqrt{l(l+1) -1}$ and $ \omega_n^{(l)} \ell =\pm ((l+1) + 2n)$, for non-negative integer $n$.

\begin{figure}[H]
        \centering
         \subfigure[]{
                \includegraphics[scale=0.55]{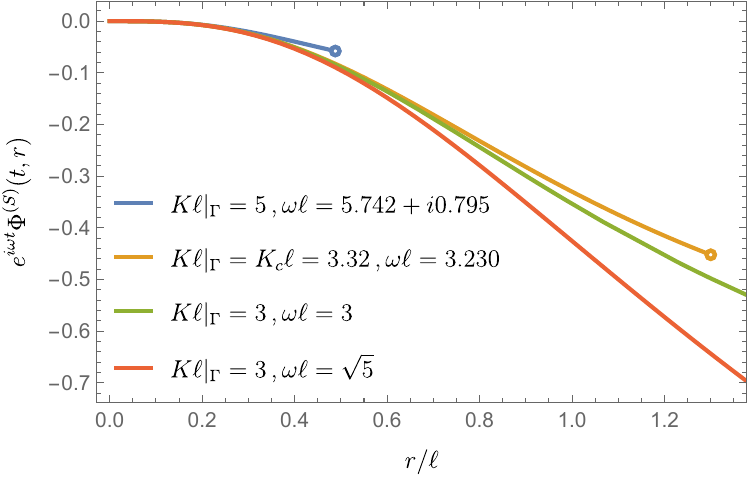}\label{fig: l2}}  \quad
                 \subfigure[]{
                \includegraphics[scale= 0.55]{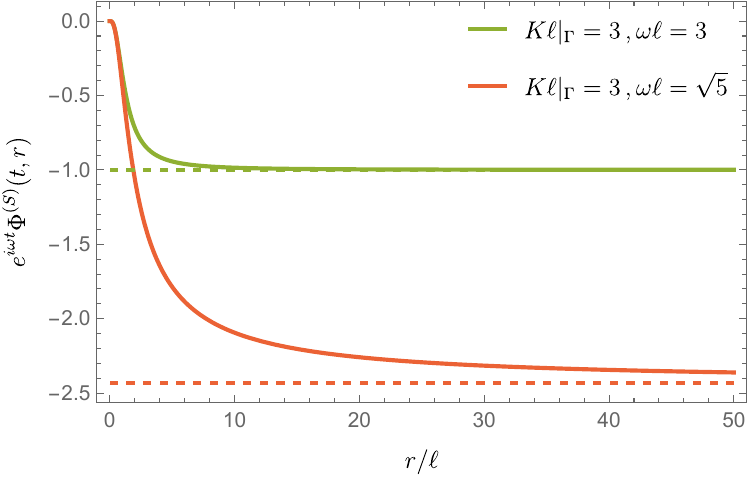} \label{fig: Re l2 scalar plot}}                           
                \caption{Radial profile of the master field with $l=2$, for different values of $K \ell$ at the boundary. In blue, we show the master field for a value of $K \ell$ that is greater than the critical; in yellow, at the critical value; and in green and red, for the asymptotic boundary of AdS$_4$. The circles indicate the position of the boundary in each case, except in the last ones where $\tfrac{\r}{\ell} \to \infty$. In all cases, the radial profiles are smooth functions of $\tfrac{r}{\ell}$. In (b), we increase the range of $\tfrac{r}{\ell}$ and only show the asymptotic form of the master field profile as the boundary approaches the conformal boundary of AdS. In dashed lines, we show the asymptotic constant values of the master field according to \eqref{asymptotic phi}.} \label{fig: Im l2 scalar plot}
\end{figure}

\textbf{Radial profiles.} It is worth noting that the radial profile of the master field at any given fixed time is smooth, even across the transition. For $l=2$, this can be seen in figure \ref{fig: Im l2 scalar plot}. When $K\ell \approx 3$, we can obtain the radial profiles analytically, finding that
\begin{eqnarray}
    e^{i \omega t} \Phi^{(S)}(t, r)|_{l=2} =
    \begin{cases}
        \frac{3 \sqrt{5} \left(4 \left(\tfrac{r}{\ell}\right)^2-3\right) \sin \left(\sqrt{5} \tan^{-1} \left(\tfrac{r}{\ell}\right) \right)+45 \left(\tfrac{r}{\ell}\right) \cos \left(\sqrt{5} \tan ^{-1} \left(\tfrac{r}{\ell}\right) \right)}{4 \left(\tfrac{r}{\ell}\right)^2} \, , \, \text{for} \,\, \omega \ell = \sqrt{5} \,, \\
        -\frac{\left(\tfrac{r}{\ell}\right)^3}{\left(\left(\tfrac{r}{\ell}\right)^2+1\right)^{3/2}} \quad , \quad \text{for} \,\, \omega \ell = 3 \,.
    \end{cases} \label{asymptotic phi 2}
\end{eqnarray}
At $r=0$ both profiles are identically zero. As $\tfrac{r}{\ell} \to \infty$, the radial profiles go to constants,
    \begin{eqnarray}
   e^{i\omega t} \Phi^{(S)}(t, r)|_{l=2} \overset{r/\ell\to \infty}{\longrightarrow}
    \begin{cases}
         3 \sqrt{5} \sin \left(\frac{\sqrt{5} \pi }{2}\right) \approx -2.43 \, , \, \text{for} \,\, \omega \ell = \sqrt{5} \,, \\
         -1 \quad , \quad \text{for} \,\, \omega \ell = 3 \,.
    \end{cases} \label{asymptotic phi}
\end{eqnarray}

\section{Linearised dynamics about planar AdS$_4$} \label{sec: planar AdS}

In this section, we consider linearised perturbations about a fixed planar AdS$_4$ spacetime, 
\begin{equation}
    ds^2 = \frac{\ell^2}{z^2}\left(dz^2-dt^2+dx^2+dy^2\right) \, ,
\end{equation}
where  $z>0$ and $(t,x,y)\in \mathbb{R}^{1,2}$ are coordinates on three-dimensional Minkowski space.

The timelike boundary, $\Gamma$, is taken to be a three-dimensional Minkowski spacetime located at $z=z_c > 0$. The spacetime region of interest will be $z\in\left(z_c,+\infty\right)$.

The background induced metric and the extrinsic curvature are given by
\begin{equation}\label{gbar2}
    \left.\bar{g}_{mn}dx^mdx^n\right|_\Gamma= \frac{\ell^2}{z_c^2}\eta_{mn}dx^m dx^n \,, \qquad \left.K_{mn}dx^m dx^n\right|_\Gamma = \frac{\ell}{z_c^2} \eta_{mn} dx^m dx^n \, .
\end{equation}
It follows that the trace of the extrinsic curvature is $K\ell = 3$. We note that, unlike the spherical case, the timelike boundary is independent of $z_c$ which means that, without any perturbation, every constant-$z$ surface leads to the same conformal class of the induced metric and trace of the extrinsic curvature. Indeed, there is no invariant meaning to $z_c$ as one can simply rescale coordinates to absorb it; it is nothing more than a bookkeeping device.

We consider perturbations of the form \eqref{metric_pert_lin},
subject to conformal boundary conditions at $\Gamma$. We fix the conformal class of the metric to be (\ref{gbar2}) and the trace of the extrinsic curvature to be $K = \tfrac{3}{\ell}$. As such, the perturbed Weyl factor, $\delta \bomega$, is defined as
\begin{equation}
    \left.ds^2\right|_\Gamma = (1+\varepsilon\,2 \delta \bomega) \frac{\ell^2}{z_c^2}\eta_{mn} dx^m dx^n \, .
\end{equation}
We require the leading perturbation of the trace of the extrinsic curvature to vanish at the boundary, $\delta K|_\Gamma = 0$. We also need to specify conformal boundary conditions for the set of allowed diffeomorphisms $\xi^\mu$. Furthermore, we require $\xi^z |_\Gamma=0$, so that allowed diffeomorphisms do not move the location of the boundary. 

In order to proceed one could follow an analogous procedure as in section \ref{sec: lin dynamics global AdS}, namely using the Kodama-Ishibashi formalism. In section \ref{sec: bulk planar AdS} we will work in the Fefferman-Graham coordinate choice so as to  compare our results with those in the AdS/CFT literature. 

\subsection{Locally diffeomorphic modes} \label{sec: diff planar}

We first consider linearised perturbations that can be expressed locally as a diffeomorphism,
\begin{equation}
    h_{\mu\nu}=\nabla_\mu\xi_\nu+\nabla_\mu\xi_\nu \,,
\end{equation}
for an arbitrary vector field $\xi^\mu$. The above perturbation automatically satisfies the linearised Einstein field equation. A straightforward calculation yields
\begin{eqnarray}\label{eqn: AdS planar metric diffeo}
\begin{cases}
    h_{zz} = \frac{2\ell^2}{z^2}\left(\partial_z \xi^z-\frac{\xi^z}{z}\right) \, , \\
    h_{zm} = \frac{\ell^2}{z^2}\left(\partial_m \xi^z + \eta_{mn} \partial_z \xi^n\right) \, , \\
    h_{mn} = \frac{\ell^2}{z^2}\left(\eta_{pn}\partial_m \xi^p + \eta_{pm}\partial_n \xi^p - \eta_{mn} \frac{2}{z}\xi^z\right)\, ,
    \end{cases}
\end{eqnarray}
where $\eta_{mn}$ is the three-dimensional Minkowski metric, and $\xi^z$ and $\xi^m$ are functions of both $z$ and $x^m$. Imposing that the perturbation \eqref{eqn: AdS planar metric diffeo} preserves the conformal structure of the induced metric \eqref{gbar2} at $z=z_c$ results in
\begin{equation}
    \left.\eta_{pn}\partial_m \xi^p + \eta_{pm}\partial_n \xi^p - \frac{2}{3}\eta_{mn}\partial_p \xi^p\right|_\Gamma = 0 \, .
\end{equation}
This is simply a conformal Killing equation for a three-dimensional Minkowski spacetime. Therefore this condition requires that the three-vector $\xi^m$ be given by
\begin{equation}\label{eqn: planar AdS xi^m}
    \xi^m(z,x^m) = \zeta^m(x^m) + \tilde{\xi}^m(z,x^m) \, , 
\end{equation}
where $\tilde{\xi}^m$ is an arbitrary three-vector that vanishes at $z=z_c$, namely $\tilde{\xi}^m (z_c,x^m)=0$, and $\zeta^m$ is a conformal Killing vector of three-dimensional Minkowski space,
\begin{equation}\label{eqn: Conf Killing 3D Mink}
    \zeta^m(x^m) = \mathfrak{a}^m +  x^n\mathfrak{b}_n{}^m{} + \left(2x^mx_n-x^px_p\delta^m_n\right)\mathfrak{c}^n + x^m\mathfrak{d}  \,,
\end{equation}
where $\mathfrak{a}^m$, $\mathfrak{b}_n{}^m$, $\mathfrak{c}^m$, and $\mathfrak{d}$ are arbitrary constants. Only $\mathfrak{c}^n$ and $\mathfrak{d}$ lead to a non-vanishing Weyl factor. Since $\tilde{\xi}^m$ vanishes at the timelike boundary, it is an allowed diffeomorphism.

The linearised trace of the extrinsic curvature is given by
\begin{equation}\label{eqn: planar AdS xi^z}
    \left.\delta K\right|_\Gamma = \left.\frac{z}{\ell} \partial^2 \xi^z\right|_\Gamma\,,
\end{equation}
where $\partial^2 \equiv \eta^{mn} \partial_m \partial_n$. This equation takes the form of a plane wave equation with $\delta K$ playing the role of an external source. Demanding that \eqref{eqn: AdS planar metric diffeo} preserves the trace of the extrinsic curvature at $z=z_c$ results in 
\begin{equation}\label{eqn: AdS planar sol xi^z}
    \xi^z(z,x^m) = \zeta(x^m) + \tilde{\xi}^z(z,x^m) \, ,
\end{equation}
where $\tilde{\xi}^z$ is an arbitrary function that vanishes as $z=z_c$, $\tilde{\xi}^z(z_c,x^m)=0$. Again, given that $\tilde{\xi}^z$ vanishes at the boundary, it is an allowed diffeomorphism. The $z$-independent function $\zeta$ is a solution to $\partial^2 \zeta=0$,
\begin{equation}\label{eqn: AdS planar sol zeta}
    \zeta(x^m) = \ell \, \text{Re} \int_{\mathbb{R}^2}\frac{d^2\bold{k}}{(2\pi)^2}\left(\zeta(\bold{k})e^{-i\omega t+i \bold{k} \cdot \bold{x}}\right) \, , \qquad \omega = |\bold{k}| \, ,
\end{equation}
where $\zeta(\bold{k})$ is an arbitrary function of $\bold{k}$. Combining both \eqref{eqn: Conf Killing 3D Mink} and \eqref{eqn: AdS planar sol zeta}, the linearised Weyl factor at $z=z_c$ can be written as
\begin{equation}\label{eqn: AdS planar linearised Weyl from xi^z}
    \delta \bomega(z_c ,x^m) = 2 x^m \mathfrak{c}_m + \mathfrak{d}-\frac{\ell}{z_c}\, \text{Re} \int_{\mathbb{R}^2}\frac{d^2\bold{k}}{(2\pi)^2}\left(\zeta(\bold{k})e^{-i\omega t+i \bold{k} \bold{x}}\right) \, , \qquad \omega = |\bold{k}|\, .
\end{equation}
We note that, as expected, allowed diffeomorphisms $\tilde{\xi}^\mu$ do not contribute to \eqref{eqn: AdS planar linearised Weyl from xi^z}. 

Thus, to uniquely specify a solution for $\delta \bomega|_{z=z_c}$ as a function of time, we need to specify initial data for $\delta \bomega$ at the intersection $\partial \Sigma$ of $\Gamma$ and the initial Cauchy surface $\Sigma$. More specifically, if we take the initial Cauchy slice to be at $t=0$, the initial dataset consists of $\{\delta \bomega,\partial_t \delta \bomega\}$ at $t=0$ and $z=z_c$. 

By setting the conformal Killing vectors \eqref{eqn: Conf Killing 3D Mink} to zero, we observe that the initial data can be expressed as
\begin{equation}\label{eqn: planar AdS Cauchy data physical diff}
    \left.\{\delta \bomega,\partial_t \delta \bomega\}\right|_{t=0,z=z_c}=\left. \left\{\frac{z^2}{4\ell^2}\left( h_{xx} + h_{yy} \right),-\frac{z}{\ell^2}h_{tz} - \frac{z}{2\ell}\left(\pi_{xx}+\pi_{yy}\right)\right\}\right|_{t=0,z=z_c} \, ,
\end{equation}
where $\pi_{xx}$ and $\pi_{yy}$ denote the linearised extrinsic curvature at $t=0$. We recall that $h_{xx}$, $h_{yy}$, $\pi_{xx}$, and $\pi_{yy}$ contribute to the standard Cauchy data of gravity. In contrast, the corner angle $\delta \beta=\left.h_{tz}\right|_{t=0,z=z_c}$  does not belong to the Cauchy data set. {One might be concerned, then, that the physical phase space is not even-dimensional or that there is a redundancy between the bulk and boundary phase space. For this reason, a better way to organise the physical Cauchy data is as follows. One can consider the standard initial dataset, namely $\mathcal{C}_\Sigma = \{h_{mn},K_{mn}\}$ on $\Sigma$, which obey the physical constraints and cannot be expressed, even locally, as diffeomorphisms of any type. Further to this, one has the boundary pair $\mathcal{C}_{\partial\Sigma} =\{\bomega, \partial_t \bomega \}$ residing at the spatial corner $\partial\Sigma$ that stems specifically from non-trivial diffeomorphisms. As such, $\mathcal{C}_{\partial\Sigma} \not\subset \mathcal{C}_\Sigma$. In asymptotically Minkowski or anti-de Sitter space we are often able to impose sufficiently stringent boundary conditions that remove $\mathcal{C}_{\partial\Sigma}$ from the physical phase space without disrupting the consistency of dynamical behaviour. For the case of a finite timelike boudary subject to conformal boundary conditions, this is no longer the case. }

Lastly, we note that by using \eqref{cby}, the linearised conformal Brown-York stress-energy tensor for \eqref{eqn: AdS planar linearised Weyl from xi^z} is given by
\begin{equation}
    T_{mn} = \left.\frac{\ell^2}{8 \pi G_N z_c}\partial_m\partial_n \delta \bomega\right|_\Gamma = -\left.\frac{\ell^2}{8 \pi G_N z_c^2}\partial_m\partial_n \zeta(x^m) \right|_\Gamma  \, .
\end{equation}
It is clear that $\partial^m T_{mn}=T^m{}_m=0$ as $\delta \bomega$ satisfies $\partial^2 \delta \bomega=0$. We note that since $K\ell=3$, we cannot directly compare to (\ref{eq: brown-york near boundary}). The contribution, as can be seen from (\ref{TmnFG}) in appendix \ref{app: near infinity}, corresponds to the divergent part of the boundary stress tensor in the standard AdS$_4$/CFT$_3$ limit. By integrating $T_{tt}$ over the spatial two-dimensional plane, we obtain the total conformal energy
\begin{equation}
    E_\text{conf} = \frac{\ell^2
    }{8 \pi G_N z_c}\int_{\mathbb{R}^2} d^2\bold{x} \, \partial_t^2 \delta \bomega(t,\bold{x}) \,.
\end{equation}
Given that $\partial^2 \delta \bomega=0$,  provided $\delta \bomega$ vanishes sufficiently fast at the boundary of $\mathbb{R}^2$ the conformal energy vanishes in the linearised approximation we are considering.

\subsection{Bulk linearised modes}\label{sec: bulk planar AdS}

Next we consider bulk modes. For this, we work in the Fefferman-Graham gauge, $h_{z\mu}=0$. It is always possible to use allowed diffeomorphisms $\tilde{\xi}^\mu$ to fix this gauge (see appendix \ref{app: FG}). Then, it is useful to parameterise the remaining components of the perturbed metric as
\begin{equation}\label{eqn: planar AdS param}
    h_{mn}(z,x^m) = \frac{\ell^2}{z^2}\gamma_{mn}(z,x^m)\, ,
\end{equation}
where $\gamma_{mn}$ is a symmetric tensor. Requiring that the perturbation \eqref{eqn: planar AdS param} preserves the conformal structure of the background induced metric at $z=z_c$ gives
\begin{equation}\label{eqn: AdS planar bdry cond 1}
    \left.\gamma_{mn}-\frac{1}{3}\eta_{mn} \gamma\right|_\Gamma = 0 \, ,
\end{equation}
where $\gamma \equiv \eta^{mn} \gamma_{mn}$ denotes the trace of $\gamma_{mn}$. Requiring that the perturbation \eqref{eqn: planar AdS param}
preserves the trace of the extrinsic curvature at $z=z_c$ yields
\begin{equation}\label{eqn: AdS planar bdry cond 2}
    \left.\partial_z \gamma\right|_\Gamma =0 \, .
\end{equation}
Using the parameterisation \eqref{eqn: planar AdS param}, the $zz$- and $zm$-components of the Einstein field equation respectively give
\begin{equation}\label{eqn: planar AdS G_zmu}
\begin{cases}
    \partial^m \partial^n \left(\gamma_{mn}-\eta_{mn} \gamma\right) + 2z^{-1}\partial_z \gamma = 0 \, , \\ 
    \partial^n \partial_z\left( \gamma_{mn}-\eta_{mn} \gamma\right) = 0 \, .
\end{cases}
\end{equation}
These equations imply that $\gamma=-\frac{1}{4}\partial^m v_m (z^2-z_c^2)+\chi$ and $
\partial^n (\gamma_{mn}-\eta_{mn}\gamma) = v_m$ for some scalar $\chi$ and vector $v_m$, which are both independent of $z$. Acting on \eqref{eqn: AdS planar bdry cond 1} with $\partial^m$, we find that $v_m = -\frac{2}{3}\partial_m \chi$. Imposing the boundary condition \eqref{eqn: AdS planar bdry cond 2}, we obtain $\partial^m v_m = - \frac{2}{3}\partial^2 \chi = 0$. 
In terms of $\gamma_{mn}$, these two equations imply that
\begin{equation} \label{eq: for appendix}
    \partial^n \gamma_{mn} = \frac{1}{3}\partial_m \gamma \, , \qquad \partial^m \partial^n \gamma_{mn} = \frac{1}{3}\partial^2 \gamma = 0 \, ,
\end{equation}
where $\gamma$ is now independent of $z$. Hence, by solving \eqref{eqn: AdS planar bdry cond 1}-
\eqref{eqn: planar AdS G_zmu}, one can decompose $h_{mn}$ as 
\begin{equation}\label{eqn: AdS planar metric param 2}
    h_{mn}(z,x^m) = \frac{\ell^2}{z^2}\left(\tilde{\gamma}_{mn}(z,x^m) + \frac{1}{3}\eta_{mn} \gamma(x^m)\right) \, ,
\end{equation}
where $\tilde{\gamma}_{mn}$ is a transverse-traceless tensor, and $\gamma$ is a $z$-independent scalar obeying $\partial^2 \gamma = 0$. The conformal boundary conditions for \eqref{eqn: AdS planar metric param 2} imply that
\begin{equation}
    \left.\tilde{\gamma}_{mn}\right|_\Gamma = 0 \, .
\end{equation}

Consider now the $mn$-components of the Einstein field equation.
By using \eqref{eqn: AdS planar metric param 2}, we obtain
\begin{equation}
    \partial^2 \tilde{\gamma}_{mn} + z^2 \partial_z \left(z^{-2}\partial_z \tilde{\gamma}_{mn}\right) = -\frac{1}{3}\partial_m \partial_n \gamma \, .
\end{equation}
We observe that, upon redefining $\tilde{\gamma}_{mn} \to \tilde{\gamma}_{mn} + \frac{1}{6}(z^2-z_c^2)\partial_m \partial_n \gamma$, the right hand side vanishes. This means that the general metric perturbation that obeys the conformal boundary conditions at $z=z_c$ is given by
\begin{equation}\label{eqn: AdS planar final exp}
    h_{mn}(z,x^m) = \frac{\ell^2}{z^2}\left(\tilde{\gamma}_{mn}(z,x^m) + \frac{1}{3}\left(\eta_{mn}+ \frac{1}{2}(z^2-z_c^2)\partial_m \partial_n  \right)\gamma(x^m)\right)  \, , \qquad  \left.\tilde{\gamma}_{mn}\right|_\Gamma = 0\, ,
\end{equation}
where the transverse-traceless tensor $\tilde{\gamma}_{mn}$ and the scalar $\gamma$ obey
\begin{equation}
\begin{cases}
    \partial^2 \tilde{\gamma}_{mn} + z^2 \partial_z \left(z^{-2}\partial_z \tilde{\gamma}_{mn}\right) = 0  \, , \\
    \partial^2 \gamma = 0 \, .
    \end{cases}
\end{equation}
The part of $h_{mn}$ that depends on $\gamma$ is a physical diffeomorphism, gauge equivalent to that analysed in the previous sub-section. Our analysis indicates that the bulk perturbation governed by the transverse-traceless tensor $\tilde{\gamma}_{mn}$ leads to a vanishing Weyl factor --- the Weyl mode is completely controlled by $\gamma$, and it is fixed uniquely by initial data localised at the boundary.

The equation for $\gamma$ can be solved straightforwardly. The general solution is given by
\begin{equation}
    \gamma(x^m) = \text{Re} \, \int_{\mathbb{R}^2}\frac{d^2 \bold{k}}{(2\pi)^2} \gamma(\bold{k}) e^{-i\omega t + i \bold{k} \cdot \bold{x}}  \, , \qquad \omega = |\bold{k}| \, ,
\end{equation}
where $\bold{k}$ is a two-dimensional vector and $\gamma(\bold{k})$ is an arbitrary function of $\bold{k}$. 

The solution for $\tilde{\gamma}_{mn}$ is also straightforward, but a bit longer, so we refer the reader to appendix \ref{app: planar perturbation} for details. The general solution for $\tilde{\gamma}_{mn}$ can be written as 
\begin{equation}
    \tilde{\gamma}_{mn}(z,x^m) = \text{Re} \int_0^\infty\frac{d q}{2\pi}\int_{\mathbb{R}^2}\frac{d^2 \bold{k}}{(2\pi)^2} \, \sum_{j=1}^2 \beta^{(j)}_{mn}(q,\bold{k}) \left(\frac{1+i q z}{1+i q z_c}e^{-iq(z-z_c)} - \frac{1-i q z}{1-i q z_c}e^{iq(z-z_c)}\right)e^{-i\omega t + i \bold{k} \cdot \bold{x}} \, , \label{eq: bulk modes planar}
\end{equation}
where $\beta^{(j)}_{mn}$ represent the two transverse-traceless polarisations of the metric perturbation and $\omega = \sqrt{|\bold{k}|^2+q^2}$. As $\omega$ is always real, the solutions are generally well behaved in time, mimicking the Dirichlet results in \cite{Andrade:2015gja}.

By using \eqref{cby}, the linearised conformal Brown-York stress-energy tensor associated with \eqref{eqn: AdS planar final exp} is given by
\begin{equation}
    T_{mn} = \left.\frac{\ell^2}{16 \pi G_N z_c^2}\left(\partial_z \tilde{\gamma}_{mn}+\frac{z_c}{3}\partial_m \partial_n \gamma\right)\right|_\Gamma \, .
\end{equation}
Once again, we can confirm that $T_{mn}$ is traceless and conserved. 

{\textbf{Asymptotic boundary limit.}} Let us consider the behavior of \eqref{eqn: AdS planar final exp} in the limit where the timelike boundary approaches the AdS$_4$ boundary.   Taking a perturbation which has the characteristic length scale of the order of $\ell$, namely $|\bold{k}|\ell\sim q\ell\sim 1$, along with $\tfrac{z_c}{\ell} \to 0$ and $\tfrac{z}{\ell} \to 0$ while keeping $\tfrac{z}{z_c}$ fixed, we obtain
\begin{equation}
    h_{mn} = \text{Re} \int_0^\infty\frac{d q}{2\pi}\int_{\mathbb{R}^2}\frac{d^2 \bold{k}}{(2\pi)^2} \, \frac{\ell^2}{z^2}\left[\frac{\eta_{mn}\gamma(\bold{k})}{3} -  \frac{z^2-z_c^2}{6}k_m k_n \gamma(\bold{k})- \sum_{j=1}^2 \frac{2i}{3}\left(z^3-z_c^3\right)q^3 \beta_{mn}^{(j)}(q,\bold{k})+ \cdots \right]e^{-i\omega t + i \bold{k}\cdot \bold{x}}\,,
\end{equation}
up to terms of order $\tfrac{z^3}{\ell^3}$. Setting $z_c=0$ and comparing with the standard Fefferman-Graham expansion, we observe that the Weyl mode, $\gamma(\bold{k})$, perturbs  $g^{(0)}_{mn}$ but not $g^{(3)}_{mn}$. It can be associated with a Fefferman-Graham gauge preserving diffeomorphism in the asymptotic boundary limit.

\subsection{Non-linear mode on AdS$_4$ black brane}

So far we have considered linearised perturbations about planar AdS$_4$. Upon implementing conformal boundary conditions, we identified a novel mode which is locally a diffeomorphism. As a final remark, and to complement the previous discussion of this section, we would like to consider the analogous setup about a black brane in AdS$_4$. The corresponding metric can be expressed in the following form
\begin{equation}\label{bb}
\frac{ds^2}{\ell^2} = \frac{dz^2}{z^2 f(z)} + \frac{1}{z^2}\left( - f(z) dt^2 + dx^2 + dy^2\right) \quad , \quad f(z) = 1 - \frac{2G_NMz^3}{\ell^4}    \,.
\end{equation}
Unlike the case of the planar AdS$_4$ background, the trace of the extrinsic curvature, $K$, on a surface of constant $z=z_c$ is now sensitive to the value of $z_c$. 

We would like to identify a hypersurface in the black brane geometry which obeys the conformal boundary conditions with constant $K$, and an induced metric conformally equivalent to $\eta_{mn}$
\begin{equation} \label{eq: flat bb bdy metric}
ds^2 |_\Gamma =  e^{{2\delta \bomega(t)}}  \frac{\ell^2 }{z_c^2} \left(- dt^2 +dx^2+dy^2\right)~.
\end{equation}
We will take $\delta \bomega(t)$ to be purely time dependent, but we will allow it to be non-linear. It is clear that $z_c$ can always be absorbed by a simple change in $\delta \bomega(t)$, so it should be viewed as a bookkeeping device.\footnote{Adding space dependence to $\delta\bomega$ is an interesting extension of this problem. It will require a more thorough analysis, beyond our scope, as the boundary mode $\delta\bomega$ would no longer be locally expressible as a diffeormophism.}

We now obtain a non-linear equation about the AdS$_4$ black brane background, analogous to (\ref{eqn: brane dynamics near AdS bdry}). 
Requiring that the timelike boundary is embedded in the AdS$_4$ black brane geometry (\ref{bb}), subject to our conformal boundary conditions, we are led to the following differential equation for $\delta\bomega$,
\begin{equation} \label{flat black brane dynamics}
    \ell^2 \partial_t^2 \delta \bomega = -\frac{3\ell^2}{z_c^2}e^{2\delta \bomega} + \frac{3G_N M z_c}{\ell^2} e^{-\delta \bomega} - 2 (\ell\partial_t \delta \bomega)^2+e^{\frac{\delta \bomega}{2}}\frac{K\ell^3}{z_c^2}\sqrt{e^{3\delta \bomega}-\frac{2G_N M z_c^3}{\ell^4}+\frac{z_c^2}{\ell^2}e^{\delta \bomega}(\ell\partial_t\delta \bomega)^2} \,.
\end{equation}

Note that upon setting $M=0$ and requiring $\delta \bomega(t)$ to be constant, this equation simply becomes $K\ell = 3$, as in the planar AdS$_4$ case. If $M$ is non-zero and we set $\delta \bomega$ to zero, the timelike boundary is at a constant $z=z_c$ slice of \eqref{bb}, and the value of $z_c$ is fixed as
\begin{equation}
    z_c^3 = \frac{\ell^4}{G_N M}\left(1-\frac{1}{9}\left(K^2\ell^2-K\ell\sqrt{K^2\ell^2-9}\right)\right) = \frac{\ell^4}{G_N M}\sqrt{\frac{2(K\ell-3)}{3}} + \mathcal{O}(K\ell-3) \, .
\end{equation}
Upon linearising \eqref{flat black brane dynamics} at finite $M$, we obtain the simpler equation
\begin{equation}\label{omegabb}
    \partial_t^2 \delta \bomega (t) = \Omega^2 \delta \bomega (t)~, \quad \text{with} \quad \Omega \equiv \frac{3 G_N M z_c^2}{\ell^2\sqrt{\ell^4-2 G_N  M z_c^3}} \,,
\end{equation}
which has solutions of the form $\delta \bomega (t) = e^{\pm \Omega \, t}$. The second order nature of (\ref{omegabb}) indicates that, as in our previous discussion, we must fix the dynamical data $\mathcal{C}_{\partial \Sigma} = \{\delta\bomega, \partial_t\delta\bomega \}|_{\partial \Sigma}$ to have a complete specification of the initial data.

Upon setting $M= 0$ in (\ref{omegabb}), such that $\Omega=0$,  we retrieve the simpler equation $\partial_t^2 \delta\bomega(t) =0$, in line with the analysis of section \ref{sec: diff planar}.  We also note that $\Omega$ is strictly real valued and diverges as the back brane horizon approaches the boundary, whereby $2 G_N  M z_c^3=\ell^4$. Such a divergence indicates that in the near-horizon region, where the metric is approximated to leading order by the Rindler spacetime, one must rescale the time coordinate to render it non-degenerate. Upon rescaling to the Rindler time coordinate, the rescaled expression for $\Omega$ is no longer divergent and our analysis naturally ties to the Rindler analysis of section 5.3 in \cite{Anninos:2024wpy}. What we are learning here, is that the Rindler boundary mode indeed non-linearises and persists away from the strict Rindler limit. 

The only non-vanishing components of the conformal Brown-York stress tensor (\ref{cby}) satisfy $T_{tt}=2T_{xx}=2T_{yy}$, with %
\begin{equation}
    T_{tt} = \frac{1}{12\pi \ell G_N} \left( K \ell \, e^{2 \delta\bomega }-3 e^{\delta\bomega /2} \sqrt{e^{3 \delta\bomega }+ e^{\delta\bomega } \left(\ell \partial_t\delta\bomega\right)^2 -\frac{2 G_N M}{\ell}} \right) \,,
\end{equation}
where we have set $\tfrac{z_c}{\ell}=1$ for notational simplicity. Upon imposing \eqref{flat black brane dynamics}, one can readily confirm that in addition to being traceless, $T_{mn}$ is also conserved with respect to boundary metric \eqref{eq: flat bb bdy metric}. The conformal energy follows immediately, by integrating the $tt$-component of the stress-tensor over a spatial slice. Interestingly, we see that for a finite conformal boundary in AdS$_4$, one can no longer decouple $\delta\bomega$ from the remaining dynamics. 

It might be tempting to view $\delta\bomega$ as a type of soft hair, along the lines of \cite{Hawking:2016msc}, which can dress the bare black brane geometry. However, in the case at hand, hair can carry physical energy. In this sense, the situation is similar to dressing a BTZ black hole with a Brown-Henneaux diffeomorphism \cite{Brown:1986nw}. A similar analysis holds for the black hole in global AdS$_4$, as detailed in appendix \ref{app: bh}.

\section{Sourcing the Weyl mode} 
\label{sec: source}

In this section we consider Euclidean AdS$_4$ subject to conformal boundary conditions. In particular, we are interested in the role of $K$ as a source for boundary operators. In ordinary AdS$_4$/CFT$_3$, the Euclidean problem produces a functional, $\mathcal{Z}$, whose arguments are the boundary values of the bulk fields. For the gravitational sector, one has a functional $\mathcal{Z}[[g_{i j}(\bold{x})]]$ of the conformal structure,  $[g_{i j}(\bold{x})]$, of a given boundary metric with coordinates $\bold{x}\in\mathbb{R}^3$. To compute the leading order contribution to $\mathcal{Z}$ one resorts to a saddle-point approximation whereby one seeks solutions to the Euclidean equations of motion that are smooth in the interior and asymptote to the specified boundary data. In the case at hand, the gravitational path integral will  produce a functional $\mathcal{Z}[[g_{ij}(\bold{x})],K(\bold{x})]$ of the conformal boundary data, as $K$ is no longer fixed to the value $K \ell = 3$. 

For an asymptotically AdS$_4$ spacetime, the configuration space near the AdS$_4$ boundary is organised in the form of a Fefferman-Graham expansion \cite{fefferman1985conformal}
\begin{equation}
\frac{ds^2}{\ell^2} = \frac{dz^2}{z^2} + \frac{1}{z^2} \left( g^{(0)}_{i j}+  z^2 g^{(2)}_{ij} + z^3 g_{ij}^{(3)} + \ldots\right) dx^i dx^j~,
\end{equation}
where $i,j$ denote indices tangential to the boundary, 
\begin{equation}
g^{(2)}_{ij} = \frac{1}{4} R[g^{(0)}_{ij}] g^{(0)}_{ij} - R_{ij}[g^{(0)}_{ij}]  \,,
\end{equation}
is minus the Schouten tensor, and $g_{ij}^{(3)}$ is traceless and transverse with respect to $g^{(0)}_{ij}$. The above series generally converges at some small but finite value of $z$ away from $z=0$. There exist residual diffeomorphisms that preserve the above asymptotic gauge. These involve a Weyl transformation of $g^{(0)}_{ij}$ accompanied by a particular transformation of the remaining coordinates. As such, the boundary data is that of a conformal metric.\footnote{In odd dimensions one must modify the discussion slightly due to anomalous transformations.} It is worth re-emphasising that for all asymptotic configurations, $K$ evaluated at the asymptotic boundary takes the same fixed value, namely $K = \tfrac{3}{\ell}$. Consequently, to impose boundary conditions for more general choices of $K$, we have to  move away from the asymptotic AdS$_4$ boundary. For instance, at a slice of constant but small $z=z_c$, we can compute the leading correction to be
\begin{equation}
K \ell = 3 + \frac{z_c^2}{4} R[g^{(0)}_{ij}] + \ldots ~,
\end{equation}
to leading order in the small $z_c$ expansion. The above expression for $K$ is now sensitive to the full boundary metric $g^{(0)}_{ij}$, and not just its conformal structure.

\subsection{Perturbed partition function}

We would like to deform $K(\bold{x})$ in an infinitesimal way, perturbing the boundary slightly away from its value $K\ell=3$ about the empty Euclidean AdS$_4$ solution. This could be taken, for instance, to be either the hyperbolic metric with flat slicing 
\begin{equation}\label{planar}
\frac{ds^2}{\ell^2} = \frac{1}{z^2} \left( dz^2 + {d\bold{x}^2} \right) ~, 
\end{equation}
where $\bold{x}\in \mathbb{R}^3$, or the hyperbolic metric with spherical slices
\begin{equation}
\frac{ds^2}{\ell^2} = \frac{dz^2}{z^2} + \frac{\ell^2}{z^2} \left( \frac{1}{4}  - \frac{z^2}{2\ell^2} + \frac{z^4}{4\ell^4} \right) d\Omega_3~,
\end{equation}
where $d\Omega_3$ is the round metric on the three-sphere, and $z\in \mathbb{R}^+$. For the sake of simplicity, in what follows we focus on (\ref{planar}).

At the level of the linearised equations of motion, we seek solutions that obey the conformal boundary conditions at $z=z_c$ that impose an induced metric conformally equivalent to the boundary metric $\delta_{ij}$, whilst the trace of the extrinsic curvature is $K = \tfrac{3}{\ell} + \,\delta K(\bold{x})$. 

In the planar case, for non-constant $\delta K(\bold{x})$, the linearised gravity solutions subject to our prescribed conformal boundary conditions enforce that the metric perturbation is locally pure gauge, that is,
\begin{equation}\label{eqn: hmunu planar ads}
    h_{\mu\nu} = \nabla_\mu \xi_\nu + \nabla_\nu \xi_\mu \, ,
\end{equation}
where $\nabla_\mu$ is the covariant derivative on the planar metric (\ref{planar}). By imposing that the induced metric at $z_c$ is conformally flat, we find
\begin{equation}
    \xi^i(z_c,\bold{x}) = 0 \, .
\end{equation}
The trace of the extrinsic curvature, to the linear order, fixes 
\begin{equation}
    \xi^z(z_c,\bold{x}) = \zeta(\bold{x}) \, , \qquad \delta K(\bold{x}) = \frac{z_c}{\ell}\partial_i \partial^i \zeta(\bold{x}) \, .
\end{equation}
We note that $\delta K(\bold{x})$ vanishes at $z_c = 0$, consistent with the fact that all Fefferman-Graham configurations have $K = \tfrac{3}{\ell}$ at the asymptotic boundary. The general diffeomorphism $\xi^\mu(z,\bold{x})$ that solves these equations is given by
\begin{equation}
    \xi^z(z,\bold{x}) =\zeta(\bold{x}) + \tilde{\xi}^z (z,\bold{x}) \, , \qquad \xi^i(z,\bold{x}) = \tilde{\xi}^i(z,\bold{x}) \, ,
\end{equation}
where $\tilde{\xi}^z (z,\bold{x})$ and $\tilde{\xi}^i (z,\bold{x})$ are residual gauge degrees of freedom obeying $\tilde{\xi}^z(z_c,\bold{x}) = \tilde{\xi}^i (z_c,\bold{x}) = 0$. Fixing the Fefferman-Graham gauge $h_{zz} = h_{zi} = 0$ imposes conditions on the remaining terms, resulting in
\begin{equation}\label{fgslicing}
\xi^z(z,\bold{x})  = -z \int \frac{d^3\bold{y}}{z_c^3}\, z_c \frac{\ell\delta K (\bold{y})}{4\pi\left|\bold{x}-\bold{y}\right|} \, , \qquad \xi^i(z,\bold{x})  = \frac{z^2-z_c^2}{2}\delta^{ij}\partial_j\int \frac{d^3\bold{y}}{z_c^3}\, z_c\frac{\ell\delta K (\bold{y})}{4\pi\left|\bold{x}-\bold{y}\right|} \, .
\end{equation}
In the strict AdS$_4$ boundary limit, we can recognize the above as a Fefferman-Graham preserving transformation (see for example equations (2.6) and (2.7) in \cite{Anninos:2010zf}). As (\ref{fgslicing}) involves a non-vanishing $\xi^z$ component, the above is a physical diffeomorphism that cannot be gauged away. In terms of the metric perturbation \eqref{eqn: hmunu planar ads}, we find
\begin{equation}\label{hK}
    h_{ij}(z,\bold{x}) = \left(\frac{2\ell^2}{z^2}\delta_{ij}+\left(1-\frac{z_c^2}{z^2}\right)\ell^2\partial_i \partial_j\right)\int \frac{d^3\bold{y}}{z_c^3}\, z_c \frac{\ell\delta K (\bold{y})}{4\pi\left|\bold{x}-\bold{y}\right|} \, .
\end{equation}
In particular, the perturbed Weyl factor of the boundary metric is given by
\begin{equation} \label{eq: bdy euclidean}
   ds^2|_{\Gamma} = \frac{\ell^2}{z_c^2} e^{2\bomega(\bold{x})} \delta_{ij} dx^i dx^j \,, \, \quad e^{\bomega(\bold{x})} \approx 1+ \delta \bomega(\bold{x}) \= 1 + \frac{z_c \ell}{4\pi} \int \frac{d^3\bold{y}}{z_c^3}\,  \frac{\delta K (\bold{y})}{\left|\bold{x}-\bold{y}\right|}\,.
\end{equation}
We note that $\delta K(\bold{x})$ controls, non-locally, the linearised Weyl mode of the boundary metric. In the $z_c\to 0$ limit while keeping $\frac{\left|d\bold{x}\right|}{z_c}$ and $\ell \delta K(\bold{x})$ fixed, the response of the Weyl factor due to $\delta K(\bold{x})$ vanishes.

Given the solution satisfying the conformal boundary conditions, we can compute the on-shell Euclidean Einstein action for the solutions of interest as a function of $\delta K(\bold{x})$. The Euclidean action is given by, %
\begin{equation}
-I_{\text{cl}} = \frac{1}{16\pi G_N}\int_{z_c}^{z_{\text{IR}}} dz\int d^3 \bold{x} \sqrt{\det g_{\mu\nu}} \left(R-2\Lambda\right)+\frac{1}{24\pi G_N} \int_{\Gamma} d^3 \bold{x} \, \sqrt{\det g_{mn}} K(\bold{x})~,
\end{equation}
where we will take $z_{\text{IR}}\to+\infty$ at the end. Given \eqref{eq: bdy euclidean}, it is easy to see that the boundary term becomes,
\begin{equation}
-I_{\text{bdy}} = \frac{1}{24\pi G_N} \int d^3 \bold{x} \, \frac{\ell^3}{z_c^3}e^{3\bomega(\bold{x})} K(\bold{x})~.
\end{equation}
As we would like to compute the two-point function of the Weyl mode, it is sufficient to expand the action to quadratic order in the metric perturbation $h_{\mu\nu}$. Using the on-shell solution \eqref{hK} and after integrating by parts (assuming that $\delta \bomega$ decays fast enough as $|\bold{x}|\to\infty$), the on-shell action is given by
\begin{equation}\label{Icl}
  - I_{\text{cl}} = \frac{3 z_{\text{IR}}\ell^2}{64 \pi z_c G_N}\int \frac{d^3 \bold{x}}{z_c^3} \ell^2\delta K(\bold{x})^2 + \frac{\ell^2}{24\pi G_N} \int \frac{d^3 \bold{x}}{z_c^3} \frac{d^3 \bold{y}}{z_c^3} \frac{z_c\ell^2\delta K(\bold{x}) \delta K(\bold{y})}{4\pi|\bold{x}-\bold{y}|} + \mathcal{O}(z_{\text{IR}}^{-1}) \, .
\end{equation}
Note that the first term, which is local in $\delta K(\bold{x})$, is divergent as $z_{\text{IR}}\to \infty$. It would be interesting to perform a more refined regularisation procedure for this term, perhaps along the lines of \cite{Castro:2024cmf}.
\subsection{Weyl factor two-point function}

Expression (\ref{Icl}) gives us, in the semiclassical limit, the generating function of correlations for the operator sourced by $\delta K(\bold{x})$. We can compute the two-point function as a simple example. This is the second functional derivative of $-I_{\text{cl}}$ with respect to $\delta K(\bold{x})$, which is given by
\begin{equation}
    - \frac{\delta^2 I_{\text{cl}}}{\delta K(\bold{x}) \delta K(\bold{y})} =  \frac{\ell^2}{12 \pi G_N} \frac{1}{z_c^6}\frac{z_c \ell^2}{4\pi\left|\bold{x}-\bold{y}\right|} \, ,
\end{equation}
for separate points $\bold{x}$ and $\bold{y}$. 
This is the two point function of a three dimensional massless scalar field. This is in line with the massless dispersion relation (\ref{eqn: bdry dispersion}) for $\delta \bomega(x^m)$ that we uncovered in our Lorentzian mode analysis, and more specifically the relation (\ref{eqn: AdS planar linearised Weyl from xi^z}) in the Lorentzian planar AdS$_4$ analysis.

\textbf{Comparison to Euclidean flat space.} It is worth  comparing the above analysis to the case of linearised gravity with a vanishing cosmological constant around $\mathbb{R}^4$, whose metric we take to be
\begin{equation}
ds^2 = dz^2 + d\bold{x}^2~.
\end{equation}
We choose the boundary to  reside at $z=0$, and as in the previous analysis, we impose that the induced metric is conformally equivalent to the  flat metric on $\mathbb{R}^3$ and the trace of the extrinsic curvature is $\delta K (\bold{x})$. A similar analysis to that for the planar AdS$_4$ geometry now yields the following linearised physical diffeomorphism
\begin{equation}\label{eqn: comparison euclidean flat}
    \xi^z (z,\bold{x}) = -\int d^3 \bold{y} \, \frac{\delta K(\bold{y})}{4\pi \left|\bold{x}-\bold{y}\right|}  + \tilde{\xi}^z (z,\bold{x}) \, , \qquad \xi^i (z,\bold{x}) = \tilde{\xi}^i (z,\bold{x}) \, , 
\end{equation}
where $\tilde{\xi}^z (z,\bold{x})$ and $\tilde{\xi}^i (z,\bold{x})$ are residual gauge degrees of freedom obeying $\tilde{\xi}^z(0,\bold{x}) = \tilde{\xi}^i (0,\bold{x}) = 0$. The corresponding metric perturbation can be computed using $h_{\mu\nu}=\partial_\mu \xi_\nu + \partial_\nu \xi_\mu$. Interestingly, the Weyl factor of the boundary metric is unaffected by $\delta K(\bold{x})$. This can be seen by noting that, for a general background geometry, the Weyl factor transforms under an infinitesimal diffeomorphism as
\begin{equation}
    \delta \omega(z,\bold{x}) = \frac{1}{3}\mathcal{D}_i \xi^i(z,\bold{x}) + \frac{1}{3}K \xi^z(z,\bold{x}) \, ,
\end{equation}
where $\mathcal{D}^i$ is the covariant derivative of the boundary metric. In the present case, $\mathcal{D}_i = \partial_i$ and $K=0$ implying that $\xi^z$ does not enter the linearised Weyl factor. Since $\xi^i(0,\bold{x})$ is zero according to \eqref{eqn: comparison euclidean flat}, then $\delta \omega(0,\bold{x}) = 0$ identically. This is in line with the observation in \cite{Anninos:2023epi} that the physical diffeomorphism that alters the corner angle $\beta$ about the standard Minkowski corner does not perturb the boundary value of the Weyl factor to linear order.

\begin{center}
\pgfornament[height=5pt, color=black]{83}
\end{center}
\vspace{5pt}

To conclude, we offer some general remarks on the current state of affairs. We view the formulation of the general problem in terms of the conformal boundary condition fixing the pair $([g_{mn}],K)$ as carrying certain merit. In particular, it is a formulation that can be extended to arbitrary value for the cosmological constant \cite{Anninos:2023epi,Anninos:2024wpy,Liu:2024ymn,Banihashemi:2024yye}, and ties naturally to the existing structure of the Fefferman-Graham expansion. Moreover, in the limit $K\ell \to \infty$ with an $S^2\times \mathbb{R}$ boundary metric, we recover the Minkowskian results of \cite{Anninos:2023epi} directly from the global AdS$_4$ analysis. 

In all cases analysed so far, one finds that a novel dynamical degree of freedom, $\bomega(x^m)$, must be included to the gravitational phase space. As such, our setup may shed light on the question of how many underlying degrees of freedom there are in a finite portion of spacetime. In this regard, it is interesting to note that the sharp distinction between the ultraviolet divergences, that are often removed by local boundary counterterms,  and the remaining quantum field theoretic data encountered in the standard AdS/CFT dictionary is no longer manifest. This feature is already illustrated  in the introduction by considering the gravitational path integral $\log \mathcal{Z}[S^3,K]$ in (\ref{Zs3}). Perhaps, from a more schematic viewpoint, what we are seeing is the necessity to consider a more local version of the holographic renormalisation group flow whereby we allow the ultraviolet cutoff scale, $\boldsymbol{\Lambda}_{\text{u.v.}}(x^m)$, to depend on the boundary spacetime coordinates. The effective action for $\bomega(x^m)$ follows from a Legendre transform of the dual field theory with respect to $\boldsymbol{\Lambda}_{\text{u.v.}}(x^m)$. 

We note that if we impose, instead, Dirichlet conditions near the AdS$_4$ boundary, the corresponding boundary Brown-York stress-tensor, $T_{mn}(x^n)$, will  acquire a non-vanishing trace, ${T_m}^m(x^m)$. From the bulk point of view, this trace is accounted for by the trace of the extrinsic curvature which is no longer fixed, in contrast to the case of the Fefferman-Graham expansion, or the conformal boundary conditions. One may ask whether there exists a new boundary mode for the Dirichlet problem. Near the Minkowski corner, the answer is partially affirmative, in that one finds large physical diffeomorphisms \cite{An:2021fcq,Anninos:2022ujl}. However, they come at the cost of a non-unique development of the equations. 

One of the main unresolved questions, left unanswered in our work so far, is the fate of the linearised modes that exhibit exponential growth about global AdS$_4$. To make progress on this issue, a non-linear analysis may be necessary. The growing modes indicate, in the very least, that the global AdS$_4$ solution (along with its $SO(3,2)$ isometry group) is no longer the vacuum state --- conformal symmetries appear to be broken. As $K\ell \to 3$, the breaking occurs at increasingly large angular momentum. To control the behaviour of the growing modes, we might consider adding further boundary terms at $\Gamma$, which would alter the dynamics of $\bomega(x^m)$, or further spacetime dependence on $K(x^m)$. For instance, a boundary cosmological constant
\begin{equation}
I_{\text{bdy}} = \lambda \int_{\Gamma} d^3x \sqrt{-\det g_{mn}} \,,
\end{equation}
adds an interaction term to the theory governing $\bomega(x^m)$. Along these lines, one may wish to consider the possibility of path integrating over the space of conformal metrics, such that the boundary theory is a type of conformal gravity, echoing ideas expressed in \cite{Banihashemi:2024yye}.

From a holographic point of view, what is missing is a clear interpretation for the meaning of $K(x^m)$ in the dual theory. Considerations of the Bekenstein-Hawking relation, as in (\ref{BH}),  appear to indicate that $K(x^m)$ is related to the number of underlying degrees of freedom in the finite-size theory. In parallel, $K(x^m)$ or its deviation away from the asymptotic value $K\ell=3$, appears to be related to the amount of conformal symmetry breaking. Relatedly, the conserved charges (\ref{eq: brown-york near boundary}) at the boundary $\Gamma$ diverge upon taking $K\ell\to 3$. In the standard AdS$_4$/CFT$_3$ literature, such divergences are related to the ultraviolet cutoff in the dual CFT$_3$. In our analysis, once the asymptotic AdS$_4$ boundary $\Gamma_\infty$ is brought in to some finite timelike boundary, $\Gamma$, the previously divergent terms are rendered finite and become sensitive to the dynamical mode $\bomega(x^m)$. The novel terms describe part of the physical content in the theory.

To make contact with concrete examples of AdS$_4$/CFT$_3$, it is worth generalising the conformal boundary conditions of \cite{An:2021fcq} to theories of supergravity. In particular, we may wish to understand the analogue condition for the whole gravitational supermultiplet. A natural proposal would be to consider the supersymmetric variation of the boundary data $([g_{mn}],K)$ and fix the whole collection at $\Gamma$. In particular, we may wish to fix the superconformal structure at $\Gamma$ along with a supersymmetric generalisation of $K(x^m)$. This, among other things such as the generalisation to other spacetime dimensions, and developing a more general string theoretic picture \cite{Silverstein:2022dfj,Ahmadain:2024uyo}, will be left for future considerations.

From a more general perspective, a sharper understanding of Euclidean gravity on a manifold with a finite size boundary may help clarify certain puzzling aspects of Euclidean quantum gravity. An interesting example that arises in this context comes from Euclidean gravity with $\Lambda>0$. Here, the Euclidean saddle corresponds to a round four-sphere. One can carve out a variety of compact three-manifolds from an $S^4$. The gravitational path integral over $S^4$ \cite{Gibbons:1976ue,Anninos:2020hfj}, though rich in structure exhibits some unusual features including (potentially) the presence of a phase \cite{Polchinski:1988ua}. On its own, the $S^4$ path integral has no external parameters that can be varied. In contrast, if we excise a solid cylinder, we obtain a portion of $S^4$ with an $S^2 \times S^1$ boundary, as  explored in \cite{Wang:2001gt,Anninos:2011zn,Anninos:2017hhn,Anninos:2018svg,Blacker:2023oan,Banihashemi:2022jys,Coleman:2021nor,Svesko:2022txo,Anninos:2024wpy,Batra:2024kjl,Silverstein:2024xnr}. It is of interest, then, to understand how features of the sphere path-integral are encoded in the excised sphere path integral. The presence of the phase, for example, may now depend on what boundary conditions are placed on the $S^2 \times S^1$ (see \cite{Maldacena:2024spf} for a related discussion). Moreover, the structure of the $S^2 \times S^1$ boundary may lend itself to an interpretation closer to that of a trace, making it easier to assess the underlying unitarity of the theory (or lack thereof). If we wish to fuse two portions of an excised sphere back together \cite{Anninos:2024wpy,Anninos:2022hqo},  we may wish to consider placing conformal boundary conditions at each boundary.

\section*{Acknowledgements}
It is a pleasure to acknowledge Tarek Anous, Michael Anderson, Frederik Denef, Daniel Harlow, Thomas Hertog, Diego Hofman, Joel Karlsson, Luis Lehner, Hong Liu, Beatrix M\"uhlmann, Oscar Reula, Olivier Sarbach, Edgar Shaghoulian, Eva Silverstein, Andrew Svesko, Toby Wiseman, and Themistokles Zikopoulos for interesting discussions. We are especially grateful to Ignacio Salazar Landea for collaboration at initial stages of the project and insightful remarks. We would like to thank all participants of the \href{https://timelikeboundaries.github.io/ingravity/}{Timelike Boundaries in Theories of Gravity} workshop for useful insights and the Gravity Theory Trust for supporting the meeting. D.A. is funded by the Royal Society under the grant ``Concrete Calculables in Quantum de Sitter".  The research of R.A. is supported by CONICET grant PIBAA 0008 and PICT grant 2021-0156. The work of D.A.G. is funded by UKRI Stephen Hawking Fellowship ``Quantum Emergence of an Expanding Universe". D.A. and D.A.G. are further funded by STFC Consolidated grant ST/X000753/1. C.M. is funded by STFC under grant number ST/X508470/1.

\appendix

\section{Linearised Dirichlet problem} \label{app: Dirichlet problem}

In this appendix we comment on linearised dynamics about different backgrounds subject to Dirichlet boundary conditions. We discuss 
uniqueness and existence properties of the perturbed solution about Minkowski, Rindler, and planar AdS in four spacetime dimensions. In all cases, we work in a Fefferman-Graham-like gauge, $h_{z\mu}=0$, for all $z \geq z_c$.

\subsection{Uniqueness and existence for Minkowski, Rindler, and planar AdS$_4$}

\textbf{Minkowski corner.} Let us first consider the linearised problem about a Minkowski corner background. Specifically, we take the background metric to be
\begin{equation}
    ds^2 = dz^2 + \eta_{mn} dx^m dx^n \, ,
\end{equation}
and we place the timelike boundary at $z=z_c$. Working in a Fefferman-Graham-like gauge where
\begin{equation}
    h_{zz}=h_{zm}=0 \, , \qquad h_{mn} = \gamma_{mn} \, ,
\end{equation}
the $zz$- and $zm$-components of the Einstein field equations read
\begin{equation}\label{eqn: Mink corner Gzmu}
\begin{cases}
    \partial^m \partial^n \left(\gamma_{mn}-\eta_{mn} \gamma\right) = 0 \, , \\ 
    \partial^n \partial_z\left( \gamma_{mn}-\eta_{mn} \gamma\right) = 0 \, ,
\end{cases}
\end{equation}
where $\gamma \equiv \eta^{mn}\gamma_{mn}$ is the trace of $\gamma_{mn}$.
Now we impose the Dirichlet boundary conditions at $z=z_c$,
\begin{equation} \label{eq: dirichlet corner}
    \left.\gamma_{mn}\right|_\Gamma = \Gamma_{mn} \, ,
\end{equation}
where $\Gamma_{mn}$ is a symmetric tensor which only depends on $x^m$. By evaluating the first equation in \eqref{eqn: Mink corner Gzmu} at $z=z_c$ and imposing the Dirichlet boundary condition \eqref{eq: dirichlet corner}, we obtain an obstruction on generic Dirichlet data,
\begin{equation}
    \partial^m \partial^n \Gamma_{mn} - \partial^2 \Gamma^m{}_m=0 \, .
\end{equation}
The above implies that the linearised Ricci scalar for the induced metric must vanish in agreement with the non-linear analysis of \cite{An:2021fcq}.

\textbf{Rindler corner.} Next, we discuss perturbations about Rindler space,
\begin{equation}
    ds^2 = dz^2 -z^2dt^2 + \delta_{ab} dx^a dx^b \, ,
\end{equation}
and place the timelike boundary at $z=z_c$. We work again in the same Fefferman-Graham-like gauge, 
\begin{equation}\label{eqn: FGlikge gauge Rindler}
    h_{zz}=h_{zt}=h_{za}=0 \, , \qquad h_{ab}=\gamma_{ab} \, , \qquad h_{ta} = z v_a \, , \qquad h_{tt} = z^2s \, ,
\end{equation}
where $\gamma_{ab}$, $v_a$, and $s$ are $z$ and $x^m$-dependent variables. 
We consider imposing Dirichlet boundary conditions at $z=z_c$,
\begin{equation}\label{eqn: Dirichlet Rindler}
    \left.h_{mn}\right|_\Gamma = \Gamma_{mn}\,,
\end{equation}
where $\Gamma_{mn}$ only depends on $(t,x^a)$. 

Combining the linearised $tz$- and $zz$- Einstein equations with the Dirichlet boundary condition (\ref{eqn: Dirichlet Rindler}) one can show that
\begin{equation}\label{eqn: Rindler obst 2}
    \left(\partial_t^3-\partial_t\right)\Gamma^a{}_a = \left.\partial_a F^a\right|_\Gamma \, , 
\end{equation}
where
\begin{equation}
F^a \equiv z^2 \partial_z v^a - z v^a - z^2 \partial_t \partial^a s + 2 z \partial_t^2 v^a - z^2 \partial_t \partial^b \left(\gamma^a{}_b-{\delta^a}_b \gamma\right)~.
\end{equation}
Note that the first term in $F^a$ contains a $z$-derivative, so $F^a$ evaluated at $z=z_c$ cannot be purely determined from the Dirichlet data, $\Gamma_{mn}$. However, upon integrating over the non-compact spatial directions, the right hand side of \eqref{eqn: Rindler obst 2} becomes a boundary integral, i.e.,
\begin{equation}\label{eqn: Rindler obst 3}
    \int_{\mathbb{R}^2}d^2x \left(\partial_t^3-\partial_t\right)\Gamma^a{}_a = \lim_{r\to\infty}\int_{S^1} rd\theta \,  F_{r} \, ,
\end{equation}
where $r$ and $\theta$ denote polar coordinates on the spatial $\mathbb{R}^2$ of the Rindler spacetime. Assuming that $F_r$ decays as $\tfrac{1}{r^2}$ or faster as $r\to \infty$, the right hand side vanishes. We  note that although ${\Gamma^a}_a$ transforms under coordinate transformations at $\Gamma$ (decaying sufficiently fast at the boundary of $\mathbb{R}^2$), its integral over $\mathbb{R}^2$ does not. We thus obtain an obstruction on generic Dirichlet data.

A similar argument can be made also for the region near a horizon. In that case, the near horizon region is approximately the product of two-dimensional Rindler space and a two-sphere, and so the left hand side of \eqref{eqn: Rindler obst 3} is replaced by an integral over a two-sphere, leading to a similar obstruction to the one found in \cite{Bredberg:2011xw,Anninos:2011zn}.

Finally, let us consider the non-uniqueness  of  linearised gravity about the Rindler background  subject to  Dirichlet boundary conditions. We work again in the Fefferman-Graham-like gauge, \eqref{eqn: FGlikge gauge Rindler}. A solution that is only dependent on $z$ and $t$ is given by
\begin{equation}\label{eqn: Rindler zero mode}
    h_{\mu\nu}dx^\mu dx^\nu = z(z-z_c)f(t)dt^2 \, ,
\end{equation}
where $f(t)$ is an arbitrary function of time. This solution is locally diffeomorphic, namely one can write \eqref{eqn: Rindler zero mode} as $h_{\mu\nu}= \nabla_\mu \xi_\nu + \nabla_\nu \xi_\mu$, with
\begin{equation}
    \xi^\mu \partial_\mu = \frac{1}{2}\int_{\mathbb{R}} \frac{d\omega }{2\pi}\,\frac{f(\omega)e^{-i\omega t}}{1+\omega^2} \left(z_c\partial_z + \left(\frac{z+(z-z_c)\omega^2}{i\omega z}\right)\partial_t\right) \, , \qquad f(\omega) = \int_{\mathbb{R}}dt \, f(t) e^{i\omega t}\,.
\end{equation}
At the boundary $z=z_c$, \eqref{eqn: Rindler zero mode} obeys the Dirichlet boundary conditions \eqref{eqn: Dirichlet Rindler} for $\Gamma_{mn}=0$. As $\xi^z$ is non-vanishing at the boundary, this solution is physical and cannot be gauged away.

The non-uniqueness of the Rindler corner follows from the fact that \eqref{eqn: Rindler zero mode} contains an arbitrary function $f(t)$.  As it is an arbitrary function of time, one can always construct a solution that is localised in time away from the initial Cauchy surface at $t=0$ and $z>z_c$, so that the solution does not leave an imprint on the initial data. As such, specifying the initial conditions together with the Dirichlet boundary conditions on the Rindler corner background will never lead to a unique solution. The same argument also holds for the region $0<z<z_c$.\footnote{This property has also been observed by T. Zikopoulos in unpublished work.}

As an explicit example, take the bump function localised on $t\in \left(4,6\right)$ and vanishing elsewhere,
\begin{equation}
    f(t) = \begin{cases}
        \alpha\,e^{-(1-(5-t)^2)^{-1}} \,,& \quad t \in (4,6)\\
        0 \,,& \quad \text{else}
    \end{cases} \, , \qquad \alpha \in \mathbb{R} \, .
\end{equation}
Plugging this back in \eqref{eqn: Rindler zero mode}, it can be checked that the perturbation vanishes at the $t=0$ initial surface.

For the Euclidean Rindler problem, we instead consider linearised gravity about the background
\begin{equation}
ds^2 = z^2 d\tau^2 + dz^2 + dx^2 + dy^2~, \quad\quad 0<z<z_c\, , \quad \quad  \tau \sim \tau + 2\pi~.
\end{equation}
Due to the periodicity of $\tau$, the Euclidean counterpart of (\ref{eqn: Rindler zero mode}) is no longer regular at the Euclidean horizon. To avoid such a complication, it is useful to consider the kernel of the linearised Einstein field equation subject to the Dirichlet boundary conditions on this background without any gauge-fixing condition \cite{Witten:2018lgb}. Let us pick an arbitrary function $\zeta(\tau)$ satisfying the periodicity condition, $\zeta(\tau) = \zeta(\tau+2\pi)$. It can be shown that by perturbing the boundary so that the physical region is $0<z<z_c-\varepsilon\, z_c\partial_\tau\zeta(\tau)$, the induced metric at the boundary remains unchanged upon reparametrising $\tau$ coordinate by $\tau\to\tau+\varepsilon\,\zeta(\tau)$. As $\zeta(\tau)$ (and $\partial_\tau \zeta(\tau)$) is continuous across $\tau=2\pi$, the boundary remains smooth. As the origin, $z=0$, is unaffected by this perturbation, the space remains regular everywhere. Since there are infinitely many $\zeta(\tau)$ obeying the periodicity condition, it follows that the kernel of the linearised Einstein field equation, without fixing any gauge, is infinite-dimensional, and hence the Dirichlet problem is not elliptic.

In terms of the metric perturbation, this is equivalent to considering the following diffeomorphism,
\begin{equation}
    \xi^\mu\partial_\mu = \left(\tilde{\xi}^\tau(\tau,z) + \zeta(\tau)\right)\partial_\tau + \left(\tilde{\xi}^z(\tau,z)-z_c \partial_\tau \zeta(\tau)\right)\partial_z \, ,
\end{equation}
where $\tilde{\xi}^\tau(\tau,z)$ and $\tilde{\xi}^z(\tau,z)$ are arbitrary functions satisfying the periodicity condition and that become zero at $z=z_c$. Then the metric perturbation $h_{\mu\nu}=\nabla_\mu \xi_\nu + \nabla_\nu \xi_\mu$ becomes
\begin{equation}
    h_{\mu\nu}dx^\mu dx^\nu = 2z\left( \tilde{\xi}^z +z\partial_\tau\tilde{\xi}^\tau+(z-z_c)\partial_\tau \zeta \right)d\tau^2 + 2 \partial_z \tilde{\xi}^z dz^2+2\left(\partial_\tau \tilde{\xi}^z+z^2\partial_z \tilde{\xi}^\tau-z_c \partial_\tau^2\zeta\right)d\tau dz\,.
\end{equation}
It is straightforward to show that $h_{\mu\nu}$ at $z=z_c$ satisfies the Dirichlet boundary conditions, $\left.h_{ij}\right|_\Gamma=0$. %
At $z=0$, one can always choose $\tilde{\xi}^\tau(\tau,z)$ and $\tilde{\xi}^z(\tau,z)$ so that the perturbation is regular. As $\xi^z|_\Gamma$ is non-vanishing, the perturbation is physical and cannot be gauged away.

\textbf{Planar AdS$_4$.} We now consider perturbations about planar AdS$_4$ in the Fefferman-Graham gauge,
\begin{equation}
    h_{zz}=h_{zm}=0\, , \qquad h_{mn}=\frac{\ell^2}{z^2}\gamma_{mn} \, .
\end{equation}
We will show, shortly, that one can always go to the above gauge at the linearised level. Then, one finds that the $zz$- and $zm$-components of the Einstein field equations, \eqref{eqn: planar AdS G_zmu}, are given by
\begin{equation}\label{eqn: dummy1}
\begin{cases}
    \partial^m \partial^n \left(\gamma_{mn}-\eta_{mn} \gamma\right) + 2z^{-1}\partial_z \gamma = 0 \, , \\ 
    \partial^n \partial_z\left( \gamma_{mn}-\eta_{mn} \gamma\right) = 0 \, ,
\end{cases}
\end{equation}
where $\gamma=\eta^{mn}\gamma_{mn}$ denotes the trace part of $\gamma_{mn}$. Now we consider imposing the Dirichlet boundary conditions at $z=z_c$, 
\begin{equation}\label{eqn: dummy2}
    \left.\gamma_{mn}\right|_\Gamma = \Gamma_{mn}\,,
\end{equation}
where $\Gamma_{mn}$ is a symmetric tensor which depend on $x^m$. Combining \eqref{eqn: dummy1} and \eqref{eqn: dummy2}, we find that
\begin{equation}\label{eqn: dummy3}
\begin{cases}
    \partial^m \gamma_{mn} = \partial^m \Gamma_{mn} + \frac{1}{4}\partial_n \left(\partial^p \partial^q \Gamma_{pq}-\partial^2 \Gamma^p{}_p\right)\left(z_c^2-z^2\right) \, , \\
    \gamma = \Gamma^m{}_m + \frac{1}{4}\left(\partial^p \partial^q \Gamma_{pq} - \partial^2 \Gamma^p{}_p\right)\left(z_c^2-z^2\right) \, , 
\end{cases}
\end{equation}
Thus, at the linearised level and unlike the case for the Minkowski \cite{An:2021fcq} or the Rindler corner, there is no obstruction to the linearised Dirichlet problem about planar AdS$_4$. This is also true for the global AdS$_4$ case \cite{Andrade:2015fna} at the linearised level.

To discuss uniqueness, it is sufficient to set $\Gamma_{mn}=0$. The equations \eqref{eqn: dummy2} and \eqref{eqn: dummy3} then imply that $\gamma_{mn}$ is a transverse-traceless tensor subject to the boundary conditions $\left.\gamma_{mn}\right|_\Gamma=0$. This is precisely $\tilde{\gamma}_{mn}$ discussed in section \ref{sec: bulk planar AdS}. Upon imposing the $mn$-components of the Einstein field equations, this perturbation leads to a unique solution. In summary, this analysis shows that, by imposing Dirichlet boundary conditions \eqref{eqn: dummy2}, the linearised initial boundary value problem about planar AdS$_4$ is well-posed.

\textbf{Existence for linearised conformal boundary conditions.} Let us briefly comment on the existence property of conformal boundary conditions at the linearised level about planar AdS$_4$. We will allow for both a small perturbation in the conformal structure and in the trace of the extrinsic curvature, such that 
\begin{equation}\label{eqn: dummy4}
    \left.\gamma_{mn}-\frac{1}{3}\eta_{mn}\gamma\right|_\Gamma = \Gamma_{mn} \, , \qquad \left.z\partial_z \gamma\right|_\Gamma = 2\delta K \ell\, ,
\end{equation}
where $\Gamma_{mn}$ is a symmetric-traceless tensor, and $\delta K$ is a scalar. Both depend only on $x^m$. One can show that the equations \eqref{eqn: dummy1} together with \eqref{eqn: dummy4} imply that
\begin{equation}\label{eqn: dummy5}
\begin{cases}
    \partial^m \gamma_{mn} = \frac{1}{3}\partial_n \varphi + \partial^m \Gamma_{mn} + \frac{\partial_n \delta K\ell}{z_c^2}\left(z^2-z_c^2\right) \, , \\
    \gamma = \varphi + \frac{\delta K \ell}{z_c^2}\left(z^2-z_c^2\right) \, , \\
    \partial^2 \varphi = \frac{6\delta K \ell}{z_c^2}+ \frac{3}{2}\partial^m \partial^n \Gamma_{mn} \, .
\end{cases}
\end{equation}
The first two equations fix completely $\partial^m \gamma_{mn}$ and $\gamma$ in terms of the boundary data $\Gamma_{mn}$ and $K$ and a $z$-independent scalar $\varphi$, which satisfies the last equation. 
These equations do not impose any constraints on the allowed boundary data $(\Gamma_{mn},K)$. Instead, we have a new degree of freedom, $\varphi$, whose dynamics is localised at the boundary. Note that by setting both $\Gamma_{mn}=0$ and $\delta K =0$, \eqref{eqn: dummy5} becomes  \eqref{eq: for appendix}.

\subsection{Permissibility of Fefferman-Graham gauge} \label{app: FG}

As a last comment, we would like to show that it is always possible to use allowed diffeomorphisms in planar AdS$_4$ to impose the Fefferman-Graham gauge. This follows from the setup in section \ref{sec: diff planar}. Consider a general linearised solution that is composed of both a bulk perturbation $\gamma_{\mu\nu}$, and a linearised diffeomorphism. Then, using \eqref{eqn: planar AdS xi^m} and \eqref{eqn: AdS planar sol xi^z}, the $h_{z\mu}$ components of the linearised metric, see \eqref{eqn: AdS planar metric diffeo}, become
\begin{equation}
    h_{zz} = \frac{2\ell^2}{z^2}\left(-\frac{\zeta}{z}+\partial_z\tilde{\xi}^z - \frac{\tilde{\xi}^z}{z} + \frac{\gamma_{zz}}{2}\right) \, , \qquad h_{zm} = \frac{\ell^2}{z^2}\left(\partial_m \zeta + \partial_m \tilde{\xi}^z + \eta_{mn} \partial_z \tilde{\xi}^n+\gamma_{zm}\right) \, .
\end{equation}
At this level, $\gamma_{zz}$ and $\gamma_{zm}$ can be arbitrary functions of $z$ and $x^m$. We observe that by choosing $\tilde{\xi}^\mu$ to be
\begin{equation}
\begin{cases}
    \tilde{\xi}^m(z,x^m) = \frac{z_c^2-z^2}{2z_c}\eta^{mn}\partial_n \zeta(x^m)-\int^z_{z_c}dz'\,\eta^{mn}\gamma_{zn}(z',x^m)+\int^z_{z_c}dz' z' \int^{z'}_{z_c} \frac{dz''}{2z''}\,\partial^m \gamma_{zz}(z'',x^m) \, , \\
    \tilde{\xi}^z(z,x^m) = \left(\frac{z}{z_c}-1\right)\zeta(x^m)-z \int_{z_c}^z \frac{dz'}{2z'}\,\gamma_{zz}(z',x^m) \, ,
    \end{cases}
\end{equation}
the metric components $h_{zz}$ and $h_{zm}$ vanish everywhere on $z\geq z_c$, including at the timelike boundary.

\section{Near AdS$_4$ timelike boundary} \label{app: near infinity}

In this appendix we collect some useful formulae regarding the intrinsic/extrinsic geometry of a timelike boundary located near the conformal boundary of an asymptotically AdS$_4$ spacetime. 

Recall first that near the infinity, one can write the metric as
\begin{equation}
    \frac{ds^2}{\ell^2} = d\rho^2 + \frac{e^{2\rho}}{\ell^2}\left(g^{(0)}_{mn} + \ell^2e^{-2\rho}g^{(2)}_{mn} + \ell^3 e^{-3\rho} g^{(3)}_{mn}  + \mathcal{O}(e^{-4\rho}) \right)dx^m dx^n \, ,
\end{equation}
where $g^{(2)}_{mn}$ is minus the Schouten tensor of $g_{mn}^{(0)}$, see (\ref{schouten}), and $g_{mn}^{(3)}$ is a transverse-traceless tensor, also with respect to $g_{mn}^{(0)}$. Let us consider placing a timelike boundary $\Gamma$ at $\rho = \rho_c \gg 1$. The induced metric then reads
\begin{equation}\label{gmn FG gauge}
    \left.g_{mn}\right|_{\Gamma}=e^{2\rho_c}\left(g^{(0)}_{mn} + \ell^2e^{-2\rho_c}g^{(2)}_{mn} + \ell^3e^{-3\rho_c} g^{(3)}_{mn} + %
    + \mathcal{O}(e^{-4\rho_c}) \right) \, ,
\end{equation}
with the inverse metric given by
\begin{eqnarray}
    \left.g^{mn}\right|_{\Gamma}= e^{-2\rho_c}\left(g^{(0)mn}-\ell^2 e^{-2\rho_c}g^{(2)mn}-\ell^3e^{-3\rho_c}g^{(3)mn} + %
      \mathcal{O}(e^{-4\rho_c}\right) \, .
\end{eqnarray}

Using $K_{mn} = \frac{1}{2\ell}\partial_\rho g_{mn}$, the extrinsic curvature is given by
\begin{equation}\label{Krho}
    \left.K_{mn}\right|_{\Gamma} = \frac{ e^{2\rho_c}}{\ell}\left(g^{(0)}_{mn} - \frac{1}{2}\ell^3e^{-3\rho_c}g_{mn}^{(3)} %
    + \mathcal{O}(e^{-4\rho_c})\right) \, .
\end{equation}
It follows that the trace of the extrinsic curvature is given by
\begin{equation}\label{Krhoc}
    \left.K\ell\right|_{\Gamma} = 3 -\ell^2 e^{-2\rho_c}\text{tr}(g^{(2)}_{mn}) +%
    \mathcal{O}(e^{-4\rho_c}) \, ,
\end{equation}
where we have used the fact that $\text{tr}g^{(3)}=0$. One straightforwardly obtains $\text{tr} \, g_{mn}^{(2)} = -R[g^{(0)}_{mn}]/4$. To compute the conformal Brown-York tensor (\ref{cby}), let us choose the representative of the conformal class of the induced metric to be
\begin{equation}\label{bar g}
    \bar{g}_{mn} = e^{-2\rho_c}\left.g_{mn}\right|_{\Gamma} = g^{(0)}_{mn} +\ell^2 e^{-2\rho_c}g^{(2)}_{mn} +\ell^3 e^{-3\rho_c} g^{(3)}_{mn} %
    + \mathcal{O}(e^{-4\rho_c}) \, ,
\end{equation}
so that the Weyl factor is given by $e^\bomega = e^{\rho_c}$. Plugging \eqref{gmn FG gauge}, \eqref{Krho}, \eqref{Krhoc}, and $e^\bomega = e^{\rho_c}$ into (\ref{cby}), the conformal Brown-York tensor can be written in the large $\rho_c$ expansion as
\begin{equation}\label{TmnFG}
    T_{mn} = e^{\rho_c} T^{(-1)}_{mn} + T^{(0)}_{mn} %
    + \mathcal{O}(e^{-\rho_c}) \, ,
\end{equation}
where the term is explicitly given by
\begin{equation}\label{TmnFG2}
    \begin{cases}
        T^{(-1)}_{mn} = \frac{\ell}{8 \pi G_N}\left(g^{(2)}_{mn} - \frac{1}{3}\text{tr}(g^{(2)}_{mn} )g^{(0)}_{mn}\right)\,, \\
        T^{(0)}_{mn}  = \frac{3\ell^2}{16 \pi G_N}g^{(3)}_{mn}\, , %
    \end{cases}
\end{equation}
By construction, $T_{mn}$ is traceless with respect to the induced metric $\left.g_{mn}\right|_{\Gamma}$. %
Provided that $\text{tr}(g^{(2)}_{mn})$ is non-vanishing, we can invert \eqref{Krhoc} to write $e^{\rho_c}$ in a small $(K\ell -3)$ expansion,
\begin{equation}\label{eqn: rho_c as Kell-3}
    \frac{e^{\rho_c}}{\ell} = \sqrt{\frac{-\text{tr}(g^{(2)}_{mn})}{K \ell-3}} %
    + \mathcal{O}(1) \,.
\end{equation}
This equation allows us to rewrite the large $\rho_c$ expansion in terms of the small $(K\ell-3)$ expansion. As an example, the induced metric \eqref{gmn FG gauge} to leading order can be written as
\begin{equation}
    \left.g_{mn}\right|_{\Gamma} = \frac{-\ell^2\text{tr}(g^{(2)}_{mn})}{K\ell-3}g^{(0)}_{mn} + \mathcal{O}(K\ell-3)^{-1/2}\,.
\end{equation}
Finally, provided $\text{tr}(g^{(2)}_{mn})$ is non-vanishing and given (\ref{eqn: rho_c as Kell-3}), we can express the Brown-York stress energy tensor in terms of the conformal boundary data to get the result shown in \eqref{eq: brown-york near boundary}.

\section{Spherically symmetric diffeomophisms on the AdS$_4$ black hole} 

In this appendix we consider the physical diffeomorphism with $l=0$ about the AdS$_4$ black hole background.

\subsection{Modes with $l=0$ in the Fefferman-Graham gauge} \label{app: FG l=0}

Here we extend the result in section \ref{l0modes} to the Fefferman-Graham gauge. To do so, we first recall that the most general spherically symmetric solution in the Fefferman-Graham gauge can be written as
\begin{equation}\label{eqn: spherically sym FG gauge}
    ds^2 = \frac{\ell^2dz^2}{z^2}-F(t,z)dt^2 + H(t,z)\ell^2 d\Omega^2_2\,,
\end{equation}
where $F(t,z)$ and $H(t,z)$ are arbitrary functions of $t$ and $z$. For the AdS$_4$ black hole solution, these functions are given by
\begin{equation}\label{eqn: AdS bh FG gauge}
    \begin{cases}
        F=\bar{F}(z)\equiv\frac{\ell^2}{z^2}+\frac{1}{2}-\frac{4G_NMz}{3\ell^2}+\frac{z^2}{16\ell^2}+...\,, \\
        H=\bar{H}(z)\equiv\frac{\ell^2}{z^2}-\frac{1}{2}+\frac{2G_NMz}{3\ell^2}+\frac{z^2}{16\ell^2}+... \, .
    \end{cases}
\end{equation}
The Schwarzchild gauge can be obtained by defining $r=\ell\sqrt{\bar{H}(z)}$. The timelike boundary is then located at $\r=\ell\sqrt{\bar{H}(z_c)}$. The trace of the extrinsic curvature of \eqref{eqn: spherically sym FG gauge} at $z=z_c$ is given by
\begin{equation}\label{eqn: K of AdS bh}
    K \ell = \left.-\frac{z \partial_z F}{2F}-\frac{z \partial_z H}{H}\right|_\Gamma \,.
\end{equation}
This equation implies that $z_c$ is implicitly a function of $K\ell$.

Now we consider metric \eqref{eqn: spherically sym FG gauge} which differs from the AdS$_4$ black hole solution by a linearised diffeomorphism. We require that this metric, at $z=z_c$, leads to a trace of the extrinsic curvature and conformal structure of the induced metric that match those of the AdS$_4$ black hole solution, namely
\begin{equation}
    \left.ds^2\right|_\Gamma = (1+2\delta\bomega)\left(-\bar{F}(z_c) dt^2 + \bar{H}(z_c)\ell^2d\Omega^2\right) \, , \qquad \left.K \ell\right|_\Gamma = \left.-\frac{z\partial_z \bar{F}}{2\bar{F}}-\frac{z\partial_z \bar{H}}{\bar{H}} \right|_\Gamma\,,
\end{equation}
where $\delta \bomega$ is a $z$-independent function describing the perturbed Weyl factor. Taking both $\tfrac{z}{\ell}\to 0$ and $\tfrac{z_c}{\ell}\to 0$ while keeping $\tfrac{z}{z_c}$ fixed, this metric admits an expansion in $\tfrac{z}{\ell}$ as
\begin{equation}\label{eqn: AdS BH FG gauge perturbed metric}
\begin{cases}
    F(t,z) &= \frac{\ell^2}{z^2}\left(1+2\delta \bomega\right) + \frac{1}{2}\left(1+\left(\frac{2z_c^2}{z^2}-1\right)2\delta\bomega\right) - \frac{4G_N M z}{3\ell^2}\left(1+\left(\frac{3z_c^3}{z^3}-1\right)\delta \bomega\right)+\mathcal{O}\left(\frac{z^2}{\ell^2}\right) \, ,  \\
    H(t,z) &= \frac{\ell^2}{z^2}\left(1+2\delta \bomega\right) - \frac{1}{2}\left(1+\frac{z_c^2}{z^2}2\delta \bomega\right) + \frac{2G_NMz}{3\ell^2}\left(1+\left(\frac{3z_c^3}{z^3}-1\right)\delta \bomega\right)+\mathcal{O}\left(\frac{z^2}{\ell^2}\right) \, .
    \end{cases}
\end{equation}
The perturbed Weyl factor $\delta \bomega$ satisfies the equation
\begin{equation}
    \ell^2\partial_t^2\delta\bomega =\left(1+\mathcal{O}\left(\frac{z_c^2}{\ell^2}\right)\right)\delta \bomega \, .
\end{equation}
In these equations, $z_c$ can be written perturbatively in terms of $K\ell-3$ as
\begin{equation}
    \frac{z_c}{\ell} = \sqrt{2(K\ell-3)} + \mathcal{O}(K\ell-3)^{3/2} \, .
\end{equation}
Setting $K\ell=3$ ($z_c=0$) and comparing with the standard Fefferman-Graham expansion, we observe that the Weyl mode, $\delta \bomega$, contributes to the $g^{(0)}_{mn}$ and $g^{(3)}_{mn}$ as
\begin{equation} \label{eq: lin g0 g3}
    g^{(0)}_{mn}= \bar{g}^{(0)}_{mn}\left(1+2\delta \bomega\right) \, , \qquad g^{(3)}_{mn}= \bar{g}^{(3)}_{mn}\left(1-\delta \bomega\right) \, ,
\end{equation}
where $\bar{g}^{(n)}_{mn}$ are those of the AdS$_4$ black hole solution. This is indeed the behavior of $g^{(0)}_{mn}$ and $g^{(3)}_{mn}$ obtained under a Fefferman-Graham gauge preserving diffeomorphism (see for instance (2.6) and (2.7) of \cite{Anninos:2010zf}).

\subsection{Dressing the global AdS$_4$ black hole} \label{app: bh}

We now perform a similar analysis to the one in section \ref{sec: non-linear diffeos}, but now for the global AdS$_4$ black hole. The background metric corresponds to 
\begin{equation}
ds^2 = -f(r) dt^2 + \frac{dr^2}{f(r)} + r^2 d\Omega_2^2 ~, \quad \text{with} \quad f(r) = 1- \frac{2 G_N M}{r} + \frac{r^2}{\ell^2} \,.
\end{equation}

The case of a boundary at a constant $r$ slice has been analysed above in section \ref{sec: conf by}. Here we would like to identify a hypersurface in the black hole geometry which obeys the conformal boundary conditions with constant $K$, and induced metric
\begin{equation} \label{eq: bh bdy metric}
ds^2 |_\Gamma =  e^{{2\delta \bomega(u)}} \left(- du^2 + \r^2 \, d\Omega_2^2 \right)~.
\end{equation}
Note that for convenience, we are rescaling the time coordinate with respect to \eqref{indmetric} and that $\r$ is fixed and part of the conformal boundary data. Requiring that $K$ is constant provides a non-linear differential equation for $\delta \bomega(u)$,
\begin{equation} \label{eq: brane dynamics bh}
    \r^2 \partial^2_u \delta \bomega(u) = K \ell e^{\frac{\delta\bomega}{2}} \sqrt{-\frac{2 \r G_N M}{\ell^2}+ \frac{\r^2}{\ell^2}  e^{\delta\bomega} \left((\r\partial_u \delta\bomega)^2+1\right)+\frac{\r ^4}{\ell^4} e^{3 \delta\bomega}}+ \frac{3 G_N M e^{-\delta\bomega}}{\r}-\frac{3 \r ^2 e^{2 \delta\bomega}}{\ell^2} - 2-2 (\r \partial_u \delta\bomega )^2 \,.
\end{equation}
Upon setting $M=0$ (and suitably rescaling time), this equation reduces to \eqref{eqn: brane eqn}. If $\delta\bomega$ is time independent, then this equation gives the trace of the extrinsic curvature of a constant $r=\r \, e^{\delta\bomega}$ slice of the black hole geometry,
\begin{equation}
    K \ell = \frac{e^{-\frac{3}{2} \delta\bomega} \left(-3 G_N \ell^2 M +2 \ell^2 \r  e^{\delta\bomega}+3 \r ^3 e^{3 \delta\bomega}\right)}{\r ^{3/2} \sqrt{-2 G_N \ell^2 M+\ell^2 \r  e^{\delta\bomega}+\r ^3 e^{3 \delta\bomega}}} \,.
\end{equation}
When $\r = \ell$ this is the $K\ell$ that appears in \eqref{omegaM}. Another interesting limit is to keep $M$ finite and linearise $\delta\bomega (u)$. In that case we obtain,
\begin{equation}\label{omegabh}
  \r^2  \partial_u^2 \delta \bomega (u) = \Omega^2 \delta \bomega (u)~, \quad \text{with} \quad \Omega \equiv \frac{\left(9 G_N^2 M^2 \ell^2 -8 G_N M \ell^2 \r +2 \ell^2 \r ^2+\r ^4\right)}{ \r^4 +\ell^2 \r^2 -2 G_N M \ell^2 \r} \,,
\end{equation}
which has solutions of the form $\delta \bomega (u) = e^{\pm \Omega  u}$. The second order nature of (\ref{omegabh}) indicates that, as in our previous discussions, the must fix the dynamical data $\mathcal{C}_{\partial \Sigma} = \{\delta\bomega, \partial_u\delta\bomega \}|_\Gamma$ to have a complete specification of the initial data. By setting $M=0$, we retrieve the exponentially growing modes of section \ref{l0modes} (appropriately rescaled by $f(\r)$ with $M=0$).

Next we can compute the conformal Brown-York stress tensor (\ref{cby}), that reads
\begin{equation} \label{eq: cby bh}
T_{uu} = \frac{2}{\ell^2}T_{\theta \theta} = \frac{2 }{\ell^2\sin^2{\theta}}T_{\phi \phi} = \frac{K \ell e^{2 \delta\bomega}-3 e^{\frac{\delta\bomega}{2}} \sqrt{-\frac{2 G_N \ell^2 M}{\r^3} + \frac{\ell^2}{\r^2}  e^{\delta\bomega} \left((\r \partial_u \delta\bomega)^2+1\right)+ e^{3 \delta\bomega}}}{12 \pi  G_N \ell}\,.
\end{equation}
Upon imposing \eqref{eq: brane dynamics bh}, this stress tensor is not only traceless, but also conserved, with respect to the metric \eqref{eq: bh bdy metric}.

\noindent \textbf{Asymptotic boundary limit.} We would like to analyse this setup as the timelike boundary $\Gamma$ gets near the conformal boundary of the AdS$_4$ black hole. In that limit, as in empty AdS, one can check that $K \ell \approx 3$, so one can consider the solution in an expansion for small $\delta K \ell \equiv K \ell - 3$. Note from \eqref{eq: brane dynamics bh} in the static limit, that $e^{\delta \bomega(u)}$ should diverge as $(K \ell -3)^{-1/2}$. 

In order to compare with the standard Fefferman-Graham expansion (where the conformal factor is constant and large), it is convenient to isolate this divergence in the conformal factor of the metric, such that,
\begin{equation}
    ds^2|_{\Gamma} = \frac{1}{K \ell -3} \left( e^{{2\delta \bomega(u)}} \left(- du^2 + \r^2 \, d\Omega_2^2 \right) \right) \,.
\end{equation}
On top of rescaling $\delta \bomega$ in \eqref{eq: cby bh}, this transformation adds an overall scaling to the stress tensor, $T_{mn} \to \frac{T_{mn}}{\sqrt{K \ell -3}}$. After this, $\delta \bomega$ needs to be finite in the $\delta K \ell \to 0$ limit, and we obtain that
\begin{eqnarray}
    e^{\delta\bomega(u)} = e^{\delta\bomega_0 (u)} + e^{\delta\bomega_1 (u)} \sqrt{\delta K \ell} + %
    \mathcal{O} \left(\delta K \ell \right) \,. 
\end{eqnarray}
Solving \eqref{eq: brane dynamics bh} order-by-order in $\delta K \ell$ gives dynamical equations for all the $\delta \bomega_n$. For brevity, here we report the first two equations,
\begin{equation}
    \begin{cases}
       \r^2 \partial_u^2 \delta\bomega_0 = e^{2 \delta \bomega_0} \frac{\r^2}{\ell^2}-\frac{1}{2}-\frac{1}{2} (\r \partial_u \delta\bomega_0)^2 \, ,\\
     \r^2 \partial_u^2  \delta\bomega_1 = \delta\bomega_1 \left(3 e^{2 \delta\bomega_0}\frac{\r^2}{\ell^2}-\frac{1}{2}-\frac{1}{2}(\r\partial_u \delta\bomega_0)^2\right)+ \r^2 (\partial_u \delta\bomega_0) (\partial_u \delta\bomega_1) \,. 
    \end{cases} \label{eq: bomega expansions}
\end{equation}
As seen in previous examples, these equations will be completely determined after we supplement initial data for $\delta \bomega$ and $\partial_u \delta \bomega$ at the corner $\partial \Sigma$.

We can also expand the Brown-York stress tensor \eqref{eq: cby bh} close to the boundary, that after the rescaling becomes
\begin{equation}
 T_{uu} = \frac{T_{uu}^{(-1)}}{\sqrt{\delta K \ell}} + T_{uu}^{(0)} %
 + \mathcal{O} \left( (\delta K \ell)^{1/2} \right) \,.
\end{equation}
Upon imposing \eqref{eq: bomega expansions}, we find that
\begin{equation} \label{eq: expansion tmn}
    \begin{cases}
        T_{uu}^{(-1)} = \frac{e^{2 \delta \bomega _0}}{12 \pi  \ell G_N} - \frac{\ell}{8 \pi  \r^2 G_N} \left(1+ (\r\partial \delta \bomega _0)^2\right) \,, \\
        T_{uu}^{(0)} = \frac{e^{-\delta \bomega _0} \ell M}{4 \pi  \r^3} + \frac{e^{-\delta \bomega _0} \left(2 e^{2 \delta \bomega _0} \delta \bomega _1+3 \ell^2 \partial_u \delta \bomega _0 \left(\partial_u \delta \bomega _0 \delta \bomega _1-\partial_u \delta \bomega _1\right)\right)}{12 \pi  \ell G_N} \,.
    \end{cases}
\end{equation}
The leading contribution is divergent in the strict $K\to \tfrac{3}{\ell}$ limit. This expression for the stress tensor near the boundary is compatible with the one in the main text, see \eqref{eq: brown-york near boundary}, upon identifying $g_{mn}^{(0)}dx^m dx^n$ with $e^{2\delta\bomega} \left(-du^2 + \r^2 d\Omega_2^2\right)$. Recall that $e^{\delta \bomega} = e^{\delta \bomega_0} + e^{\delta\bomega_1} \delta K \ell + \cdots$. Then the leading contribution to the stress tensor will come exclusively from $\delta \bomega_0$ and perfectly matches the divergent term in \eqref{eq: brown-york near boundary}.

The next term $T_{uu}^{(1)}$ receives contributions from two different sources. The first term is proportional to the mass of the black hole $M$. The scaling with $e^{-\delta\bomega_0}$ is compatible with the expected scaling for $g^{(3)}_{mn}$, see also the linearised analysis in \eqref{eq: lin g0 g3}. The second term is exactly the first subleading contribution that comes from the leading term in \eqref{eq: brown-york near boundary} upon expanding $\delta\bomega$.

One can continue this analysis order-by-order systematically, finding that when the timelike boundary is sufficiently away from the conformal boundary of AdS$_4$, $M$ and $\delta \bomega$ no longer decouple.

\section{Locally diffeomorphic modes for any $l$} \label{sec: phys_diffeos_l}

In this appendix we show that there are no physical diffeomorphisms for $l\geq 2$ satisfying conformal boundary conditions (\ref{gbar_global}) and (\ref{eq: K function of r}) about global AdS$_4$ (\ref{globalAdS4}). Consider a general-$l$ metric perturbation in the scalar sector, which is locally a diffeomorphism, $ h_{\mu\nu} = \nabla_\mu \xi_\nu + \nabla_\nu \xi_\mu $. Then we can parameterise the vector field $\xi^\mu$,
\begin{equation}
    \xi^r = r \zeta_r \,\mathbb{S}_l \, , \qquad \xi^t = r^{-2} \zeta_t \, \mathbb{S}_l \, , \qquad \xi^\theta = r^{-2} \zeta_\sigma \, \partial_\theta \mathbb{S}_l \, , \qquad \xi^\phi = r^{-2} \zeta_\sigma \, \frac{\partial_\phi \mathbb{S}_l}{\sin^2\theta} \, ,
\end{equation}
for some $(t,r)$-dependent functions $\zeta_r$, $\zeta_t$, and $\zeta_\sigma$. %

Now we consider inserting a timelike boundary at $r=\r$. For a generic $\r$, the induced metric and the trace of the extrinsic curvature at the boundary are given by
\begin{eqnarray}
    \left.ds^2\right|_{\Gamma} &=& -\left(1+\frac{\r^2}{\ell^2} + \varepsilon\,2\left(\frac{\r^2}{\ell^2}\zeta_r + \left(\frac{1}{\ell^2}+\frac{1}{\r^2}\right)\partial_t \zeta_t\right)\,\mathbb{S}_l\right)dt^2 + 2 \left(\partial_t \zeta_\sigma-\left(\frac{1}{\ell^2}+\frac{1}{\r^2}\right)\zeta_t \right)dt d\mathbb{S}_l \nonumber \\
    & & + \left(\r^2 + \varepsilon\, 2 \r^2\zeta_r\,\mathbb{S}_l\right)d\Omega^2 + \varepsilon\,2\zeta_\sigma \tilde{\nabla}_i\tilde{\nabla}_j \mathbb{S}_l\, d\Omega^i d\Omega^j \,\Bigg|_\Gamma\, , \label{eq: diffeo_any_d_metric}\\
    \left.K\ell\right|_{\Gamma} &=& \frac{2\ell^2+3\r^2}{\ell \r\sqrt{1+\frac{\r^2}{\ell^2}}} + \varepsilon\,\frac{\ell \, \mathbb{S}_l}{\r \left(1+\frac{\r^2}{\ell^2}\right)^{3/2}}\left(\r^2 \partial_t^2 \zeta_r + \left(\frac{\r^2}{\ell^2}(l^2+l-1)+(l^2+l-2)\right)\zeta_r\right)\Bigg|_\Gamma \, .
\end{eqnarray}
By requiring the perturbed induced metric \eqref{eq: diffeo_any_d_metric} to be conformal to \eqref{gbar_global}, we obtain that all three $\zeta_\sigma, \zeta_t,$ and $\zeta_r$ should identically vanish. So, in general, there are no physical diffeomorphisms in the scalar sector for $l\geq 2$. One can show, also, that diffeomorphisms with $l\ge 2$, built from vectorial spherical harmonics $\mathbb{V}_i$ are never physical.

Nonetheless, a notion of approximate diffeomorphisms emerges when we consider $\r \gg \ell$. In that case, by keeping $\zeta_r$, $\zeta_t$, and $\zeta_\sigma$ finite, the induced metric and the trace of the extrinsic curvature to first non-trivial order in the perturbative parameter are given by
\begin{eqnarray}\label{eqn: large r global ads induced metric}
    \left.ds^2\right|_\Gamma &=& \left.\left(1+ \varepsilon\, 2 \zeta_r \, \mathbb{S}_l\right)\left(- \frac{\r^2}{\ell^2}dt^2 + \r^2 d\Omega^2\right)\right|_\Gamma + \mathcal{O}\left(1 \right) \, , \\ \left.K\ell\right|_\Gamma &=& 3 + \left.\frac{\ell^2}{\r^2}\left(\frac{1}{2}+ \varepsilon \, \Big(\ell^2 \partial_t^2 \zeta_r + \left(l^2+l-1\right)\zeta_r\Big)\,\mathbb{S}_l\right)\right|_\Gamma + \mathcal{O}\left({\r^{-4}}\right) \, .\label{eqn: large r global ads trace of K}
\end{eqnarray}
It follows that the induced metric is automatically conformally equivalent to the cylinder in this limit. Note that $\zeta_t$ and $\zeta_\sigma$ do not appear in this analysis. Requiring that $\xi^\mu$ preserves the trace of the extrinsic curvature, we obtain a differential equation for $\zeta_r$ at the boundary,
\begin{equation}
  \left.  \partial_t^2 \zeta_r + \left(\frac{l(l+1)-1}{\ell^2}\right) \zeta_r \right|_\Gamma = 0 \, .
\end{equation}
Thus, we can write the general diffeomorphism that preserves conformal boundary conditions as
\begin{equation}
    \xi^r(t,\r,\theta,\phi) = \sum_{l\in\mathbb{N}^0}\r \left(e^{- i \omega^{(l)} t}f_{+,l}(\r) + e^{i \omega^{(l)} t}f_{-,l}(\r)\right) \mathbb{S}_l \, ,
\end{equation}
where 
\begin{equation}\label{masslessdisp}
    \omega^{(l)} \ell = \sqrt{l(l+1)-1} \,,
\end{equation} 
and $f_{+,l}(\r)$ and $f_{-,l}(\r)$ are constants of integration. Since the orthogonal component $\xi^r$ is non-vanishing on the boundary, this solution is physical and cannot be gauged away. Note that these frequencies coincide with the lowest frequencies found in section \ref{sec: bulk modes}.

From \eqref{eqn: large r global ads induced metric}, we can also compute the perturbation of the Weyl factor at the boundary,
\begin{equation}\label{eqn: any l diff sol}
    \delta \boldsymbol{\omega} |_{r=\r} = \varepsilon \,  \sum_{l\in\mathbb{N}} \left(e^{- i \omega^{(l)} t}f_{+,l}(\r) + e^{i \omega^{(l)} t}f_{-,l}(\r)\right) \mathbb{S}_l \, .
\end{equation}
For $l\geq 1$, these solutions do not exhibit exponential growth at late times. Only the $l=0$ sector leads to an exponentially growing solution.

We can also compute the conformal Brown-York stress-energy tensor for these near-diffeomorphic modes, for which we obtain,
\begin{equation} \label{conformal BY near diffeos}
    T_{mn} = \left.\frac{\ell}{8\pi G_N}\left(\frac{\bar{R}}{3} \bar{g}_{mn}-\bar{R}_{mn}+\left(\bar{\mathcal{D}}_m \bar{\mathcal{D}}_n+\frac{\bar{R} }{2}\bar{g}_{mn} -\bar{R}_{mn}\right) \delta \bomega\right)\right|_\Gamma \,,
\end{equation}
where $\bar{\mathcal{D}}_m$ and $\bar{R}_{mn}$ are the covariant derivative and Ricci tensor with respect to the metric
\begin{equation}
ds^2=-\frac{\r^2}{\ell^2}dt^2+\r^2 d\Omega_2^2~.
\end{equation}
One can readily confirm that $T_{mn}$ is traceless and covariantly conserved upon imposing \eqref{eqn: any l diff sol}.

Computing the subleading correction reveals that the large $\r$ approximation should breakdown when $\omega^{(l)}  \sim \mathcal{O}(\r)$. In terms of $K\ell$, we have $\tfrac{\ell}{\r} \sim \sqrt{K\ell-3}$, such that the above analysis breaks down when $\omega^{(l)} \ell \sim (\sqrt{K\ell-3})^{-1/2}$.



\section{Complex frequencies at large $l$} \label{app: large l}

In section \ref{sec: bulk modes}, we found that the behavior of the complex frequencies depended on the position of the boundary given by $K \ell$. When $K\ell \to \infty$, the complex frequencies at large $l$ asymptote the flat space behavior $\omega \r \approx \pm l \pm i 0.34 l^{1/3}$ \cite{Anninos:2023epi,Liu:2024ymn}. When $K\ell \to 3$, we used the WKB approximation to show that there were no complex frequencies. In this appendix, we provide numerical evidence of the intermediate regime.

For any fixed $K\ell$ and $l$, we numerically search for the complex frequencies. Then we vary $l$ for large values in between $l=30$ and $l=120$ and we keep track of the values of the complex frequencies. We find reasonable to assume the following ansatz for the frequencies at large $l$,
\begin{equation}
    \omega \ell (l) \approx \pm \, \omega_{\text{Re}} \, l^{\alpha_\text{Re}} \pm  i \, \omega_{\text{Im}} \, l^{\alpha_{\text{Im}}} \,. 
\end{equation}

\begin{figure}[h!]
        \centering
         \subfigure[$\alpha_{\text{Re}}$]{
                \includegraphics[scale=0.55]{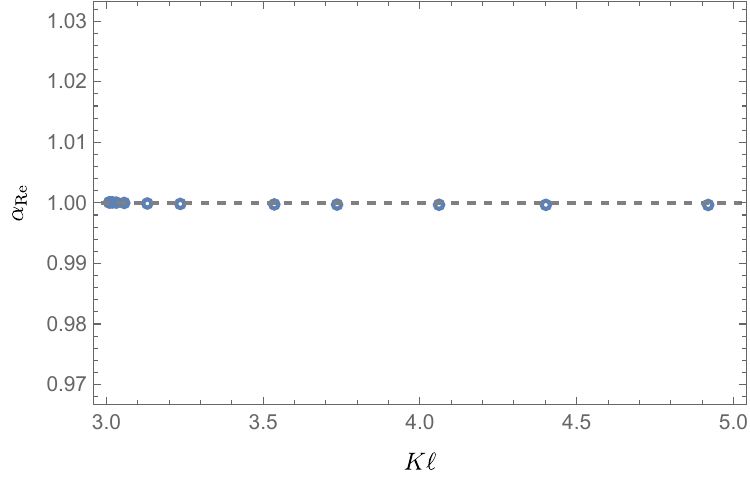}\label{fig: appA2}}  \quad
                 \subfigure[$\omega_{\text{Re}}$]{
                \includegraphics[scale= 0.55]{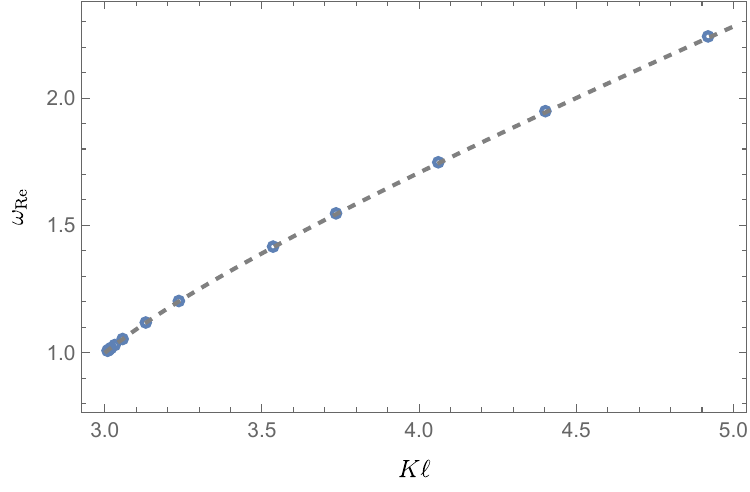} \label{fig: appB2}}                           
                \caption{Behavior of the real part of the complex frequencies at large $l$. In (a), in dashed gray, we show $\alpha_{\text{Re}}=1$. In (b), we show in dashed gray the analytic result \eqref{eq: app re}, which perfectly matches the numerical results for all the values of $K\ell$ analysed.} \label{fig: Re largel}
\end{figure}

The real part behaves in a universal manner, finding that $\alpha_{\text{Re}} \approx 1$ for any $K \ell$. We provide a numerical plot for both $\alpha_{\text{Re}}$ and $\omega_{\text{Re}}$ in figure \ref{fig: Re largel}. It is interesting to note that $\alpha_{\text{Re}}$ and $\omega_{\text{Re}}$ can be obtained analytically by doing a large $l$ expansion of \eqref{eqn: deltaK scalar}. The hypergeometric function can be approximated using a steepest descend method \cite{paris1,paris2}, after which we obtain $\alpha_{\text{Re}}=1$ and,
\begin{equation} \label{eq: app re}
    \omega_{\text{Re}} = \sqrt{\frac{K\ell}{8} \left(\sqrt{K^2\ell^2-8}+K\ell\right)-\frac{1}{2}} = \begin{cases}
        1 + (K\ell - 3) + \mathcal{O} (K\ell-3)^2 \, , \, \text{for} \, K\ell \to 3 \,, \\
        \frac{K\ell}{2}-\frac{1}{K\ell}+\mathcal{O}\left(\frac{1}{K\ell}\right)^3 \, ,  \, \text{for} \, K\ell \to \infty \,.
        \end{cases}
    \end{equation}

In the limit of $K\ell \to \infty$ this agrees with the flat space result. Moreover, it agrees with the numerical results for all values of $K \ell$ analysed, as seen in figure \ref{fig: appB2}.

\begin{figure}[h!]
        \centering
         \subfigure[$\alpha_{\text{Im}}$]{
                \includegraphics[scale=0.55]{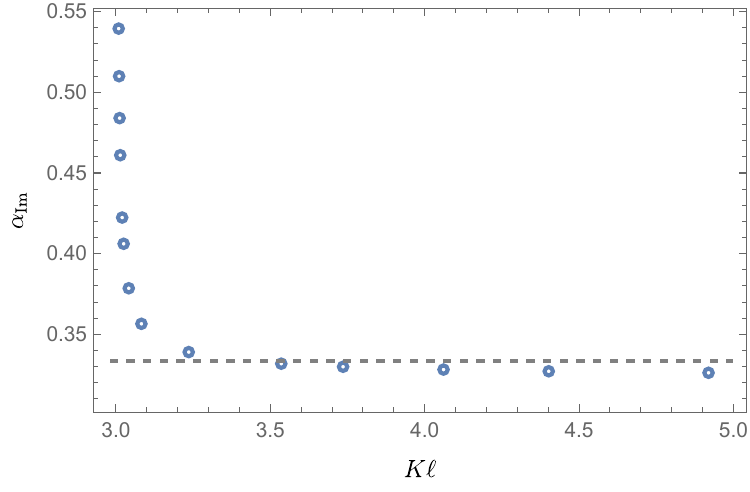}\label{fig: appA}}  \quad
                 \subfigure[$\omega_{\text{Im}}$]{
                \includegraphics[scale= 0.55]{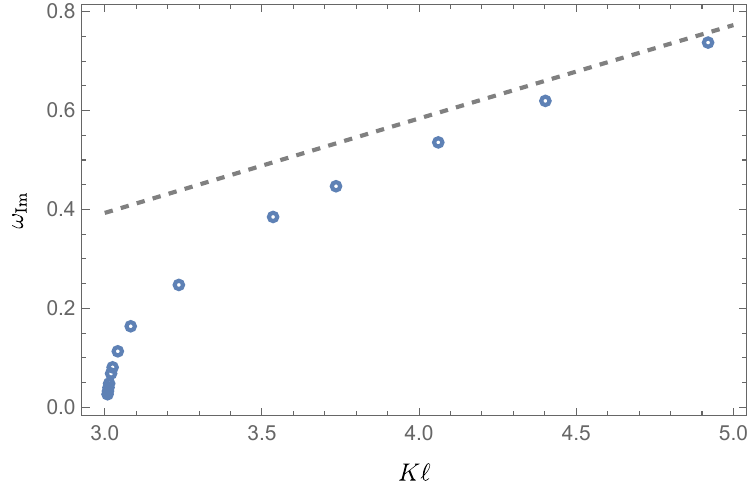} \label{fig: appB}}                           
                \caption{Behavior of the imaginary part of the complex frequencies at large $l$. In dashed gray, we show the flat space asymptotic value at large $l$. In (a), the small discrepancy at large $K\ell$ can be atributed to numerical error, while in (b), the coefficient becomes linear at large $K\ell$ due to the fact that frequencies scale with $\r$ instead of $\ell$ in the flat space limit.} \label{fig: Im largel}
\end{figure}

In contrast, the imaginary part behaviour appears to strongly depend on the position of the boundary. In figure \ref{fig: Im largel}, we plot both $\omega_{\text{Im}}$ and $\alpha_{\text{Im}}$ for a range of $K \ell$. Note that while $\alpha_{\text{Im}}$ grows as $K\ell \to 3$, $\omega_{\text{Im}}$ decays. One can check that, at least for the values of $K\ell$ and $l$ analysed, the full imaginary part $\omega_{\text{Im}} \, l^{\alpha_{\text{Im}}}$ does not seem to grow larger than the real part as the boundary moves towards the boundary of AdS$_4$, consistent with the WKB analysis of section \ref{sec: bulk modes}. It would be desirable to get an analytical understanding of the behavior of these frequencies for any value of $K \ell$, at large $l$. Also notice that these frequencies do not appear in flat AdS$_4$. In that case,  the reason might be that $K \ell$ is fixed to $K \ell =3$. It would be interesting to see if the complex frequencies appear as perturbations about the black brane geometry, where $K \ell$ is allowed to change. We leave these for future work.

\section{Bulk perturbations in planar AdS$_4$} \label{app: planar perturbation}

In this appendix we give an explicit derivation of the expression for the bulk metric perturbation \eqref{eq: bulk modes planar} about planar AdS$_4$. To find a solution for $\tilde{\gamma}_{mn}$, it is useful to consider the mode expansion
\begin{equation}
    \tilde{\gamma}_{mn}(z,x^m) = \sum_{j=1}^2 \beta^{(j)}_{mn}\varphi(z)e^{ik_mx^m} \, ,
\end{equation}
where $\beta^{(j)}_{mn}$ represent the two transverse-traceless polarisations of gravity, and $\varphi(z)$ is a solution to
\begin{equation}\label{eqn: planar AdS eqn for phi}
    q^2 \varphi + z^2 \partial_z \left(z^{-2}\partial_z \varphi\right) = 0 \, , \qquad q \equiv \sqrt{-k^m k_m} \, .
\end{equation}
The conformal boundary condition simplifies to 
\begin{equation}\label{eqn: planar AdS bdry cond for phi}
    \varphi(z_c)=0 \, .
\end{equation}
The asymptotic behaviour of $\tilde{\gamma}_{mn}$ or equivalently $\varphi$ as $z\to\infty$ depends on the value of $q$. Let us consider separately the case where $q =0$, $q$ is a positive imaginary number, and $q$ is a positive real number.

\textbf{Polynomial solutions.} Let us first consider the case where $q=0$. This is equivalent to making $k^m$ a null vector. The solution to \eqref{eqn: planar AdS eqn for phi} is given by
\begin{equation}
    \varphi(z) = c_1 z^3 + c_2 \, ,
\end{equation}
where $c_1$ and $c_2$ are arbitrary constants. Requiring that $h_{mn}$ is finite as $z\to \infty$ fixes $c_1 = 0$. The boundary condition \eqref{eqn: planar AdS bdry cond for phi} then sets $c_2 = 0$. Hence, there is no $\tilde{\gamma}_{mn}$ with $q =0$ that obeys the conformal boundary conditions.

\textbf{Exponentially growing/decaying solutions.} Consider the case where $\alpha\equiv iq \in \mathbb{R}_{>0}$. This is equivalent to make $k^m$ a spacelike vector. The solution to \eqref{eqn: planar AdS eqn for phi} is given by
\begin{equation}
    \varphi(z) = c_1 (1+\alpha z)e^{-\alpha z} + c_2 (1-\alpha z)e^{\alpha z} \, ,
\end{equation}
where $c_1$ and $c_2$ are arbitrary constants. Requiring that $h_{mn}$ is finite as $z\to\infty$ fixes $c_2 = 0$. The boundary condition \eqref{eqn: planar AdS bdry cond for phi} sets also $c_1=0$. This is similar to the case where $q=0$, namely there is no $\tilde{\gamma}_{mn}$ with purely imaginary $q$ that obeys the conformal boundary conditions.

\textbf{Oscillating solutions.} Consider the case where $q \in \mathbb{R}_{>0}$. This is equivalent to making $k^m$ a timelike vector. The solution to \eqref{eqn: planar AdS eqn for phi} is given by
\begin{equation}
    \varphi(z) = c_1 (1+i q z)e^{-i q z} + c_2 (1-iq z)e^{i q z} \, ,
\end{equation}
where $c_1$ and $c_2$ are real constants. As this solution is oscillating in the $z$-direction, the regularity of $h_{mn}$ as $z\to \infty$ requires certain normalisability condition on $\varphi$ after integrating the solution over $q$. Imposing the boundary condition \eqref{eqn: planar AdS bdry cond for phi}, we find
\begin{equation}
    \frac{c_1}{c_2} = -\frac{1-i q z_c}{1+i q z_c}e^{2iqz_c} \, .
\end{equation}
Therefore, the general $\tilde{\gamma}_{mn}$ that obeys the conformal boundary conditions is given by
\begin{equation}
   \tilde{\gamma}_{mn}(z,x^m) =  \text{Re} \, \int_0^\infty\frac{d q}{2\pi}\int_{\mathbb{R}^2}\frac{d^2 \bold{k}}{(2\pi)^2} \, \sum_{j=1}^2 \beta^{(j)}_{mn}(q,\bold{k}) \left(\frac{1+i q z}{1+i q z_c}e^{-iq(z-z_c)} - \frac{1-i q z}{1-i q z_c}e^{iq(z-z_c)}\right)e^{-i\omega t + i \bold{k} \bold{x}} \, ,
\end{equation}
where $\omega = \sqrt{|\bold{k}|^2+q^2}$.

\bibliographystyle{JHEP}
\bibliography{bibliography}

\end{document}